\documentclass[12pt,preprint,twocolumn]{aastex6}
\usepackage{graphicx}    
\pdfoutput=1
\usepackage[latin1]{inputenc}
\usepackage{amsmath}
\usepackage{amsfonts}
\usepackage{amssymb}
\usepackage{multirow}
\usepackage{rotating}
\usepackage{booktabs}
\usepackage{color}
\usepackage{ragged2e}
\usepackage{setspace}
\def \fps@figure{htbp}
\makeatother
\def \aap  {A\&A}

\def \apjs  {ApJS}

\def \apjl {ApJL}

\DeclareGraphicsExtensions{.ps,.pdf,.png}
\makeatletter

\def \sersic  {S\'{e}rsic}

\def \re      {$R_e$}
\def \rkron   {$R_{\rm Kron}$}

\def\oiii{[\ion{O}{3}]}
\def\hst{{\it HST}}
\def\farcs{\hbox{$.\mkern-4mu^{\prime\prime}$}}
\def\asec{$^{\prime\prime}$}

\begin{document}

\title[Host Galaxies of Obscured Quasar]{The Role of Major Mergers and Nuclear 
Star Formation in Nearby Obscured Quasars}

\author{Dongyao Zhao\altaffilmark{1}, Luis~C.~Ho\altaffilmark{1,2}, Yulin Zhao\altaffilmark{1,2}, Jinyi Shangguan\altaffilmark{1,2,3}, and Minjin Kim\altaffilmark{4,5}}
\altaffiltext{1}{The Kavli Institute for Astronomy and Astrophysics, Peking University, 5 Yiheyuan Road, Haidian District, Beijing 100871, China}
\altaffiltext{2}{Department of Astronomy, School of Physics, Peking University, 5 Yiheyuan Road, Haidian District, Beijing 100871, China}
\altaffiltext{3}{Max-Planck-Institut f\"{u}r extraterrestrische Physik, Gie{\ss}enbachstr. 1, D-85748 Garching, Germany}
\altaffiltext{4}{Department of Astronomy and Atmospheric Sciences, Kyungpook National University, Daegu 41566, Republic of Korea}
\altaffiltext{5}{Korea Astronomy and Space Science Institute, Daedeokdae-ro 776, Yuseong-gu, Daejeon 34055, Republic of Korea}

\date{Accepted ??. Received ??}

\label{firstpage}

\begin{abstract}
We investigate the triggering mechanism and the structural properties of 
obscured luminous active galactic nuclei from a detailed study of the rest-frame $B$ and $I$ {\it Hubble Space Telescope}\ images of 29 nearby ($z\approx 0.04-0.4$) 
optically selected type 2 quasars.  Morphological classification reveals that
only a minority ($34\%$) of the hosts are mergers or interacting galaxies.
More than half ($55\%$) of the hosts contain regular disks, and a substantial 
fraction ($38\%$), in fact, are disk-dominated ($B/T\lesssim 0.2$) late-type 
galaxies with low \sersic\ indices ($n < 2$), which is characteristic of pseudo bulges. The prevalence of bars in the spiral host galaxies may be sufficient 
to supply the modest fuel requirements needed to power the nuclear activity in 
these systems.  Nuclear star formation seems to be ubiquitous in the central regions, leading to 
positive color gradients within the bulges and enhancements in the central 
surface brightness of most systems. 
\end{abstract}


\section{Introduction} 
\label{sec:introduction}
The ubiquitous presence of supermassive black holes in the centers of 
galaxies and the tight correlations between black hole mass and bulge stellar mass 
(\citealt{Magorrian98, KormendyHo13}) and velocity dispersion
(\citealt{FerrareseMerritt00}; \citealt{Gebhardt00}) have often been 
attributed to a close connection between the growth of the central black hole (through 
accretion) and the growth of the host galaxy (through star formation).
The exact nature of this connection, however, is still under debate.
In this regard, active galactic nuclei (AGNs) are of great importance to 
understand the physical link between black holes and their host galaxies, as 
AGNs are powered by intense accretion of material onto the central black hole. 
Strong outflows from AGNs can quench star formation efficiently, and may be 
responsible for establishing the empirical correlations between black hole mass and
host galaxy properties (e.g., \citealt{DiMatteo05}; \citealt{Springel05}).

One of the most crucial but yet unknown factors on clarifying the role of nuclear activity in the coevolution of black holes and their hosts is how AGNs are triggered.  For AGNs 
with low to moderate luminosities (e.g., Seyfert galaxies with bolometric 
luminosities $L_{\rm bol}\lesssim 10^{45}$ erg s$^{-1}$), observational and 
theoretical studies suggest that various internal processes can trigger mass 
accretion to the central black hole (e.g., \citealt{HopkinsHernquist09}; 
\citealt{Hopkins14}). However, these mechanisms are insufficient to explain 
the ignition of more powerful AGNs with $L_{\rm bol}> 10^{45}$ erg s$^{-1}$.
It seems unlikely that the gas reservoir on kpc scales can lose sufficient 
angular momentum to feed luminous quasars (\citealt{Jogee06}).  

In numerical simulations, gas-rich major mergers are suggested as a 
promising mechanism to trigger luminous AGNs (e.g., \citealt{Hopkins08,
AlexanderHickox12}). The morphological signatures of major mergers, 
such as close pairs, double nuclei, disturbed morphologies, tidal tails, and
shells and bridges are expected to be visible up to $\sim 0.5-1.5$ Gyr 
after the merger. Given that the AGN lifetime is thought to be $\lesssim 100$ 
Myr (\citealt{MartiniWeinberg01, YuTremaine02, Martini04}), the features of 
morphological disturbance should be observable in their host galaxies if 
luminous AGNs are triggered by gas-rich major mergers.   

A number of observational studies have examined the morphologies of the host 
galaxies of luminous AGNs to test the major-merger scenario.  A high frequency 
of distortions in the morphologies of AGN host galaxies has been reported 
in a number of studies using quasars selected from various methods (e.g., 
radio quasars: \citealt{RamosAlmeida11, RamosAlmeida12}; optically unobscured 
quasars: \citealt{Veilleux09}), 
seemingly consistent with the conventional hypothesis.  On the other hand, 
there are no shortage of studies that reach the opposite conclusion, that major
 mergers play only an insignificant role in triggering luminous AGNs (e.g., 
radio quasars: \citealt{Dunlop03, Floyd04}; X-ray quasars: \citealt{Cisternas11,
Villforth14}; optically unobscured quasars: \citealt{Mechtley16}).

According to the gas-rich major-merger scenario, luminous AGNs should be
highly obscured during the early stages of the merger because of the 
enhanced concentration of gas and dust from the progenitor 
galaxies.  In the aftermath of the merger event, these highly obscured, 
luminous AGNs---classified as ``type 2'' quasars---should be morphologically 
highly disturbed.  When the obscuration clears and ``type 1'' quasars emerge 
toward the late stages of the evolution, it is unclear the extent to which
the morphological signatures of the merger process still remain visible.  
Thus, type 2 quasars are the more promising targets to test the role of 
gas-rich major mergers in triggering AGNs, and the overall major merger-driven 
framework of black hole-galaxy coevolution. 

In terms of morphological studies of AGN host galaxies, obscured sources enjoy 
another strong advantage compared with their unobscured counterparts because 
of their absence of a bright nucleus.  The host galaxies of type 1 AGNs can be
extraordinarily difficult to study because of the dominating influence of 
their strong central point source (e.g., \citealt{Kim08a, Kim08b,Kim17}; \citealt{Mechtley16}).  Even basic morphological classifications---not to 
mention of more quantitative structural parameters---can be challenging to 
obtain.  By contrast, the nuclear obscuration of type 2 AGNs serves as a 
natural coronagraph to block the blinding nucleus, thereby affording a cleaner
view of the detailed internal structures of the host galaxy.

Although various theoretical studies have long predicted the existence of 
obscured luminous AGNs, a limited number of type 2 quasars were known until 
large samples were discovered in the last decade
(e.g., \citealt{Zakamska03, MartinezSansigre05, Reyes08, Alexandroff16}). 
At high redshifts ($z\approx 2$), \cite{Donley18} demonstrate that major 
mergers play a dominant role in triggering and fuelling infrared-selected, 
luminous obscured AGNs. At intermediate redshifts ($z \approx 0.5$), morphological
hints of interactions have also been found to be prevalent in the host 
galaxies of optically selected type 2 quasars (e.g., \citealt{VillarMartin11, 
Wylezalek16}), further supporting the major merger scenario. However, the situation is less clear at lower redshifts ($z \lesssim 0.3$). \citet{Bessiere12} reported a significant fraction (75\%) of type 2 quasars showing evident features of morphological disturbance. Nevertheless, elliptical host galaxies were seen to 
be dominant ($\sim 70$\%) in type 2 samples by other studies, such as those of \citet{Zakamska06} and \citet{VillarMartin12}. 

This work reports deep, high-resolution, rest-frame optical images obtained 
with the \textit{Hubble Space Telescope} (\hst) of a sample of 29 nearby 
($z\approx 0.04-0.4$) type 2 quasars.  Although the sample is modest, our
observations represent the most extensive, detailed study to date of the host 
galaxies of obscured quasars in the local Universe.  The \hst\ images, taken 
in rest-frame $B$ and $I$, enable us not only to investigate the morphological 
properties of the host galaxies but also to derive crude constraints on their 
stellar populations. We analyze the morphologies, photometric structures, 
colors, and stellar masses of the host galaxies, paying special emphasis on 
their bulges.  We examine whether major mergers are causally connected to 
AGN activity.

The paper is organized as follows. Section~\ref{sec:sample_obs} introduces the sample and describes the \hst\ observations and data reduction.  Section~\ref{sec:Data_analyse} presents the image analysis of the host galaxies, including morphological classification, 
structure decomposition, color map construction, and stellar mass estimation. Results and discussions are presented in Section~\ref{sec:results}.  We summarize our main conclusions in Section~\ref{sec:conculde}.  This work adopts the following cosmological parameters: $\Omega_m= 0.286$, $\Omega_{\Lambda}= 0.714$, and $H_0 = 69.6$ km s$^{-1}$ Mpc$^{-1}$ (\citealt{Bennett14}).

\section{Data}
\label{sec:sample_obs}

\subsection{Sample Selection}
Our type 2 quasars were originally selected to complement a matching study
of low-redshift type 1 quasars selected from the Palomar-Green survey of 
\citet{SchmidtGreen83} to examine the evolutionary connection between these two populations, which will be reported in an upcoming paper.  The comparison 
sample of Palomar-Green quasars consists of 87 objects with $z < 0.5$ (\citealt{Boroson92}).  To this end, we randomly selected 87 type 2 quasars, matching the Palomar-Green sample in 
terms of redshift and \oiii\ 
$\lambda$5007 luminosity,  from the catalog of 887 type 2 quasars published by \citet{Reyes08}\footnote{Reyes et al. (2008) chose a 
luminosity cut of $L_{\rm [O~III]} > 10^{8.3}\, L_\odot$ to define type 2 
quasars, but a fraction ($\sim 17\%$) of the sources in the catalog have 
luminosities below this limit because of recent recalibration of the 
SDSS spectrophotometry.  In this study, we adopt the extinction-corrected \oiii\ luminosities of \citet{KongHo18}.}. Type 2 quasars in this catalog were identified from the Sloan Digital Sky Survey (SDSS; \citealt{York00}) Data Release 6 spectroscopic database (\citealt{Adelman08}).  Our selection assumes that type 2 quasars have the same intrinsic AGN luminosity as type 1 quasars for a given $L_{\rm [O~III]}$, and that $L_{\rm [O~III]}$ is related to the bolometric luminosity of the AGN (\citealt{Heckman05, LaMassa09, Dicken14}).
 
\begin{figure}
\raggedright{\includegraphics[scale=1.1]{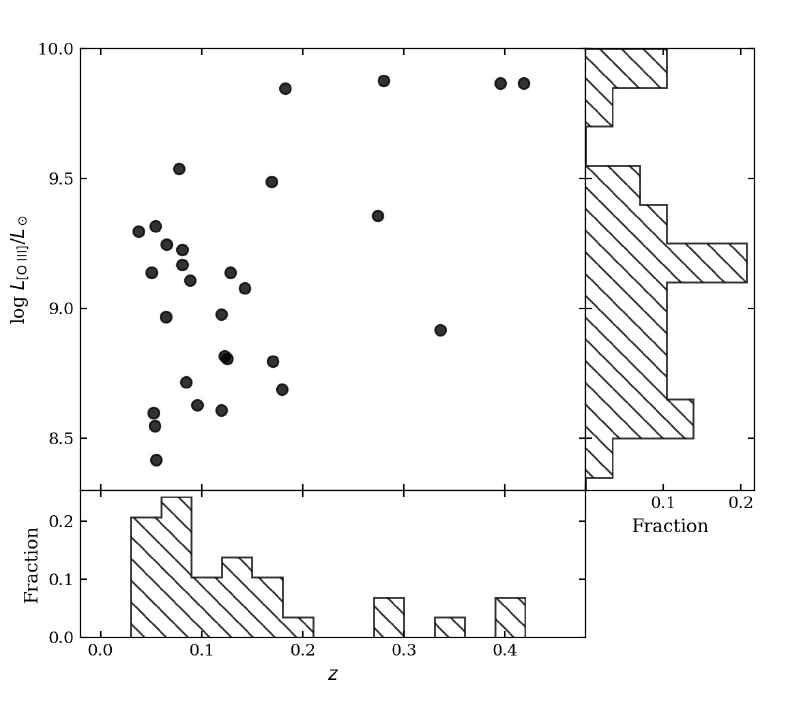}}
\caption{The distribution of extinction-corrected \oiii\ luminosity (\citealt{KongHo18}) and redshift for our sample of 29 type 2 quasars.  The sample is nearby (median $z\approx 0.1$) and has a median $\log (L_{\rm [O~III]} / L_{\odot}) = 9.11$.}
\label{fig:Loiiiz12}
\end{figure}

As the observations were conducted in the 
``snapshot'' mode of \hst, only $\sim 1/3$ of the original sample of type 2 
quasars was observed successfully, yielding a sample of 29 objects 
(Table~\ref{tab:obs_info}).  The final sample spans a redshift range of 0.04 
to 0.4, with a median value of $z = 0.12$, and extinction-corrected \oiii\ 
luminosities from $\log (L_{\rm [O~III]}/L_\odot) = 8.42$ to 9.88, with a 
median value of 9.11 (Fig.~\ref{fig:Loiiiz12}). The 29 objects have a similar distribution of redshifts, \oiii\ luminosities, and optical magnitudes as the original sample of 87 objects. They also match the distribution of these 
quantities in the parent catalog of \citet{Reyes08} at $z<0.5$.

\subsection{\hst\ WFC3 Observations }
\label{sec:WFC3_obs}

The observations were conducted using the WFC3 camera 
between November 2012 and July 2014 (proposal ID 12903; PI: Luis C. Ho). Each object was observed with a blue filter and a red filter using the UVIS or IR channel.  The bandpasses  were carefully 
chosen from the large suite of available WFC3 filters, with two considerations: to 
match approximately rest-frame $B$ and $I$, and to avoid strong emission lines.
Hereinafter, we denote the bluer filter (F438W, F475W, F555W) as $B_{\rm WFC3}$ and the redder filter (F814W, F105W, F110W, F125W) as $I_{\rm WFC3}$. 

For the UVIS channel, each quasar was observed with three long exposures, using the three-point dithering pattern \texttt{WFC3-UVIS-DITHER-LINE-3PT}. For the IR channel, four long exposures were taken using the four-point dithering 
pattern \texttt{WFC3-IR-DITHER-BOX-MIN}. We used subarrays to minimize the
readout time and buffer size, resulting in a field-of-view (FoV) of 
$67 \times 67$ arcsec$^2$ and $40 \times 40$ arcsec$^2$ for the IR and UVIS 
channels, respectively.  These FoVs, which correspond to $\sim 145$ and $87$ 
kpc at the median redshift of the sample, are sufficiently wide to cover the 
outskirts of the host galaxies for detecting extended features and to achieve 
accurate sky measurement.  Total exposure times, which varied between 147 and 
780 s, were set to reach a surface brightness limit of $\mu \approx 25$ mag 
arcsec$^{-2}$, a depth that previous studies had demonstrated can yield robust 
detections of faint outer structures (e.g., \citealt{Kim08a,Greene08,Jiang11}). 
Table~\ref{tab:obs_info} gives a summary of the observations.

\begin{deluxetable*}{ccccccccc}
\fontsize{9.5}{9}\selectfont

\tablecaption{\label{tab:obs_info} Observational Information}
\tablewidth{0pt}

 \tablehead{
 \colhead{Object} & \colhead{$z$} & \colhead{$D_L$} & \colhead{$E(B-V)$}& \colhead{$\log L_{\rm [O~III]}$} & \colhead{$\log L_{\rm [O~III]}$} &\colhead{Filter}& \colhead{ExpTime}& \colhead{ObsDate} \\[0.5ex]
 \colhead{  } & \colhead{   } & \colhead{(Mpc)}       & \colhead{(mag)}     & \colhead{($L_\odot$)}           & \colhead{($L_\odot$)}           &\colhead{$(I_{\rm WFC3}/B_{\rm WFC3}$)}& \colhead{(s)}    & \colhead{(yy-mm-dd)} \\[0.5ex]
 \colhead{(1)} & \colhead{(2)} & \colhead{(3)} & \colhead{(4)}& \colhead{(5)} & \colhead{(6)} &\colhead{(7)}& \colhead{(8)}& \colhead{(9)}}

\startdata 
SDSS J011935.63$-$102613.1   &  0.125    &  586      & 0.0371          & 8.43    & 8.81     & F105W/F475W    &  147/780     &    2013-06-19        \\[0.5ex]
SDSS J074751.56+320052.1     &  0.280    & 1438      & 0.0700          & 8.95    & 9.88     & F110W/F555W    &  147/780     &    2014-04-27        \\[0.5ex]
SDSS J075329.93+230930.7     &  0.336    & 1775      & 0.0633          & 8.59    & 8.92     & F110W/F555W    &  147/780     &    2014-02-13        \\[0.5ex]
SDSS J075940.95+505024.0     &  0.054    &  243      & 0.0415          & 8.83    & 9.32     & F814W/F438W    &  470/300     &    2013-09-09        \\[0.5ex]
SDSS J080252.92+255255.5     &  0.081    &  368      & 0.0329          & 8.86    & 9.23     & F105W/F438W    &  147/705     &    2013-09-13        \\[0.5ex]
SDSS J080337.32+392633.1     &  0.065    &  295      & 0.0456          & 8.12    & 9.25     & F105W/F438W    &  147/705     &    2014-01-12        \\[0.5ex]
SDSS J080523.29+281815.8     &  0.128    &  604      & 0.0476          & 8.62    & 9.14     & F105W/F475W    &  147/570     &    2013-01-03        \\[0.5ex]
SDSS J081100.20+444216.3     &  0.183    &  888      & 0.0418          & 8.27    & 9.85     & F105W/F475W    &  147/540     &    2013-01-21        \\[0.5ex]
SDSS J084107.06+033441.3     &  0.274    & 1404      & 0.0334          & 8.77    & 9.36     & F110W/F555W    &  147/780     &    2014-01-21        \\[0.5ex]
SDSS J084344.99+354941.9     &  0.054    &  241      & 0.0360          & 8.14    & 8.55     & F814W/F438W    &  260/320     &    2013-02-23        \\[0.5ex]
SDSS J090754.07+521127.5     &  0.085    &  386      & 0.0159          & 8.23    & 8.72     & F105W/F438W    &  147/720     &    2013-03-12        \\[0.5ex]
SDSS J091819.66+235736.4     &  0.419    & 2303      & 0.0443          & 9.58    & 9.87     & F125W/F555W    &  147/780     &    2013-04-14        \\[0.5ex]
SDSS J093625.36+592452.7     &  0.095    &  439      & 0.0200          & 8.35    & 8.63     & F105W/F438W    &  147/780     &    2013-04-28        \\[0.5ex]
SDSS J103408.59+600152.2     &  0.050    &  225      & 0.0093          & 8.85    & 9.14     & F814W/F438W    &  156/130     &    2013-10-09        \\[0.5ex]
SDSS J105208.19+060915.1     &  0.052    &  232      & 0.0310          & 8.20    & 8.60     & F814W/F438W    &  325/390     &    2013-02-25        \\[0.5ex]
SDSS J110213.01+645924.8     &  0.077    &  352      & 0.0319          & 8.45    & 9.54     & F105W/F438W    &  147/660     &    2013-08-03        \\[0.5ex]
SDSS J111015.25+584845.9     &  0.143    &  678      & 0.0096          & 8.88    & 9.08     & F105W/F475W    &  147/780     &    2014-03-21        \\[0.5ex]
SDSS J113710.77+573158.7     &  0.395    & 2152      & 0.0097          & 9.61    & 9.87     & F125W/F555W    &  147/780     &    2014-03-21        \\[0.5ex]
SDSS J115326.42+580644.5     &  0.064    &  290      & 0.0249          & 8.48    & 8.97     & F105W/F438W    &  147/630     &    2014-03-19        \\[0.5ex]
SDSS J123804.81+670320.7     &  0.179    &  871      & 0.0190          & 8.26    & 8.69     & F105W/F475W    &  147/780     &    2012-12-31        \\[0.5ex]
SDSS J125850.77+523913.0     &  0.055    &  246      & 0.0141          & 8.26    & 8.42     & F814W/F438W    &  325/390     &    2013-01-11        \\[0.5ex]
SDSS J130038.09+545436.8     &  0.088    &  403      & 0.0180          & 8.94    & 9.11     & F105W/F438W    &  147/660     &    2013-08-06        \\[0.5ex]
SDSS J133542.49+631641.5     &  0.169    &  816      & 0.0187          & 8.47    & 9.49     & F105W/F475W    &  147/780     &    2013-08-10        \\[0.5ex]
SDSS J140541.21+402632.5     &  0.080    &  366      & 0.0135          & 8.78    & 9.17     & F105W/F438W    &  147/660     &    2014-08-07        \\[0.5ex]
SDSS J140712.94+585120.4     &  0.170    &  823      & 0.0107          & 8.27    & 8.80     & F105W/F475W    &  147/780     &    2013-05-29        \\[0.5ex]
SDSS J144038.09+533015.8     &  0.038    &  166      & 0.0115          & 8.94    & 9.30     & F814W/F438W    &  188/132     &    2013-11-24        \\[0.5ex]
SDSS J145019.18$-$010647.4   &  0.120    &  559      & 0.0459          & 8.42    & 8.61     & F105W/F475W    &  147/660     &    2013-06-15        \\[0.5ex]
SDSS J155829.36+351328.6     &  0.119    &  558      & 0.0246          & 8.77    & 8.98     & F105W/F475W    &  147/540     &    2012-10-27        \\[0.5ex]
SDSS J162436.40+334406.7     &  0.122    &  573      & 0.0227          & 8.56    & 8.82     & F105W/F475W    &  147/660     &    2013-09-07        \\[0.5ex]
\enddata

\vspace{0.2cm}
\begin{spacing}{1.2}
\fontsize{9.5}{9}\selectfont
\justifying \textbf{Notes.} Column (1): Object name. Column (2): Redshift. Column (3): Luminosity distance. Column (4): Galactic extinction. Column (5): Observed [O~III] luminosity \citep{Reyes08}.  Column (6): Extinction-corrected [O~III] luminosity \citep{KongHo18}. Column (7): WFC3 filter.  Column (8): Exposure time for 
$I_{\rm WFC3}$ and $B_{\rm WFC3}$. Column (9): Date of observations.
\end{spacing}
\vspace{0.5cm}
\end{deluxetable*}

\subsection{Data Reduction} 
\label{sec:data_reduction}

We use \texttt{AstroDrizzle} to combine the dithered images to generate 
cosmic ray-removed science images.  The 
pixel scale is set to $0\farcs06$ for the IR channel and $0\farcs03$ for the
UVIS channel so that it Nyquist samples the point-spread function (PSF), which 
has a full width at half maximum (FWHM) of $\sim 0\farcs13$ and $0\farcs07$
for the IR and UVIS channel, respectively. 
Fortunately, none of the central pixels near the nucleus was saturated.
Although \texttt{AstroDrizzle} performs sky subtraction, further adjustments 
of the sky level were made during the two-dimensional (2-D) image fitting 
process using {\tt GALFIT} (\citealt{Peng02, Peng10}; see 
Section~\ref{sec:Galfit}).

A robust model of the PSF is crucial for accurate image decomposition, even 
for the hosts of type 2 AGNs.  Ideally, the PSF can be constructed from bright 
stars observed simultaneously in the science images, but in general this is 
not possible in our program because of the relatively small FoV of our subarray
images.  While synthetic {\tt TinyTim} (\citealt{Krist11}) PSFs are commonly 
used as a substitute (e.g., \citealt{Kim17}), our experience (Huang et al. 2019) indicates that empirical PSFs generated from stacked WFC3 images of 
multiple bright, unsaturated, isolated stars observed with the same filter,
in the same dither pattern, but at different times perform significantly 
better than synthetic PSFs.  Hence, our analysis uses a library of empirical 
PSFs of high signal-to-noise (S/N) created from stacking individual stars. The total number of stars used to generate the stacked PSF differs from filter to filter, with the average being a few tens.

\section{Analysis}
\label{sec:Data_analyse}

We first inspect the images to visually classify the morphologies of the host 
galaxies.  We then quantify their structural parameters through detailed 2-D 
image decomposition.  We generate color maps and color profiles, which enable 
us to derive stellar masses and explore the stellar population of the host 
galaxies. 

\subsection{Morphology Classification}

The high resolution and sensitivity of the WFC3 images, coupled with the 
low redshifts of our sample,  allow us to perform quite reliable visual classifications of 
the galaxy morphologies. We distinguish five broad types: merging/disturbed, 
unbarred spirals, barred spirals, lenticulars, and ellipticals.  In this work, we regard spirals and lenticulars as disk galaxies, and we consider spirals as late-type.

\begin{figure*}
\centering
\includegraphics[scale=0.19]{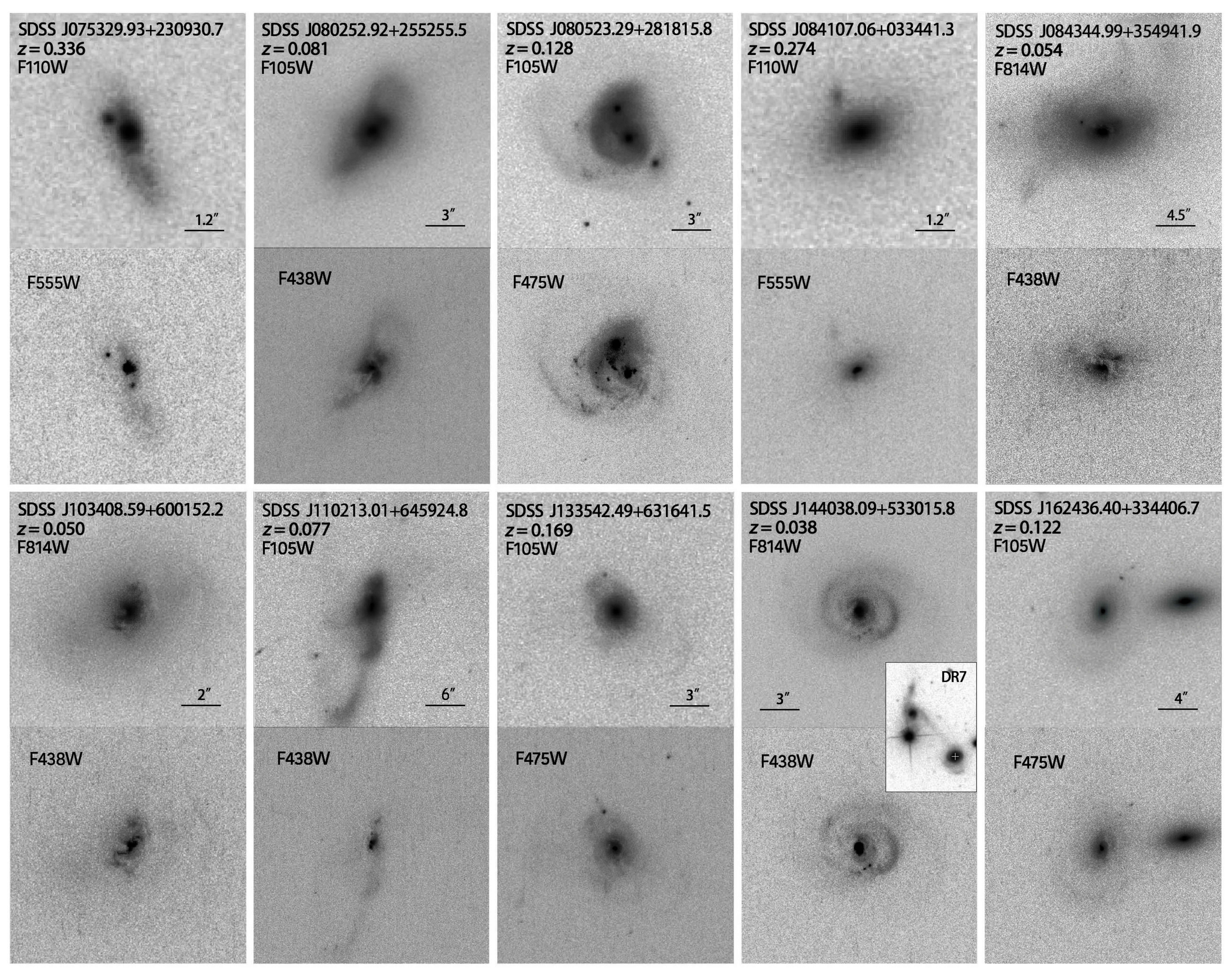}  
\caption{Morphological classifications of our sample. The objects are 
classified into five types: merging/disturbed, barred spiral, unbarred spiral, 
lenticular, and elliptical. The $I_{\rm WFC3}$ and $B_{\rm WFC3}$ images of 
each quasar are shown in the upper and lower panel, respectively, displayed 
using a logarithmic scale with the same FoV.  The objects illustrated here 
have merging/disturbed host galaxies. A SDSS image with larger FoV 
is shown as an inset for SDSS J144038.09+533015.8 to demonstrate its 
interaction with another galaxy.}
\label{fig:morph_class}
\end{figure*}

\renewcommand{\thefigure}{\arabic{figure} (Cont.)}
\addtocounter{figure}{-1}

\begin{figure*}
\centering
\includegraphics[scale=0.22]{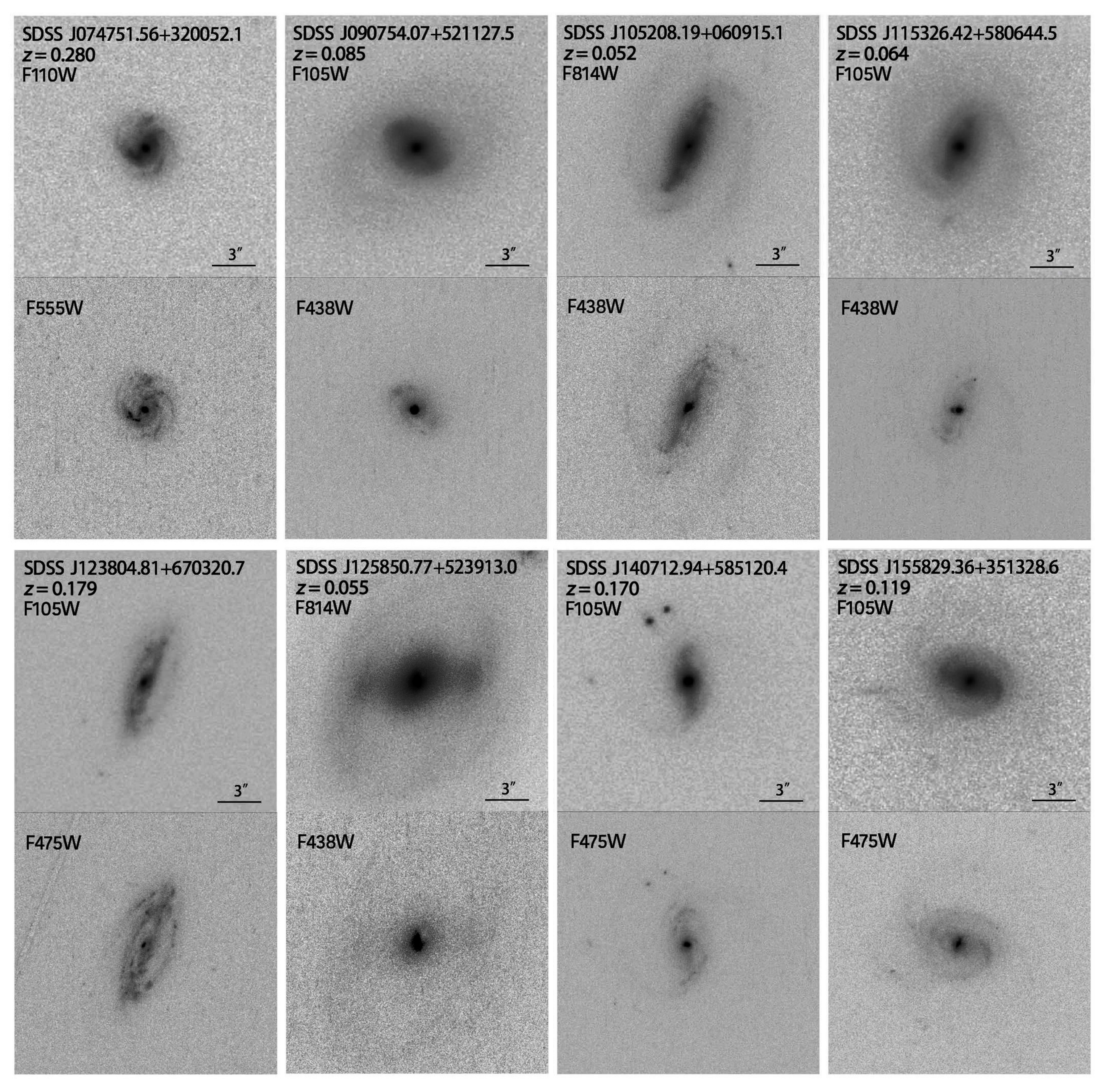}  
\caption{Host galaxies with morphology of barred spirals.}
\end{figure*}
\renewcommand{\thefigure}{\arabic{figure} (Cont.)}
\addtocounter{figure}{-1}

\begin{figure*}
\centering
\includegraphics[scale=1.3]{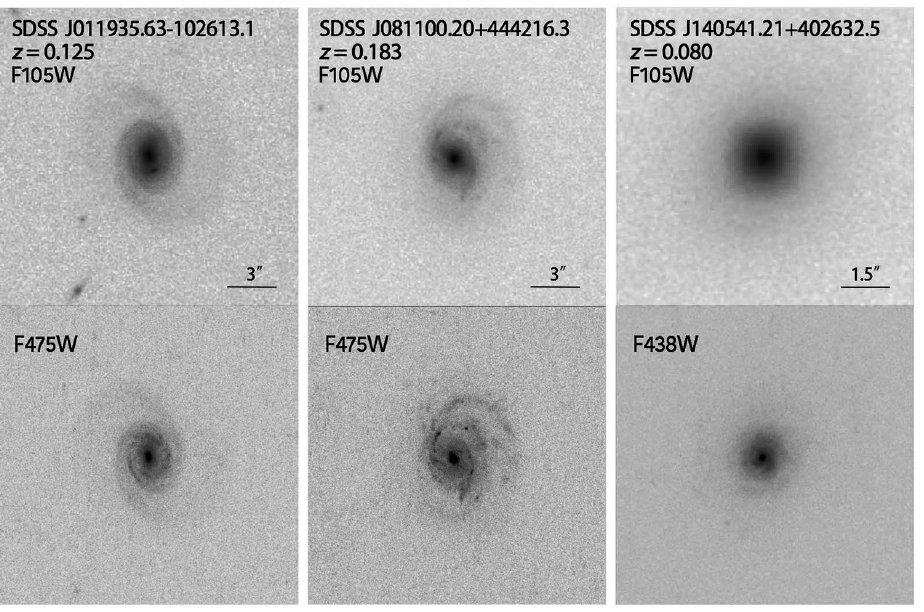}  
\caption{Host galaxies with morphology of unbarred spirals.}
\end{figure*}
\renewcommand{\thefigure}{\arabic{figure} (Cont.)}
\addtocounter{figure}{-1}

\begin{figure*}
\centering
\includegraphics[scale=0.19]{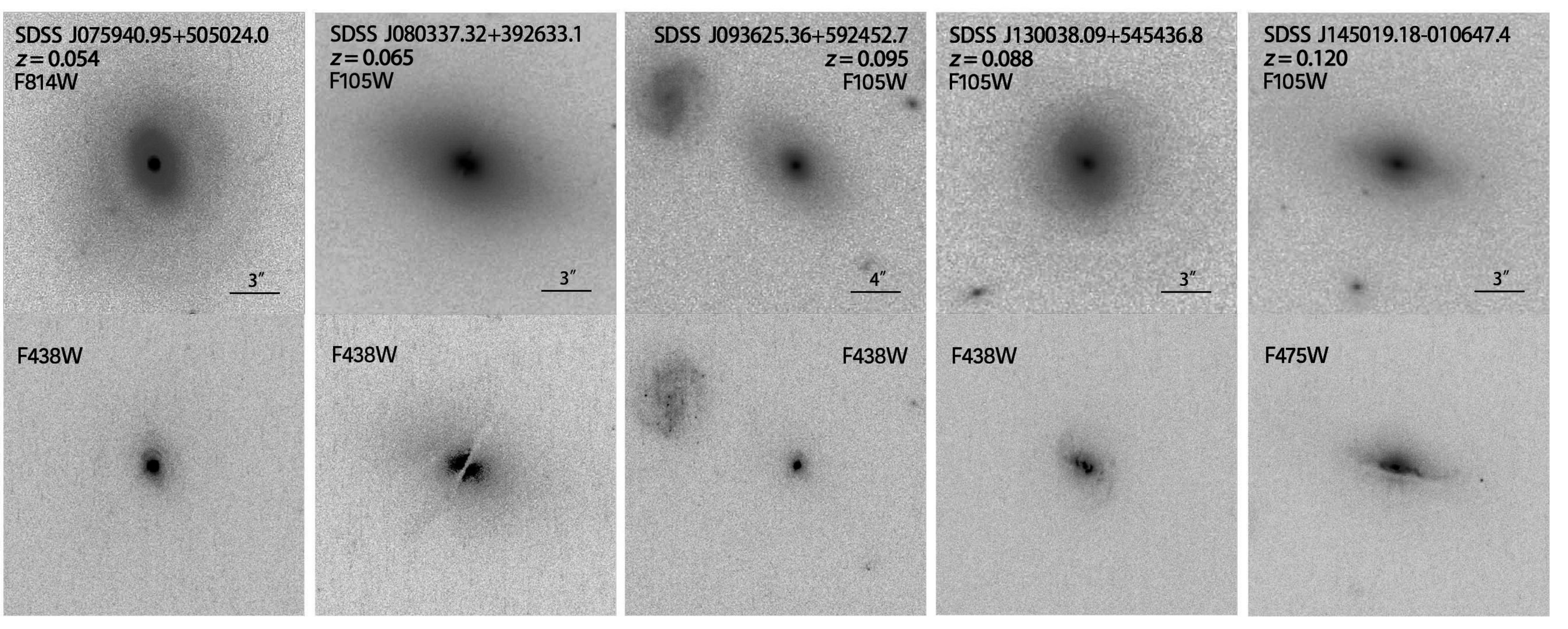}  
\caption{Host galaxies with morphology of lenticulars.}
\end{figure*}
\renewcommand{\thefigure}{\arabic{figure} (Cont.)}
\addtocounter{figure}{-1}

\begin{figure*}
\centering
\includegraphics[scale=0.3]{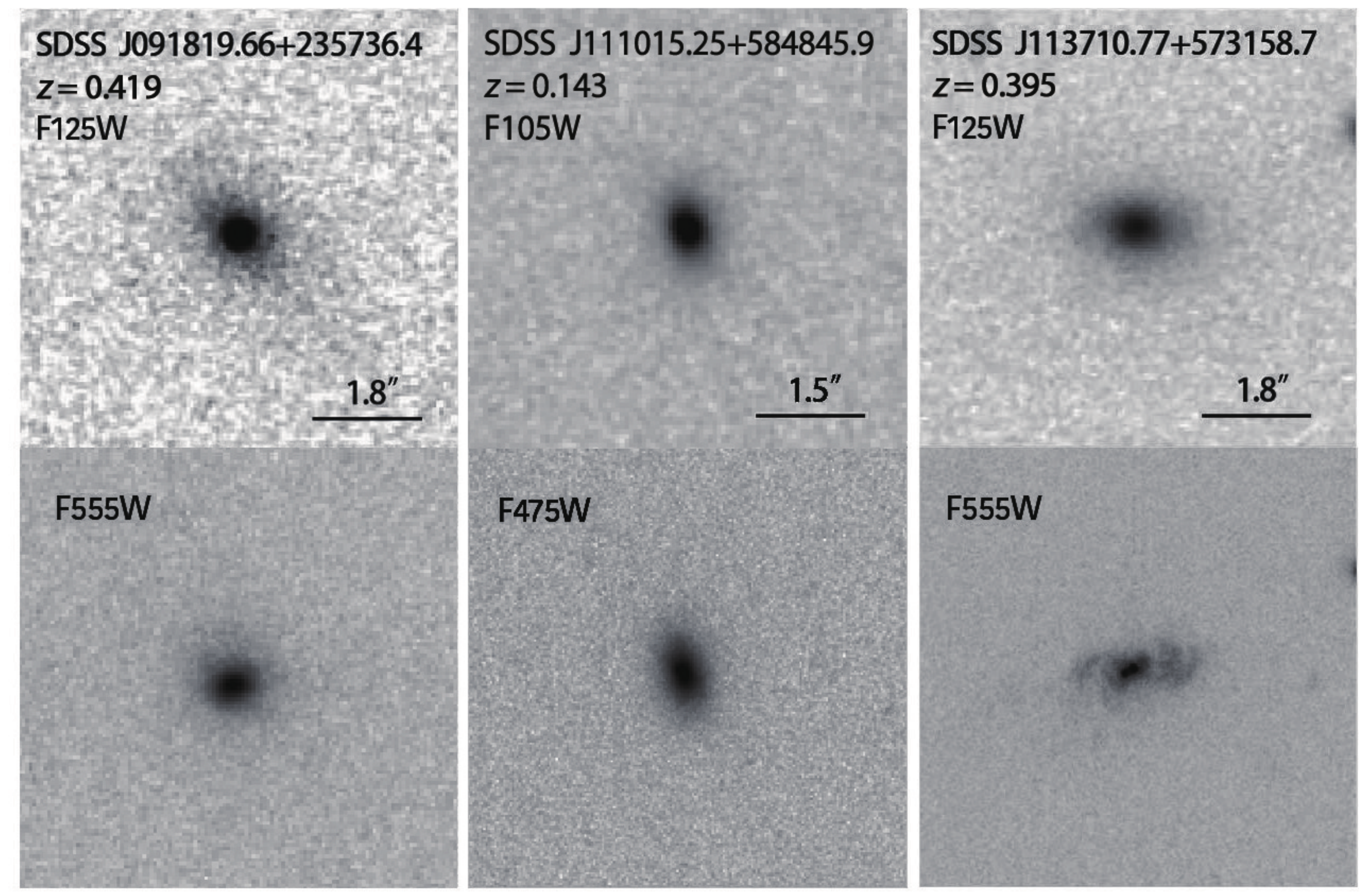}  
\caption{Host galaxies with morphology of ellipticals.}
\end{figure*}
\renewcommand{\thefigure}{\arabic{figure}}

We summarize the classifications of the 29 objects as follows 
(Figure~\ref{fig:morph_class}): 10 can be considered merging/disturbed 
because they exhibit obvious signs of interactions, distorted features, or 
otherwise reside in host galaxies in close pairs; three are found in unbarred 
and eight in barred spirals; five are hosted by lenticulars; and the remaining 
three are in ellipticals. The morphological type of each quasar can be found in Table~\ref{tab:galfit_bestfit}.

Both the merging/disturbed system SDSS J162436.40+334406.7 and the lenticular galaxy SDSS J093625.36+592452.7 have an apparent close companion. Based on available redshift measurements, the companion of SDSS J162436.40+334406.7 is genuinely associated with it, thus indeed constituting a merging system. However, the apparent companion of SDSS J093625.36+592452.7 is a foreground galaxy at $z=0.04$.  While the WFC3 images of SDSS J144038.09+533015.8 suggest that the host galaxy is an isolated spiral, a larger FoV SDSS image (see inset 
panel in Figure~\ref{fig:morph_class}) reveals a 
long stellar bridge connecting the quasar to a disturbed companion, which 
prompted us to classify the host as merging/disturbed.   We inspected large-FoV
SDSS images for all the other objects in the sample and found no other 
examples of potential companions.

Surprisingly, the majority of the sample (55\%) exhibit unambiguous large-scale 
disks, with a significant fraction (38\%) hosting clear late-type morphologies 
in the form of spiral arms and bars.  Only approximately one-third of the host 
galaxies reside in close pairs or show obvious signatures of ongoing or 
recent interactions.  If the ellipticals can be considered merger products, 
then in total $\sim 45\%$ of the sample are or have been associated with 
mergers of one type or another.  Section~\ref{sec:discuss_merging} discusses
the implications of these findings in relation to quasar triggering mechanisms.

\begin{figure}
\raggedright{\includegraphics[scale=1.1]{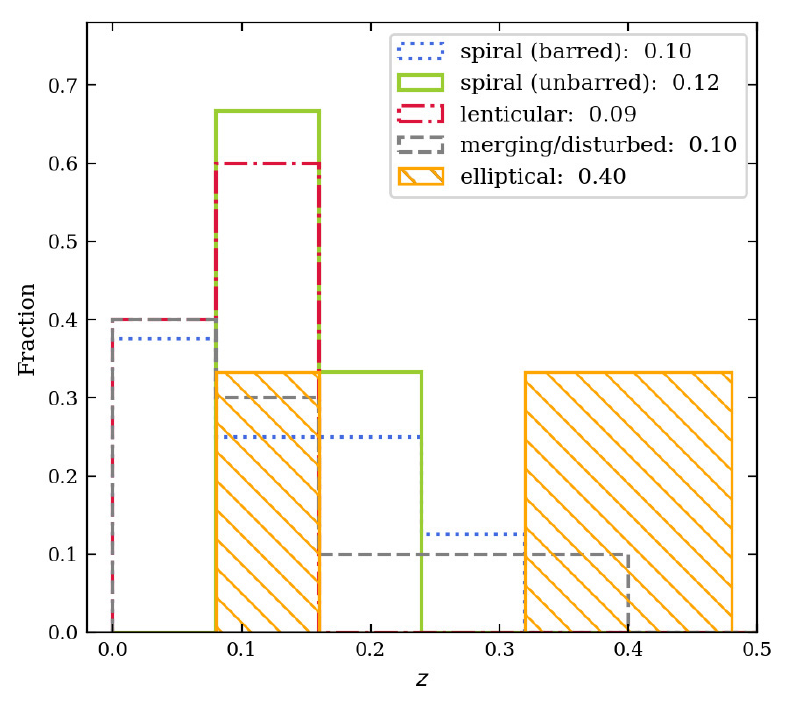}}
\caption{Redshift distribution of host galaxies with different morphologies. 
The median redshift of each subsample is shown. Quasars with elliptical 
morphology have higher redshift ($z\approx 0.4$) than others ($z\approx 0.1$).}
\label{fig:z_distrib}
\end{figure}

\begin{figure*}
\center{\includegraphics[scale=0.26]{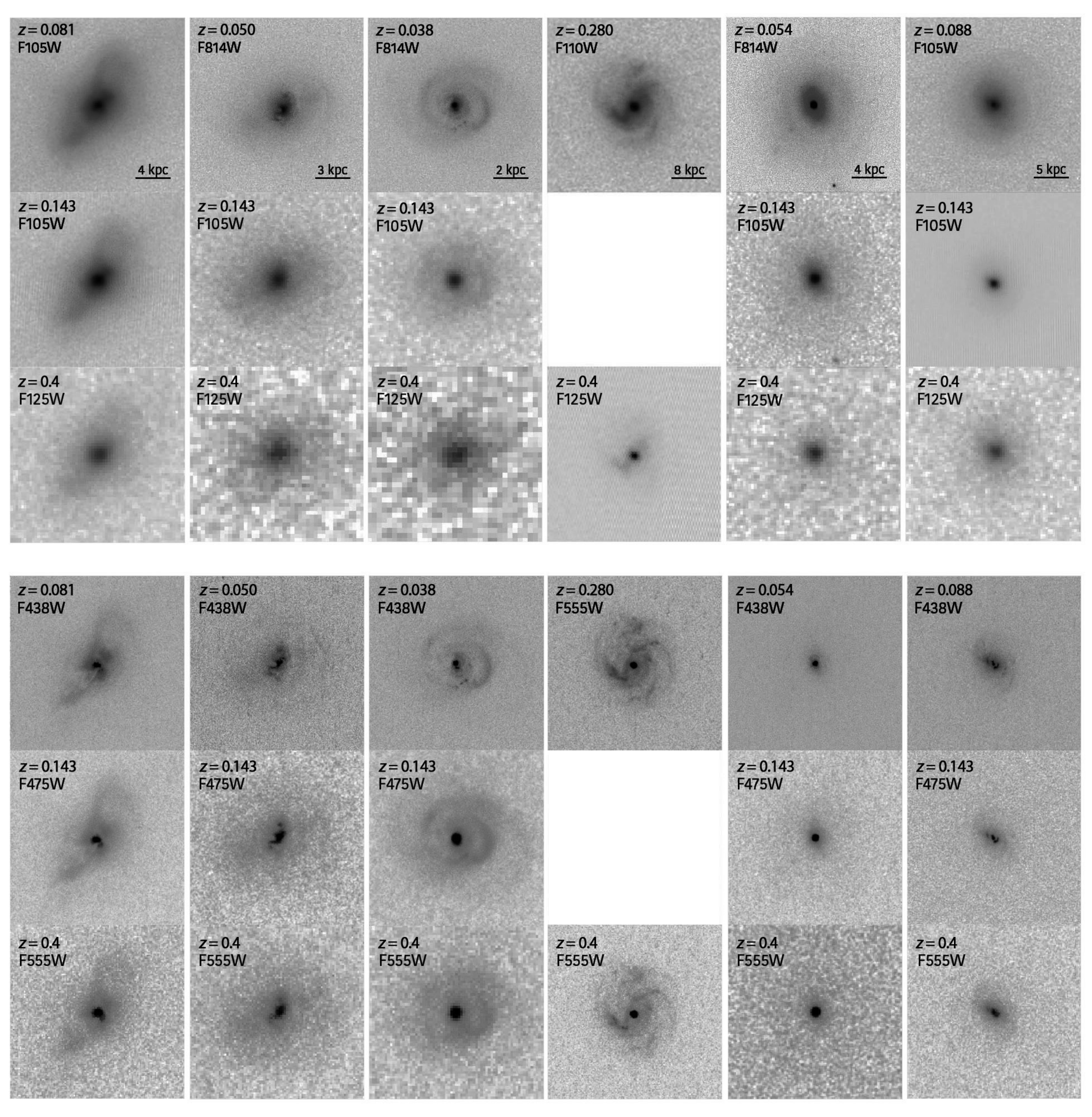}}
\caption{Simulated galaxies created with the {\tt FERENGI} code after shifting 
low-$z$ quasars with their original filters (first row in the upper/bottom 
panel) to $z=0.143$ with F105W/F475W filter (second row in the upper/bottom 
panel) and to $z=0.4$ with F125W/F555W filter (third row in the upper/bottom 
panel). Note that the quasar at $z=0.28$ can only be shifted to $z=0.4$, so 
that no simulated image exists at $z=0.143$.}
\label{fig:shift_to_highz}
\end{figure*}

Figure~\ref{fig:z_distrib} shows the redshift distribution of the host galaxies 
with different morphological types.  The galaxies with disks and 
morphologically disturbed features generally concentrate toward lower 
redshifts ($z \approx 0.1$) than those classified as ellipticals (median 
$z\approx 0.4$).  Have extended features of low surface brightness been missed 
in these more distant objects?  

To test whether surface brightness dimming is the main cause of the 
classification of the elliptical galaxies, we use the {\tt FERENGI} code 
(\citealt{Barden08}) to generate mock, redshifted images of galaxies using the 
actual observed images of lower redshift objects.  To match the luminosity of 
quasars with elliptical hosts ($\log L_{\rm [O~III]}/L_\odot= 8.88$, 9.58, 
9.61), we choose nearby counterparts with similar \oiii\ luminosities covering 
the range of extended morphologies (merging/disturbed, barred and unbarred 
spirals, lenticulars).  The code takes into account the cosmological corrections
for size, surface brightness, bandpass shifting, and $k$-correction.  
{\tt FERENGI} treats the cosmological evolution of the stellar
population, crudely parameterizing the luminosity evolution as $dM/dz=-1$ 
(\citealt{Ilbert05}).  We assume that the mock high-$z$ quasars are located at 
$z=0.143$ and observed with the filter pair F475W/F105W, or at $z=0.4$ 
and observed with the filter pair F555W/F125W.

In the set of simulated IR images (upper panel in 
Figure~\ref{fig:shift_to_highz}), many of the original morphological details 
are lost, and most of the galaxies would be incorrectly classified as ellipticals 
or lenticulars when viewed in F105W at $z=0.143$ or in F125W at $z=0.4$.  
However, the simulated UVIS F475W and F555W images do a much better job in 
retaining the original structural information (bottom panel in 
Figure~\ref{fig:shift_to_highz}).  Therefore, so long as images in both 
filters are available, the morphological classification is unlikely to be 
biased in our study. We conclude that the morphological classifications of the three elliptical hosts in our sample should be secure.

\subsection{Structural Decomposition }

We analyze the images using {\tt GALFIT} V3.0 (\citealt{Peng10}), a non-linear least-squares fitting 
code that uses Levenberg-Marquardt minimization to decompose the major 
structural components of galaxies.  We allow  {\tt GALFIT} to generate its 
own $\sigma$ (weight) image. The sky value is fixed to a constant determined 
from the average background value of five source-free $150\times 150$ pixel$^2$ 
regions, and the uncertainty of the sky is the standard deviation of the five 
measurements.  The sky measurement is measured separately for each 
image of each filter. 
To minimize contamination from nearby sources, we masked all 
objects beyond 1.5 times  the Kron radius\footnote{We use the following definition of Kron radius: \rkron$\,=2.5r_1$, where $r_1$ is the first moment of the light distribution \citep{Kron80,BA96}.  For an elliptical light distribution, this is, strictly speaking, the semi-major axis.} from the target quasar. 
Additionally, objects that are more than 2.5 mag fainter than the target quasar are masked out regardless of their position because they will hardly affect the fit 
for the primary target.  Unmasked close companions  
are simultaneously fit with the target galaxy.  Prominent dust lanes and
regions need to be masked in some objects (e.g., SDSS J080337.32+392633.1, 
J145019.18$-$010647.4, and J084344.99+354941.9).

\subsubsection{Best-fit Models}
\label{sec:Galfit}
We fit bulges with the \citet{Sersic68} profile. We set an 
upper limit of $n=8$ for the S\'ersic index, which is close to the largest 
values seen in the most luminous ellipticals (e.g., \citealt{Kormendy09}).
Moreover, values larger than $n=8$ are often associated with poor model fits (\citealt{Barden12}).  We adopt an exponential profile ($n=1$) for the disk component.Spiral arms, when clearly visible, are modeled by coordinate rotation and bending modes provided by {\tt GALFIT} V3.0 (\citealt{Peng10}; additional examples can be found in \citealt{GaoHo17} and \citealt{Gao19}). Three lenticular galaxies (SDSS J075940.95+505024.0, 
J093625.36+592452.7, and J130038.09+545436.8) exhibit an inner lens. \citet[see also \citealt{Gao18}]{GaoHo17} demonstrate that neglecting inner lenses
will bias the derived bulge parameters significantly. Therefore, we model the inner lens as an independent component with either a \sersic\ or an
exponential profile, depending on which gives the better fit.

\begin{figure*}
\center{\includegraphics[scale=0.47]{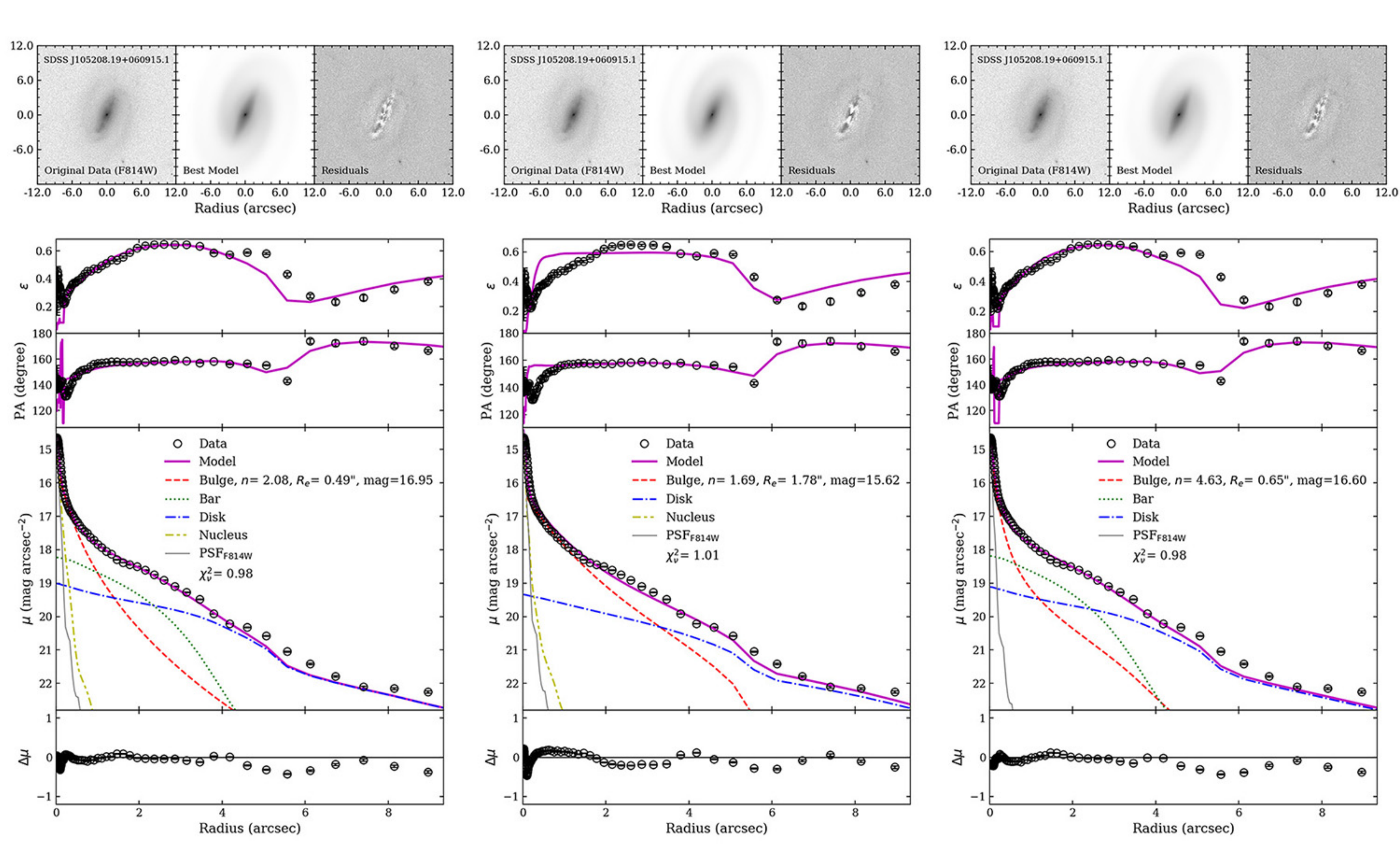}}
\caption{Comparison of the best-fit parameters for different models of the bulge of the spiral host galaxy of SDSS J105208.19+060915.1, to show the effects of the bar and AGN nucleus. For each panel, the original $I_{\rm WFC3}$ image, 2-D model, and residual images are illustrated in the upper panel, and the radial distributions of ellipticity, position angle, surface brightness, and residuals are shown in the bottom panel. The PSF is plotted with arbitrary amplitude. Left panel illustrates the best-fit model that includes both the bar and the nucleus. The overall structure is well-modeled, and the parameters have reasonable values. Middle panel shows the results from the model without a bar component. Although the \sersic\ index $n$ appears reasonable, \re\ and magnitude of bulge are overestimated; moreover, the ellipticity and residual profiles show the necessity of an additional component around 2\asec$-4$\asec. The fit without a nucleus component, shown in the right panel, gives a bulge \sersic\ index that is larger than expected for late-type galaxies. Therefore, both a bar and a nucleus are essential to obtained reasonable structural parameters for this object.  }
\label{fig:Bar_effect}
\end{figure*}

The bar, when present, needs to be properly included to avoid incurring large 
errors on the derived properties of the bulge (\citealt{Laurikainen04, 
Laurikainen05, Gadotti08,GaoHo17}). Following common practice 
(e.g., \citealt{Freeman66, deJong96}), we adopt a fixed \sersic\ $n=0.5$ profile for 
the bar component.  Figure~\ref{fig:Bar_effect} demonstrates the effect of the 
bar component for one of the objects in the sample. Without a bar component 
(middle panel), the size, brightness, and ellipticity of the bulge appear to be
substantially overestimated due to the influence of the bar. The poor match of 
the ellipticity profile as well as the large residuals at $r\approx 2$\asec$-4$\asec\ 
further attest to the inadequacy of the model without a bar.  

For host galaxies classified as merging/disturbed, especially for 
those with substantial disturbance, the bulge component cannot always be 
distinguished clearly from the other irregular components.  We take as the bulge the 
prominent central component, which is usually well fit with a 
single \sersic\ profile (with $n \leqslant 8$). We use Fourier modes (Peng 
et al. 2010) to model asymmetric structures such as lopsided features and tidal tails. 

Although the active nucleus should be deeply obscured in type 2 quasars, a 
fraction of its light can still scatter out (\citealt{AntonucciMiller85}) 
and thereby modify the innermost light profile of the galaxy.  A significant 
fraction of our sample requires an additional compact, 
nuclear component---modeled as an unresolved PSF component---to achieve a 
satisfactory fit for the bulge.  Absent the nuclear component, the \sersic\ 
index of the bulge can reach unrealistically high values that are 
inconsistent with the values expected for the morphological types of the host 
galaxies.  An example is illustrated in the right panel of 
Figure~\ref{fig:Bar_effect}.  The three-component (bulge, bar, and disk) model 
yields a bulge \sersic\ index of $n=4.63$, which would be unprecedented for such an 
obviously late-type galaxy (e.g., \citealt{Balcells03,Vika15}).  The origin of 
the large \sersic\ index is clear: the very central region contains a sharp 
spike, which has the effect of mimicking a large $n$.  After including an 
additional nucleus component (left panel of Figure~\ref{fig:Bar_effect}), the 
bulge \sersic\ index drops to a much more reasonable value of $n=2.08$.  This 
suggests that accounting for a nuclear component, probably due to scattered 
light, is essential to deriving accurate photometric properties of the bulge. 
A nuclear component seems to be required in 14 objects, of which six are spirals, two are ellipticals, and six are merging/disturbed hosts.  The nucleus typically 
contributes $\lesssim 10$\% of the total brightness of the galaxy (average of 
8\% in $I_{\rm WFC3}$ and 9\% in $B_{\rm WFC3}$).  Note that, with absolute 
magnitudes of $-18$ to $-21$, these central components are unlikely to be nuclear star clusters, which have typical absolute magnitudes of $M_I \approx -10$ to $-14$ (\citealt{Boker02}).

The $I_{\rm WFC3}$-band images are significantly deeper than the 
$B_{\rm WFC3}$-band images.  The redder bandpass is also intrinsically more 
sensitive to the dominant, older stellar component of the host, and, of course,
is less affected by dust extinction.  We first determine the best-fit model 
using the $I_{\rm WFC3}$-band image. 
Then we fix all structural parameters of the sub-components (e.g., \re, position angle, ellipticity, central position) to solve only for their brightnesses in the $B_{\rm WFC3}$ band.  The mask and sky level of the $B_{\rm WFC3}$-band image are determined in the same manner as the $I_{\rm WFC3}$-band image.

The best-fit models from the $I_{\rm WFC3}$-band images for the sample are given in Appendix A, and the final parameters are  summarized in Table~\ref{tab:galfit_bestfit}. The fits are generally good, with reduced $\chi^2 \approx 1$. Quasars that are classified as barred and unbarred spirals tend to have less centrally concentrated ($n \lesssim 2$) and smaller ($R_e \lesssim 0.6$\asec) bulges, with $B/T$ generally less than $0.2$. In contrast, the ellipticals and bulges of lenticular and merging/disturbed hosts have much more concentrated ($n>2$), larger ($R_e > 0.6$\asec), and more dominant ($B/T>0.2$) spheroids. We will discuss in detail the implication of the bulge properties with different morphologies in Section~\ref{sec:results_SF}.

\subsubsection{Uncertainties of Best-fit Parameters}

Three main factors contribute to the uncertainties of the structural parameters:
uncertainties in sky determination, variations of the PSF, and assumptions of 
the model construction.  Uncertainties in sky determination do not affect much 
components of high surface brightness, such as the bulge, but they do impact 
the lower surface brightness, extended structures, such as the disk and tidal 
features.  To study the impact of sky determination, we repeat the fits by 
perturbing the sky level one standard deviation above and below the mean 
value.  The impact of PSF variations was assessed by stacking different 
combinations of stars to generate variants of the empirical PSFs, and then 
repeating the fits. 

By far the largest source of uncertainty comes from 
the assumptions that unavoidably need to be made when constructing simplified
2-D models to fit the intrinsically complex structure of galaxies.  This 
problem was recently investigated by \citet{GaoHo17}, who studied the impact 
of including various morphological components (e.g., inner/outer lenses, 
bars, disk breaks, spiral arms) on the derived parameters of  galaxy bulges.
\citet{GaoHo17} showed that inner lenses and bars present the dominant source 
of uncertainty for the bulge parameters, whereas outer spiral arms have a 
marginally effect.
Therefore, for the disk galaxies in our sample, we explicitly treat bars and, 
if present, lenses and nuclei.  For completeness, we also include spiral arms, 
even though they are not essential for the robust measurements of the bulge.  
The best-fit models reproduce properly the observed surface brightness
profiles of most galaxies.  Following \citet{GaoHo17}, we adopt an average uncertainty of 0.1 mag for the bulge luminosity, 
and 10\% for other structural parameters of the bulge.  Ellipticals are 
obviously less complex, but our single-component fits may be an 
oversimplification (\citealt{Huang13}). We nominally assign their uncertainties  half of the above uncertainties adopted for bulges. The model uncertainty for the hosts with merging/disturbed morphologies is difficult to ascertain.  For concreteness, we simply assume that their uncertainties are twice those for 
bulges.  The final uncertainties for the structural parameters are the 
quadrature sum of the uncertainties from the sky, PSF, and model decomposition.

\begin{table*} 
\fontsize{7}{8.4}\selectfont

\newcolumntype{R}{>{\raggedright \arraybackslash} X}
\newcolumntype{S}{>{\centering \arraybackslash} X}
\newcolumntype{T}{>{\raggedleft \arraybackslash} X}

\begin{sideways}
\begin{minipage}{\textheight}
\caption{\label{tab:galfit_bestfit} Best-fit Parameters of Host Galaxies}
\begin{tabular}{ccccccccccccccccl}
\hline\hline

                                 & \multicolumn{9}{c}{$I_{\rm WFC3}$}                                                                                                   & \multicolumn{6}{c}{$B_{\rm WFC3}$}                                                              &           \\

\cmidrule(lr){2-10}\cmidrule(lr){11-16}

 Object      &Filter       &$n_{\rm bulge}$                       &$R_{e,\rm bulge}$                &Bulge                      &Disk                          &Bar                            &Nucleus                       &Total       &$B/T$           &Filter                 &Bulge                      &Disk                    &Bar                            &Nucleus                    &Total           &Morphology \\
             &             &                          &(\asec)                        &(mag)                      &(mag)                         &(mag)                          &(mag)                         &(mag)       &                &                       &(mag)                      &(mag)                   &(mag)                          &(mag)                      &(mag)           &           \\
(1)          &(2)          &(3)                       &(4)                        &(5)                        &(6)                           &(7)                            &(8)                           &(9)         &(10)            &(11)                   &(12)                       &(13)                    &(14)                           &(15)                       &(16)            &(17)       \\[0.5ex]

\hline

J011935  &   F105W     &  1.85$^{+0.19}_{-0.20}$    & 0.78$^{+0.08}_{-0.08}$    & 17.03$^{+0.09}_{-0.09}$    & 16.58$^{+0.01}_{-0.01}$      &           ....                & 19.52$^{+0.09}_{-0.03}$       & 15.97      & 0.27           & F475W        & 19.94$^{+0.12}_{-0.11}$    & 18.85$^{+0.02}_{-0.02}$      &           ....                & 21.42$^{+0.13}_{-0.09}$     & 18.51         & spiral (unbarred)    \\[0.5ex]
J074751  &   F110W     &  0.53$^{+0.11}_{-0.12}$    & 0.09$^{+0.04}_{-0.05}$    & 18.83$^{+0.10}_{-0.10}$    & 16.98$^{+0.03}_{-0.01}$      & 18.70$^{+0.03}_{-0.02}$       &           ....                & 16.62      & 0.13           & F555W        & 20.92$^{+0.11}_{-0.11}$    & 19.18$^{+0.02}_{-0.01}$      & 21.65$^{+0.03}_{-0.02}$       &           ....              & 18.89         & spiral (barred)        \\[0.5ex]
J075329  &   F110W     &  2.25$^{+0.46}_{-0.47}$    & 0.80$^{+0.16}_{-0.16}$    & 18.20$^{+0.19}_{-0.19}$    & 19.66$^{+0.01}_{-0.03}$      &           ....                & 20.05$^{+0.02}_{-0.02}$       & 17.95      & 0.79           & F555W        & 21.05$^{+0.22}_{-0.21}$    & 22.31$^{+0.03}_{-0.02}$      &           ....                & 23.98$^{+0.14}_{-0.01}$     & 20.76         & merging     \\[0.5ex]
J075940\tablenotemark{b}  &   F814W     &  1.04$^{+0.12}_{-0.20}$    & 0.12$^{+0.01}_{-0.02}$    & 16.60$^{+0.09}_{-0.09}$    & 15.48$^{+0.03}_{-0.03}$      & 16.38$^{+0.01}_{-0.01}$       &           ....                & 14.85      & 0.20           & F438W        & 18.17$^{+0.12}_{-0.09}$    & 17.62$^{+0.06}_{-0.07}$      & 18.84$^{+0.05}_{-0.04}$       &           ....              & 16.91         & lenticular     \\[0.5ex]
J080252  &   F105W     &  4.53$^{+0.91}_{-0.92}$    & 1.65$^{+0.33}_{-0.33}$    & 14.16$^{+0.16}_{-0.14}$    & 14.76$^{+0.02}_{-0.00}$      &           ....                &           ....                & 13.67      & 0.63           & F438W        & 17.25$^{+0.19}_{-0.17}$    & 17.69$^{+0.01}_{-0.00}$      &           ....                &           ....              & 16.70         & merging     \\[0.5ex]
J080337  &   F105W     &  4.57$^{+0.46}_{-0.46}$    & 4.82$^{+0.52}_{-0.52}$    & 13.68$^{+0.07}_{-0.07}$    & 15.16$^{+0.04}_{-0.04}$      &           ....                &           ....                & 13.43      & 0.80           & F438W        & 18.68$^{+0.15}_{-0.12}$    & 17.19$^{+0.03}_{-0.04}$      &           ....                &           ....              & 16.94         & lenticular     \\[0.5ex]
J080523\tablenotemark{a}  &   F105W     &  2.22$^{+0.45}_{-0.49}$    & 0.87$^{+0.18}_{-0.18}$    & 16.87$^{+0.18}_{-0.17}$    & 15.40$^{+0.01}_{-0.00}$      &           ....                & 18.05$^{+0.05}_{-0.07}$       & 15.15      & 0.21           & F475W        & 20.23$^{+0.22}_{-0.21}$    & 17.91$^{+0.04}_{-0.02}$      &           ....                & 19.94$^{+0.01}_{-0.09}$     & 17.79         & merging        \\[0.5ex]
J081100  &   F105W     &  4.03$^{+0.48}_{-0.56}$    & 0.41$^{+0.05}_{-0.06}$    & 16.97$^{+0.12}_{-0.10}$    & 16.07$^{+0.02}_{-0.01}$      &           ....                &           ....                & 15.61      & 0.18           & F475W        & 20.05$^{+0.11}_{-0.10}$    & 18.30$^{+0.02}_{-0.03}$      &           ....                &           ....              & 18.10         & spiral (unbarred)       \\[0.5ex]
J084107  &   F110W     &  4.08$^{+0.85}_{-0.87}$    & 1.46$^{+0.30}_{-0.30}$    & 17.07$^{+0.17}_{-0.17}$    &           ....               &           ....                & 19.65$^{+0.19}_{-0.06}$       & 17.07      & 1.00           & F555W        & 19.39$^{+0.19}_{-0.19}$    &           ....               &           ....                & 23.30$^{+0.02}_{-0.08}$     & 19.39         & merging      \\[0.5ex]
J084344  &   F814W     &  3.98$^{+0.89}_{-1.23}$    & 4.18$^{+1.57}_{-1.99}$    & 14.67$^{+0.25}_{-0.42}$    & 15.12$^{+0.10}_{-0.24}$      &           ....                &           ....                & 14.12      & 0.60           & F438W        & 16.97$^{+0.18}_{-0.17}$    & 17.66$^{+0.08}_{-0.08}$      &           ....                &           ....              & 16.51         & disturbed     \\[0.5ex]
J090754  &   F105W     &  1.89$^{+0.21}_{-0.21}$    & 0.24$^{+0.02}_{-0.02}$    & 16.42$^{+0.09}_{-0.08}$    & 15.49$^{+0.01}_{-0.01}$      & 16.22$^{+0.01}_{-0.01}$       & 19.46$^{+0.48}_{-0.20}$       & 14.77      & 0.22           & F438W        & 19.39$^{+0.10}_{-0.11}$    & 18.44$^{+0.10}_{-0.11}$      & 19.56$^{+0.06}_{-0.05}$       & 23.24$^{+0.12}_{-0.05}$     & 17.82         & spiral (barred)         \\[0.5ex]
J091819  &   F125W     &  1.79$^{+0.14}_{-0.25}$    & 1.10$^{+0.08}_{-0.07}$    & 18.75$^{+0.08}_{-0.06}$    &           ....               &           ....                & 19.26$^{+0.14}_{-0.06}$       & 18.75      & 1.00           & F555W        & 20.95$^{+0.06}_{-0.06}$    &           ....               &           ....                & 21.01$^{+0.15}_{-0.01}$     & 20.95         & elliptical     \\[0.5ex]
J093625\tablenotemark{b}  &   F105W     &  2.92$^{+0.29}_{-0.31}$    & 0.17$^{+0.02}_{-0.02}$    & 16.81$^{+0.09}_{-0.09}$    & 16.01$^{+0.01}_{-0.01}$      & 17.56$^{+0.02}_{-0.04}$       &           ....                & 15.42      & 0.28           & F438W        & 19.77$^{+0.10}_{-0.10}$    & 19.39$^{+0.10}_{-0.11}$      & 17.90$^{+0.01}_{-0.02}$       &           ....              & 17.51         & lenticular     \\[0.5ex]
J103408  &   F814W     &  3.42$^{+0.79}_{-1.55}$    & 3.99$^{+1.10}_{-1.56}$    & 14.45$^{+0.35}_{-1.52}$    & 14.82$^{+0.18}_{-0.64}$      &           ....                &           ....                & 13.87      & 0.59           & F438W        & 16.31$^{+0.17}_{-0.16}$    & 17.35$^{+0.10}_{-0.11}$      &           ....                &           ....              & 15.96         & disturbed     \\[0.5ex]
J105208  &   F814W     &  2.08$^{+0.22}_{-0.22}$    & 0.49$^{+0.05}_{-0.05}$    & 16.95$^{+0.09}_{-0.09}$    & 15.39$^{+0.03}_{-0.03}$      & 16.37$^{+0.03}_{-0.05}$       & 19.59$^{+0.05}_{-0.07}$       & 14.85      & 0.14           & F438W        & 19.48$^{+0.10}_{-0.10}$    & 17.38$^{+0.06}_{-0.07}$      & 18.86$^{+0.07}_{-0.04}$       & 20.49$^{+0.07}_{-0.05}$     & 17.01         & spiral (barred)         \\[0.5ex]
J110213\tablenotemark{a}  &   F105W     &  2.06$^{+0.41}_{-0.47}$    & 1.58$^{+0.32}_{-0.32}$    & 15.11$^{+0.16}_{-0.16}$    & 15.01$^{+0.01}_{-0.00}$      &           ....                & 16.57$^{+0.16}_{-0.03}$       & 14.30      & 0.48           & F438W        & 19.08$^{+0.19}_{-0.19}$    & 17.76$^{+0.00}_{-0.01}$      &           ....                & 20.44$^{+0.08}_{-0.09}$     & 17.48         & merging      \\[0.5ex]
J111015  &   F105W     &  4.56$^{+0.24}_{-0.45}$    & 0.13$^{+0.01}_{-0.01}$    & 17.94$^{+0.05}_{-0.04}$    &           ....               &           ....                &           ....                & 17.94      & 1.00           & F475W        & 19.59$^{+0.09}_{-0.07}$    &           ....               &           ....                &           ....              & 19.59         & elliptical     \\[0.5ex]
J113710  &   F125W     &  5.97$^{+0.38}_{-0.40}$    & 0.97$^{+0.08}_{-0.08}$    & 17.16$^{+0.05}_{-0.05}$    &           ....               &           ....                &           ....                & 17.16      & 1.00           & F555W        & 19.81$^{+0.06}_{-0.06}$    &           ....               &           ....                &           ....              & 19.81         & elliptical     \\[0.5ex]
J115326  &   F105W     &  1.75$^{+0.19}_{-0.27}$    & 0.19$^{+0.02}_{-0.02}$    & 16.66$^{+0.11}_{-0.08}$    & 15.34$^{+0.01}_{-0.01}$      & 16.54$^{+0.01}_{-0.01}$       & 18.04$^{+0.26}_{-0.02}$       & 14.81      & 0.18           & F438W        & 19.31$^{+0.10}_{-0.10}$    & 17.95$^{+0.07}_{-0.07}$      & 19.81$^{+0.09}_{-0.07}$       & 20.22$^{+0.01}_{-0.04}$     & 17.54         & spiral (barred)         \\[0.5ex]
J123804  &   F105W     &  1.31$^{+0.13}_{-0.13}$    & 0.20$^{+0.02}_{-0.02}$    & 19.12$^{+0.10}_{-0.10}$    & 16.93$^{+0.00}_{-0.00}$      & 16.91$^{+0.00}_{-0.00}$       &           ....                & 16.10      & 0.06           & F475W        & 22.17$^{+0.12}_{-0.12}$    & 20.13$^{+0.02}_{-0.00}$      & 18.83$^{+0.02}_{-0.04}$       &           ....              & 18.51         & spiral (barred)         \\[0.5ex]
J125850  &   F814W     &  1.62$^{+0.16}_{-0.33}$    & 0.61$^{+0.06}_{-0.06}$    & 16.57$^{+0.11}_{-0.09}$    & 14.88$^{+0.05}_{-0.05}$      & 16.43$^{+0.03}_{-0.02}$       &           ....                & 14.48      & 0.15           & F438W        & 18.85$^{+0.12}_{-0.10}$    & 17.24$^{+0.13}_{-0.15}$      & 19.60$^{+0.16}_{-0.14}$       &           ....              & 16.92         & spiral (barred)         \\[0.5ex]
J130038\tablenotemark{b}  &   F105W     &  3.82$^{+0.42}_{-0.41}$    & 0.71$^{+0.09}_{-0.08}$    & 15.90$^{+0.09}_{-0.10}$    & 15.79$^{+0.01}_{-0.02}$      & 16.86$^{+0.09}_{-0.09}$       &           ....                & 14.90      & 0.40           & F438W        & 18.43$^{+0.10}_{-0.10}$    & 19.85$^{+0.65}_{-1.85}$      & 19.16$^{+0.09}_{-0.09}$       &           ....              & 17.80         & lenticular     \\[0.5ex]
J133542  &   F105W     &  3.01$^{+0.61}_{-0.61}$    & 1.69$^{+0.34}_{-0.34}$    & 16.06$^{+0.16}_{-0.16}$    &           ....               &           ....                & 18.58$^{+0.05}_{-0.01}$       & 16.06      & 1.00           & F475W        & 18.33$^{+0.19}_{-0.18}$    &           ....               &           ....                & 21.93$^{+0.17}_{-0.03}$     & 18.33         & disturbed     \\[0.5ex]
J140541  &   F105W     &  1.69$^{+0.17}_{-0.20}$    & 0.50$^{+0.05}_{-0.05}$    & 15.83$^{+0.09}_{-0.08}$    & 16.16$^{+0.02}_{-0.01}$      &           ....                & 17.60$^{+0.07}_{-0.01}$       & 15.19      & 0.45           & F438W        & 18.16$^{+0.10}_{-0.10}$    & 18.65$^{+0.09}_{-0.10}$      &           ....                & 19.70$^{+0.11}_{-0.03}$     & 17.63         & spiral (unbarred)    \\[0.5ex]
J140712  &   F105W     &  1.86$^{+0.19}_{-0.19}$    & 0.19$^{+0.02}_{-0.02}$    & 17.28$^{+0.09}_{-0.09}$    & 16.80$^{+0.01}_{-0.01}$      & 17.44$^{+0.03}_{-0.03}$       &           ....                & 15.94      & 0.29           & F475W        & 19.66$^{+0.13}_{-0.10}$    & 19.00$^{+0.04}_{-0.04}$      & 20.39$^{+0.04}_{-0.04}$       &           ....              & 18.35         & spiral (barred)        \\[0.5ex]
J144038  &   F814W     &  2.04$^{+0.42}_{-0.45}$    & 0.19$^{+0.04}_{-0.05}$    & 16.35$^{+0.17}_{-0.16}$    & 14.34$^{+0.03}_{-0.03}$      &           ....                & 17.96$^{+0.16}_{-0.04}$       & 14.18      & 0.14           & F438W        & 17.54$^{+0.19}_{-0.18}$    & 16.09$^{+0.04}_{-0.04}$      &           ....                & 18.64$^{+0.03}_{-0.08}$     & 15.84         & merging     \\[0.5ex]
J145019  &   F105W     &  4.22$^{+0.54}_{-0.50}$    & 1.81$^{+0.39}_{-0.28}$    & 15.38$^{+0.14}_{-0.12}$    & 16.88$^{+0.36}_{-0.20}$      &           ....                &           ....                & 15.14      & 0.80           & F475W        & 18.51$^{+0.12}_{-0.10}$    & 18.89$^{+0.05}_{-0.06}$      &           ....                &           ....              & 17.93         & lenticular     \\[0.5ex]
J155829  &   F105W     &  0.80$^{+0.08}_{-0.09}$    & 0.31$^{+0.03}_{-0.03}$    & 18.10$^{+0.09}_{-0.09}$    & 16.25$^{+0.01}_{-0.01}$      & 17.11$^{+0.03}_{-0.03}$       & 19.48$^{+0.07}_{-0.00}$       & 15.72      & 0.11           & F475W        & 19.99$^{+0.10}_{-0.10}$    & 18.30$^{+0.05}_{-0.05}$      & 19.82$^{+0.05}_{-0.05}$       & 21.67$^{+0.13}_{-0.03}$     & 17.89         & spiral (barred)        \\[0.5ex]
J162436  &   F105W     &  1.93$^{+0.39}_{-0.41}$    & 0.79$^{+0.16}_{-0.16}$    & 16.84$^{+0.18}_{-0.18}$    & 15.88$^{+0.01}_{-0.00}$      &           ....                & 19.14$^{+0.08}_{-0.02}$       & 15.51      & 0.29           & F475W        & 19.27$^{+0.20}_{-0.19}$    & 18.59$^{+0.04}_{-0.04}$      &           ....                & 21.31$^{+0.12}_{-0.05}$     & 18.13         & merging     \\[0.5ex]
         
\hline 
\end{tabular}
\begin{spacing}{1.2}
\fontsize{8.5}{9}\selectfont
\justifying \textbf{Notes.} 
Column (1): Galaxy name. 
Column (2): Filter used, corresponding to $I_{\rm WFC3}$. 
Column (3)$-$(8): Best-fit {\tt GALFIT} parameters.
Column (9): Total magnitude of the host, excluding the nucleus, if present.
Column (10): $B/T$ ratio. 
Column (11): Filter used, corresponding to $B_{\rm WFC3}$. 
Column (12)$-$(15): Best-fit {\tt GALFIT} parameters.
Column (16): Total magnitude of the host, excluding the nucleus, if present.
Column (17): Morphological classification.
\end{spacing}
\tablenotetext{a}{The magnitude of the disk in Column (6) and (13) is the sum of the magnitudes of all the components except the bulge and nucleus component.}
\tablenotetext{b}{This galaxy has an inner lens; the magnitude given in Column (7) and (14) refer to the lens instead of the bar.}
\end{minipage} 
\end{sideways}
\end{table*}

\subsection{Colors}
\label{sec:cm_create}
The \hst\ observations were designed specifically to provide at least 
rudimentary color information with the filter combination of rest-frame $B$ 
and $I$.  In view of the possibility that the central region of the galaxy 
might be contaminated mildly by scattered light from the AGN (Section~\ref{sec:Galfit}), 
we perform the color analysis on the images after subtracting the best-fit 
nucleus component, if present.  We can generate color images in a 
straightforward manner for the six targets that were observed with the same 
(UVIS) detector. The majority (23/29), however, were observed with two detectors
with different pixel scales and spatial resolutions.  We rebin the UVIS images 
to match the pixel scale of the IR images, and we convolve the images taken in one 
filter with the corresponding PSF of the other filter.
The color maps are generated by $(B-I)_{\rm WFC3}= -2.5\log (f_{B_{\rm WFC3}}/f_{I_{\rm WFC3}}) + {\rm ZP}_{B_{\rm WFC3}} - {\rm ZP}_{I_{\rm WFC3}}$, where $f_{B_{\rm WFC3}}$ and $f_{I_{\rm WFC3}}$ are the counts in $B_{\rm WFC3}$ and $I_{\rm WFC3}$, respectively, and ${\rm ZP}_{B_{\rm WFC3}}$ and ${\rm ZP}_{I_{\rm WFC3}}$ are the corresponding zero points.  We apply the {\tt IRAF} task {\tt ellipse} to the color images to derive radial color profiles.

\begin{figure*}
\centering
\includegraphics[scale=0.21]{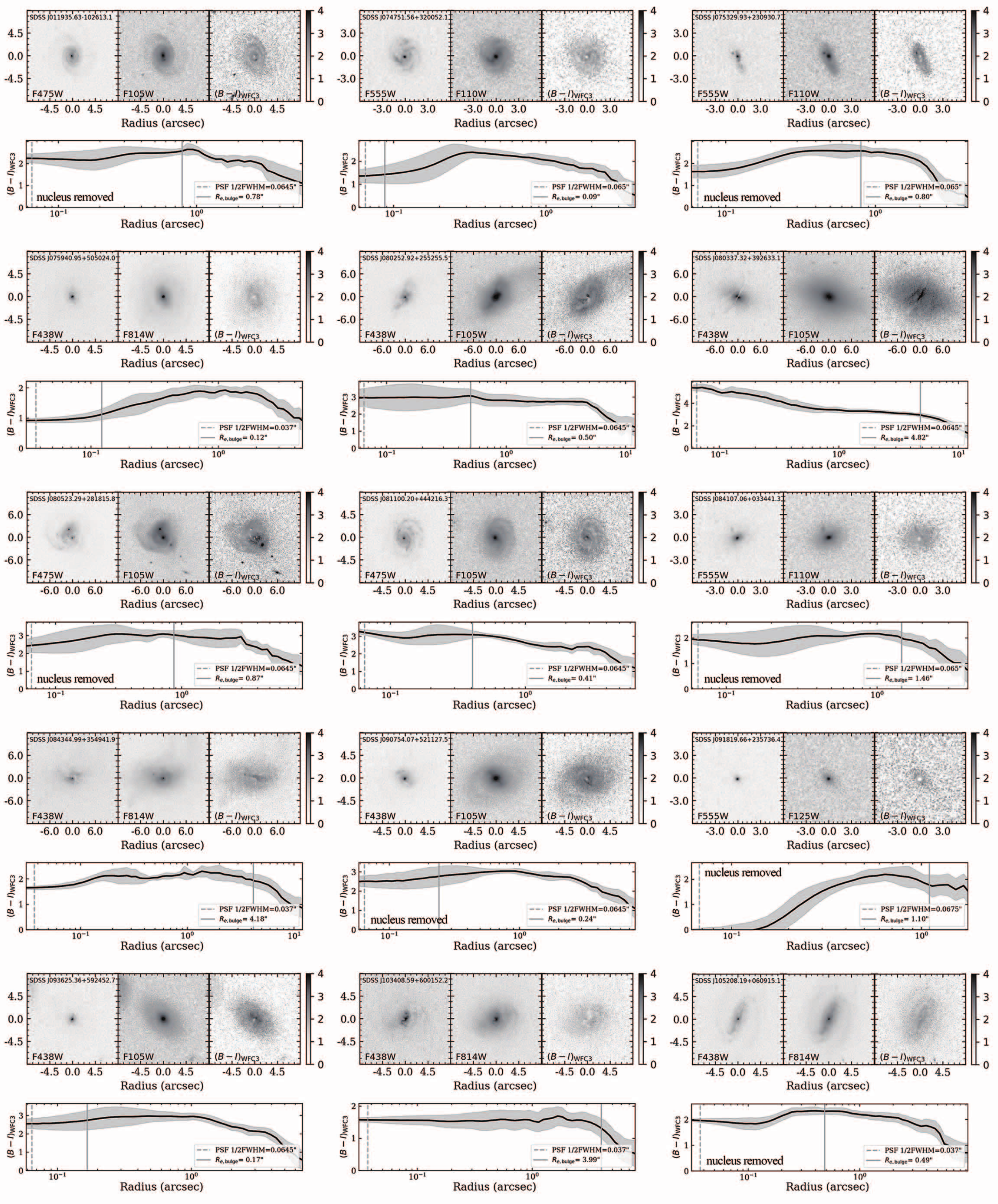}  
\caption{Color maps and color profiles of the host galaxies. The original 
$I_{\rm WFC3}$ and $B_{\rm WFC3}$  images are rebinned onto the same pixel 
scale and convolved with each other's PSF. A nucleus component, if needed from 
the best-fit {\tt GALFIT} model, has been removed.  The upper panel shows the 
$B_{\rm WFC3}$ image, the $I_{\rm WFC3}$ image, and the $(B-I)_{\rm WFC3}$ 
color map, with color bar on the far right end. The bottom panel shows the 
color radial profile (black solid curve) measured from the center of the 
galaxy out to 1.5 times the Kron radius. The grey shaded area shows the 
root-mean-square scatter. The dashed vertical line marks 1/2 FWHM of the PSF, 
and the vertical solid line shows the best-fit effective radius of the bulge.}
\label{fig:color_map}
\end{figure*}

Figure~\ref{fig:color_map} gives the color maps and color profiles. 
For each object, the upper panel shows the $B_{\rm WFC3}$ image, the 
$I_{\rm WFC3}$ image, and the $(B-I)_{\rm WFC3}$ color map.
The color radial profile is shown in the bottom panel. We will discuss the results in Section~\ref{sec:results_SF}.

\renewcommand{\thefigure}{\arabic{figure} (Cont.)}
\addtocounter{figure}{-1}

\begin{figure*}
\centering
\includegraphics[scale=0.21]{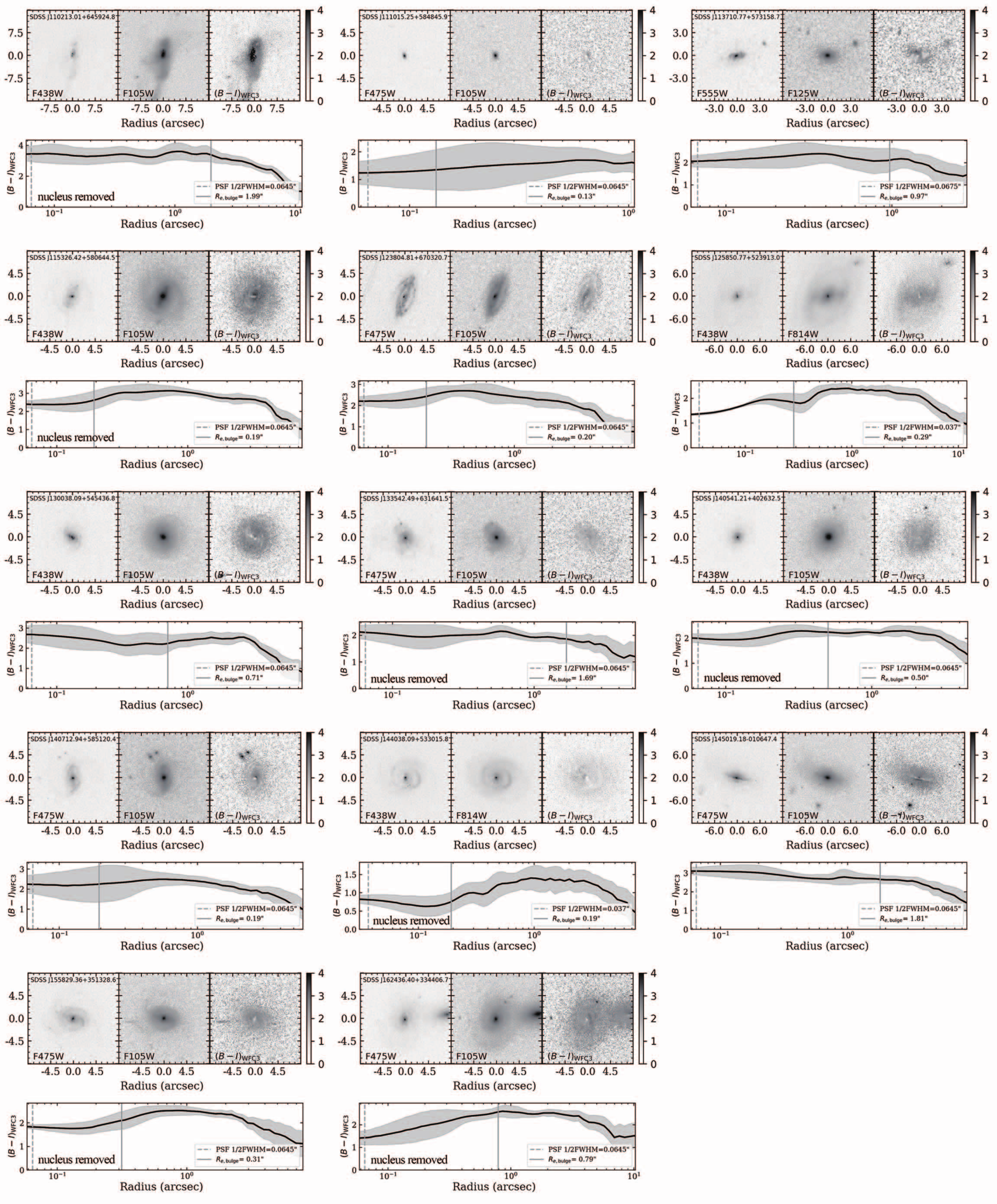}  
\caption{Color maps and color profiles of  host galaxies.}
\end{figure*}
\renewcommand{\thefigure}{\arabic{figure}}

\subsection{Bulge Stellar Masses}
\label{sec:mass_bulge}

We convert the $I_{\rm WFC3}$ magnitude of the bulge ($M_{I,{\rm bul}}$) to 
the bulge stellar mass ($M_{\rm bul}$) using a mass-to-light ratio ($M/L$) inferred from the 
rest-frame $(B-I)$ color. Adopting the color-based conversion from 
\citet{Bell01}, 

\begin{equation}
\label{eq:M/Lratio}
\log \left( \frac{ M_{\rm bul}}{M_\odot} \right)  = -0.4(M_{I,{\rm bul}} 
- M_{I,\odot}) +  0.439(B-I) - 0.594,
\end{equation} 

\noindent
where $M_{I,\odot}= 4.08$ is the $I$-band absolute magnitude of the Sun 
(\citealt{BinneyMerrifield98}).  This relation assumes a \citet{Kroupa01} 
stellar initial mass function. The stellar masses have typical errors of $\sim 0.1$ dex due to uncertainties in stellar population (\citealt{Bell03}; \citealt{Conroy09}).

We use the SDSS Data Release 7 (DR7) spectrum of each object to calculate the 
$k$-correction and the color conversion from the $B_{\rm WFC3}$ and 
$I_{\rm WFC3}$ filters to conventional $B$ and $I$ filters. Strictly speaking, the SDSS spectrum, taken using a 3\asec\ fiber, covers the central $\sim 5.6$ kpc of the host galaxy (for median $z \approx 0.1$), much larger than the size scale of the bulge. Nevertheless, we will use it and confirm later that the stellar mass derived from the spectrum well represents the bulge mass. 

\begin{figure*}
\center{\includegraphics[scale=1.45]{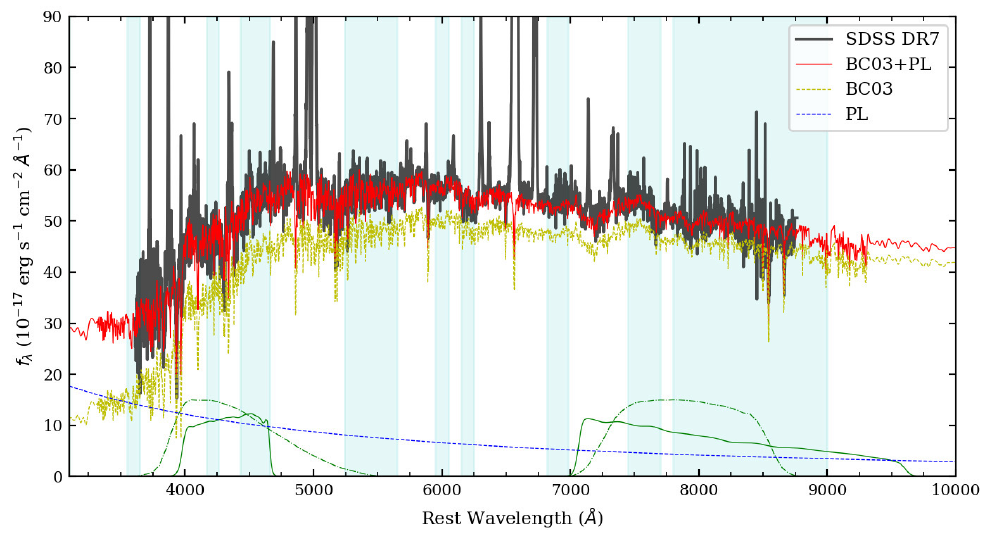}}
\caption{An example of fitting the rest-frame, Galactic extinction-corrected SDSS spectrum (black solid curve) of SDSS J105208.19+060915.1 to obtain a sufficiently broad stellar continuum. The shaded cyan areas show the parts of the spectrum used in our fits that avoid strong emission lines. The model consists of four BC03 stellar population synthesis components. A power-law AGN component is included to be consistent with the {\tt GALFIT} fit. The red solid curve is the total best-fit continuum containing all BC03 components and the power law; the yellow dashed curve is the stellar component (linear sum of the four BC03 components); the blue dashed line is the power-law component. The bandpasses of the WFC3 filters are presented as green solid curves, and those of the $B$ and $I$ filters are shown as green dash-dotted curves.}
\label{fig:SDSS_spec_fit}
\end{figure*}

Note that the spectral range of the SDSS spectrum is not wide enough to cover the $I_{\rm WFC3}$ bandpass. We, therefore, derive the stellar spectrum by fitting the SDSS spectrum (see Figure~\ref{fig:SDSS_spec_fit} for an example) with a \citet[BC03]{BC03} stellar population synthesis model consisting of four stellar components with ages in the range [0.04, 0.3], [0.3, 1.0], [0.0, 3.0], and 8.0 Gyr, assuming solar metallicity. Strong emission lines are excluded from the fit. Meanwhile, Galactic extinction is removed adopting $R_V = 3.1$ and the Milky Way extinction law of \citet{Cardelli89}.  As previously noted, contamination from nuclear scattered light is not entirely negligible for 14 of the quasars. For consistency, a power-law continuum representing the AGN component is included in the fits for these objects. The SDSS spectra of 24 targets have sufficiently high S/N ($\gtrsim15$) that their best-fit BC03 model yields a reduced $\chi^2 \approx 1$. For the remaining five objects with spectra of lower quality (S/N $<7$), we simply adopt a model consisting of an old (10 Gyr) and a young (2.1 Gyr) population (\citealt{CanalizoStockton13}). 

In view of the fact that the SDSS spectrum generally covers larger regions of 
the host galaxy than the scale of bulge, we consider the SDSS spectrum to be 
representative of the whole host galaxy and derive an alternative estimate of 
the bulge stellar mass.  We first obtain the total stellar mass of the host 
galaxy ($M_*$) using its integrated $I$-band magnitude and $(B-I)$ color by Equation~\ref{eq:M/Lratio}, and 
then we attribute a fraction $B/T$ thereof to the bulge ($M_{\rm bul,B/T}$).  
These alternative estimates of the bulge masses derived from $B/T$ agree well 
with those derived from bulge magnitudes: 
$\langle \log M_{\rm bul}- \log M_{\rm bul,B/T} \rangle= 0.03\pm 0.16$ dex. 

Another check on our stellar masses can be made using the subset of 19 objects 
that overlap with the MPA-JHU 
catalog\footnote{{\tt www.mpa-garching.mpg.de/SDSS/DR7/}} of spectral 
measurements from SDSS DR7.  The stellar masses in this catalog were derived
from spectral energy distribution fits to the DR7 photometry\footnote{See 
details in {\tt www.mpa-garching.mpg.de/SDSS/DR7/mass\_comp}.}  The
median total stellar masses from DR7 ($M_{*, \rm DR7}$) agree well with our 
estimates: $\langle \log M_*-\log M_{*, \rm DR7}\rangle = -0.01\pm 0.13$ dex. 

Table~\ref{tab:smass} lists the stellar masses derived in this work. 
For simplicity, in the following discussion, we simply adopt the bulge masses ($M_{\rm bul}$) 
estimated from Equation~(\ref{eq:M/Lratio}).  We note that none of the 
conclusions of this paper depends on which of the two bulge masses we choose.

\begin{deluxetable*}{ccccDDl}
\tabletypesize{\normalsize}
\fontsize{9.5}{10.}\selectfont
\tablecaption{\label{tab:smass} Stellar Masses and Color Gradients}
\tablehead{
\colhead{Object}& \colhead{$\log M_*$} & \colhead{$\log M_{\rm bul,\textit{B/T}}$}& \colhead{$\log M_{\rm bul}$}  &\multicolumn2c{$\nabla(B-I)_{\rm in}$}  &\multicolumn2c{$\nabla(B-I)_{\rm out}$}  &\colhead{Morphology} \\
\colhead{  } & \colhead{($M_\odot$)} & \colhead{($M_\odot$)}       & \colhead{($M_\odot$)}     \\
\colhead{(1)} & \colhead{(2)} & \colhead{(3)} & \colhead{(4)}  & \multicolumn2c{(5)}   & \multicolumn2c{(6)}     & \colhead{(7)}
}
\decimals
 
\startdata 
J011935.63      &10.71           &10.29                  &10.45          & 0.64      &$-$0.72            & spiral (unbarred)    \\[0.5ex]
J074751.56      &11.03           &10.14                  &10.06          & 1.24        & 1.65            & spiral (barred)        \\[0.5ex]
J075329.93      &11.00           &10.90                  &10.92          & 0.83      &$-$2.99            & merging     \\[0.5ex]
J075940.95      &10.67           & 9.97                  & 9.75          & 0.73        & 0.79            & lenticular     \\[0.5ex]
J080252.92      &11.27           &11.07                  &11.10          & 0.63      &$-$0.66            & merging     \\[0.5ex]
J080337.32      &11.30           &11.20                  &11.16        &$-$1.27      &$-$3.91            & lenticular     \\[0.5ex]
J080523.29      &11.12           &10.43                  &10.75          & 1.17      &$-$0.86            & merging        \\[0.5ex]
J081100.20      &11.18           &10.64                  &10.89          & 0.60      &$-$1.65            & spiral (unbarred)       \\[0.5ex]
J084107.06      &11.03           &11.03                  &11.03          & 0.64      &$-$2.43            & merging      \\[0.5ex]
J084344.99      &11.05           &10.83                  &10.79          & 0.07      &$-$2.44            & disturbed     \\[0.5ex]
J090754.07      &10.92           &10.26                  &10.23          & 0.90        & 1.29            & spiral (barred)         \\[0.5ex]
J091819.66      &10.26           &10.26                  &10.26          & 1.70      &$-$1.76            & elliptical     \\[0.5ex]
J093625.36      &10.28           & 9.73                  &10.11          & 1.15        & 1.17            & lenticular     \\[0.5ex]
J103408.59      &10.99           &10.76                  &10.66        &$-$0.10      &$-$2.20            & disturbed     \\[0.5ex]
J105208.19      &10.65           & 9.81                  & 9.97          & 0.20      &$-$0.60            & spiral (barred)         \\[0.5ex]
J110213.01      &10.95           &10.62                  &10.90          & 0.35      &$-$2.57            & merging      \\[0.5ex]
J111015.25      & 9.70           & 9.70                  & 9.70          & 1.34        & 1.23            & elliptical     \\[0.5ex]
J113710.77      &10.91           &10.91                  &10.91          & 0.39      &$-$1.57            & elliptical     \\[0.5ex]
J115326.42      &10.56           & 9.82                  & 9.79          & 1.02        & 1.48            & spiral (barred)         \\[0.5ex]
J123804.81      &10.93           & 9.73                  &10.01          & 1.52        & 0.13            & spiral (barred)         \\[0.5ex]
J125850.77      &10.96           &10.12                  &10.05          & 0.75        & 0.82            & spiral (barred)         \\[0.5ex]
J130038.09      &10.86           &10.46                  &10.30          & 0.48        & 0.63            & lenticular     \\[0.5ex]
J133542.49      &10.88           &10.88                  &10.88          & 0.36      &$-$1.47            & disturbed     \\[0.5ex]
J140541.21      &10.58           &10.32                  &10.27          & 0.41      &$-$0.05            & spiral (unbarred)    \\[0.5ex]
J140712.94      &10.97           &10.44                  &10.43          & 1.69        & 1.54            & spiral (barred)        \\[0.5ex]
J144038.09      &11.01           &10.14                  & 9.94        &$-$0.29        & 1.45            & merging     \\[0.5ex]
J145019.18      &11.14           &11.04                  &11.10          & 0.03      &$-$1.90            & lenticular     \\[0.5ex]
J155829.36      &10.71           & 9.76                  & 9.64          & 0.59        & 1.06            & spiral (barred)        \\[0.5ex]
J162436.40      &10.99           &10.46                  &10.38          & 1.21      &$-$0.13            & merging     \\[0.5ex]
\enddata

\vspace{0.2cm}
\begin{spacing}{1.2}
\fontsize{9.5}{9}\selectfont
\justifying \textbf{Notes.} 
Column (1): Galaxy name. 
Column (2): Total stellar mass of host galaxy. 
Column (3): Bulge mass calculated from $M_*$ and $B/T$. 
Column (4): Bulge mass calculated from $M/L$ ratio of Equation~(\ref{eq:M/Lratio}).
Column (5): Color gradients of 
the galaxy inner regions. The units are $\Delta$mag per dex in radius (arcsec).
Column (6): Color gradients of 
the galaxy outer regions.
Column (7): Morphological classification.
\end{spacing}
\vspace{0.5cm}
\end{deluxetable*}

\section{Results and Implications}
\label{sec:results}

\subsection{Evidence of Nuclear Star Formation}
\label{sec:results_SF}

Black hole accretion and star formation are often suggested to go hand in hand 
(\citealt{Springel05, Hopkins06}), but the verdict from observations is somewhat
mixed. While some studies of the stellar population of AGN host galaxies find 
nuclear and starburst activity to be broadly synchronized (e.g., 
\citealt{Tadhunter11, Bessiere14, Bessiere17}), others maintain that AGN 
activity significantly lags behind star formation (e.g., \citealt{Wild10,
CanalizoStockton13}). The color information from this study provides some 
fresh insights into this issue, from the point of view of luminous, obscured 
quasars that should be experiencing both intense star formation and black 
hole growth.

The color maps shown in Figure~\ref{fig:color_map} illustrate that in most of the hosts the central regions of the bulge are generally bluer than their outer regions.  This holds for most of the merging/disturbed hosts 
and most of the disk galaxies.  The few that do not follow this trend may be 
affected by dust reddening (i.e. some of the disturbed systems and two of the 
lenticulars with prominent inner dust lanes).  Consistent with the expected behavior of galaxies, almost all the global color profiles initially exhibit a negative color gradient in their outermost regions (become redder toward smaller radii).  However, upon reaching the central regions, roughly on the scale of the effective radius of the bulge, most of the color profiles flatten or even turn over at smaller radii.

To quantify this effect, we derive the color gradients of the inner 
and outer regions of the host galaxies (see Table~\ref{tab:smass}), and we 
directly compare them with a 
control sample of inactive galaxies measured in the same manner from the 
Carnegie-Irvine Galaxy Survey (CGS; \citealt{Ho11}).  The inner region 
is defined as the range from half of the PSF FWHM (to avoid possible AGN 
contamination, even after subtracting the nucleus) to the effective radius of 
the bulge (\re); the outer region extends between \re\ and 2.5\re.  Following 
 \citet{Li2011}, the color gradient, corrected for Galactic extinction, 
is calculated as the slope of the color profile, representing the change in 
color per dex in radius:
\begin{equation}
\label{eq:color_grad}
\nabla(\rm color)_{\rm in/out}= \dfrac{({\rm color})_{\textit{r}_1}-({\rm color})_{\textit{r}_0}}{\log {\textit{r}_1} - \log {\textit{r}_0}},
\end{equation} 
\noindent
where $r_1$ and $r_0$ correspond to \re\ and 0.5 FWHM of PSF for the inner 
region, and 2.5\re\ and \re\ for the outer region.   As in \citet{Li2011},
color gradients larger than 0.1 are considered positive, between $-0.1$ 
and 0 are flat, and smaller than $-0.1$ are negative. A positive slope 
indicates that the galaxy is getting bluer toward the center.
 
\begin{figure}
\centering
\includegraphics[scale=0.84]{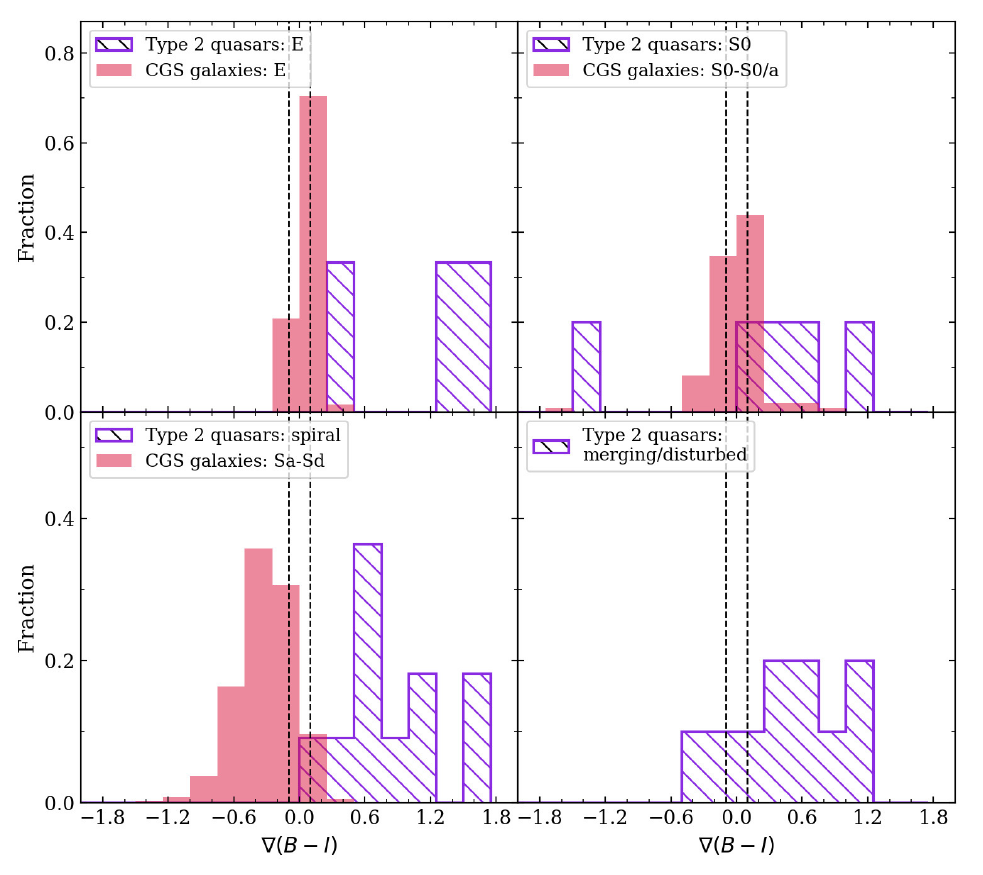}  
\caption{Color gradients of the inner regions (radius $< R_e$) of type 2 quasars (hatched purple histograms), divided by morphology. The inner gradients of normal (inactive) CGS galaxies with similar morphologies (\citealt{Li2011}) are shown for comparison as the filled red histograms; no CGS comparison is given for the merging/disturbed category. The vertical dashed lines in each panel mark the adopted boundaries for negative ($\nabla(B - I) < -0.1$), flat ($-0.1 \leqslant \nabla(B - I) \leqslant 0.1$), and positive ($\nabla(B - I) >0.1$) color gradients. A positive color gradient means that the color becomes bluer toward the center. The central regions of most quasars have positive color gradients, opposite to the trend in normal galaxies.}
\label{fig:colorgrad_comp}
\end{figure}

Regardless of the sign of the outer color gradients, most of our objects exhibit \textit{positive} inner $B-I$ color gradients. Figure~\ref{fig:colorgrad_comp} compares the inner color gradients of our sample, split by morphological types, relative to a comparison sample drawn from CGS  (\citealt{Li2011}).
The two samples are clearly different.  Apart from two S0 hosts with 
central dust lanes, all the early-type hosts of type 2 quasars (ellipticals and S0s) have positive 
inner color gradients.  By comparison, 85\% of the ellipticals and 64\% of the 
S0s in CGS have flat inner color profiles.  The contrast is even more pronounced for spiral galaxies, for which {\it all}\ spiral AGN hosts have clearly positive inner color gradients, whereas negative gradients characterize the majority ($65-88$\%, depending on the exact morphological type) of the bulges of inactive spirals.
Even merging/disturbed systems hosting type 2 quasars, whose environments expected to be dusty, mostly exhibit positive inner color gradients.  The host galaxies of type 2 quasars have a preponderance of blue central regions compared to normal galaxies of similar Hubble type, consistent with enhanced ongoing or recent star formation.

\begin{figure}
\raggedright{\includegraphics[scale=0.91]{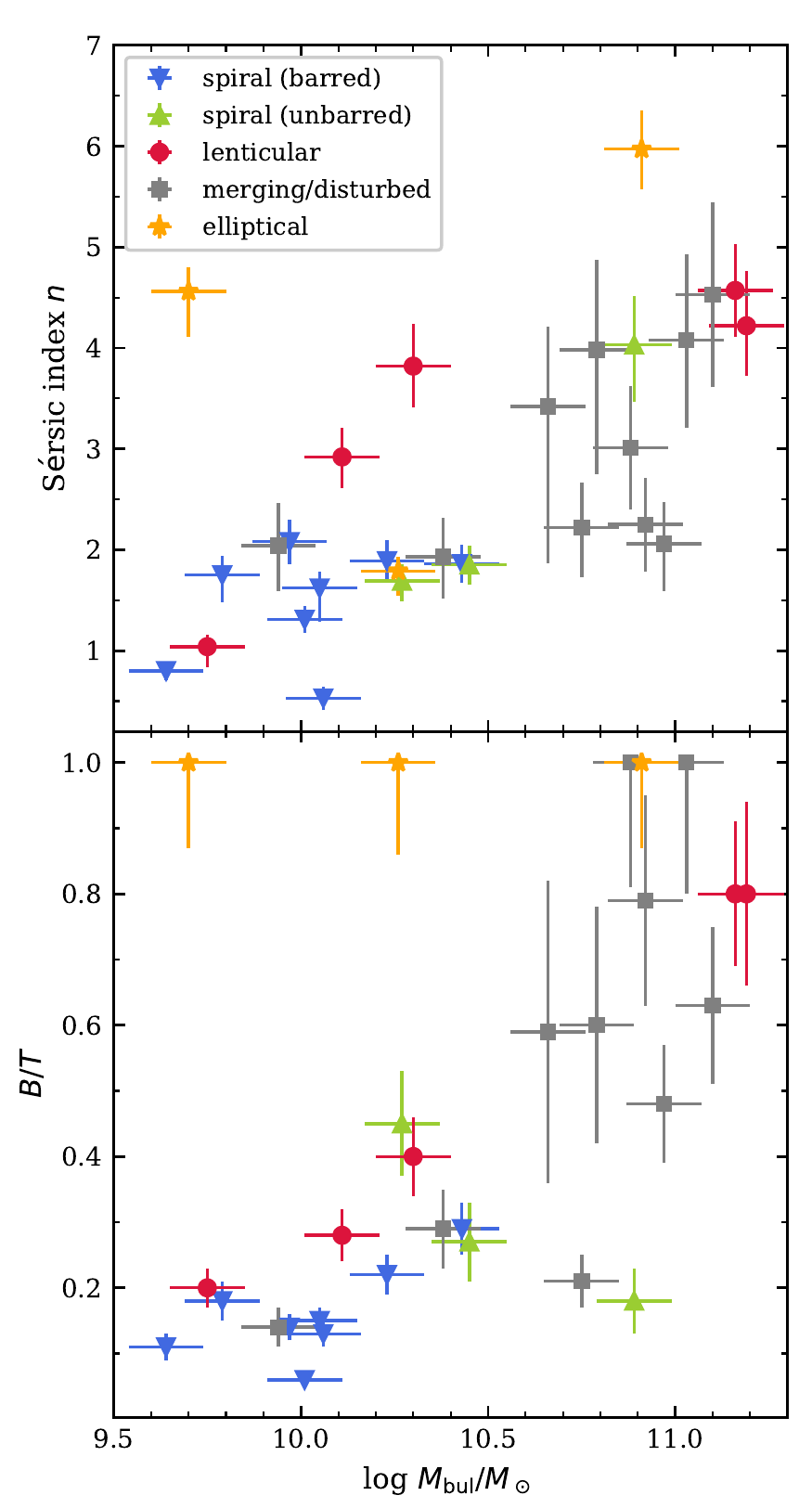}}
\caption{Dependence of bulge S\'ersic index $n$ (top) and $B/T$ (bottom) with 
bulge stellar mass.  The symbols are color-coded by morphological type
of the host.}
\label{fig:structure_bulgemass}
\end{figure}

Additionally, less direct but nevertheless compelling evidence for enhanced central star 
formation in type 2 quasars comes from inspection of the detailed structural 
parameters of the host galaxies.  Figure~\ref{fig:structure_bulgemass} 
illustrates the distributions of \sersic\ index ($n$) and bulge-to-total ratio 
($B/T$) as a function of bulge stellar mass ($M_{\rm bul}$) for the sample. Not
unexpectedly, the merging/disturbed hosts (grey symbols) tend to have more 
massive ($M_{\rm bul} \gtrsim 10^{10.5}\,M_\odot$), more prominent ($B/T 
\gtrsim 0.2$) bulges with relatively large \sersic\ indices (median $n=2.63$),
akin to classical bulges and consistent with the expectation that major mergers
build classical bulges (\citealt{KormendyKennicutt04, FisherDrory08}).  The 
undisturbed early types in the sample (lenticulars and ellipticals; red and 
orange symbols, respectively) span a wide range in mass, but their \sersic\ 
indices are large (median $n=4.02$), again consistent with classical bulges.
The late-type hosts are strikingly different.  Most of the barred and unbarred 
spirals (blue and green symbols) have bulges of lower mass ($M_{\rm bul} 
\lesssim 10^{10.5}\,M_\odot$), lower prominence ($B/T \lesssim 0.2$), and 
characteristically lower \sersic\ indices ($n \lesssim 2$).  They conform to
the typical properties of pseudo bulges (\citealt{KormendyKennicutt04,
FisherDrory08}).  

\begin{figure}
\raggedright{\includegraphics[scale=0.9]{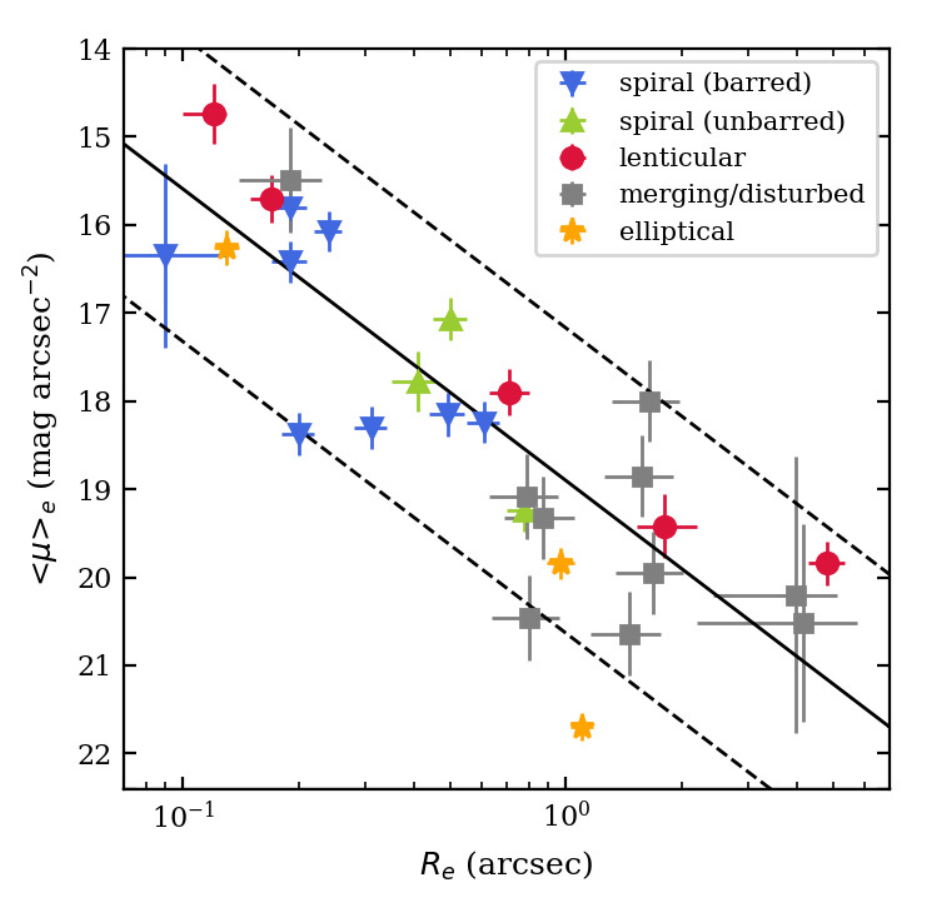}}
\caption{Kormendy relation for the bulges of the host galaxies, color-coded 
by morphological type.  The solid line is the best-fit $\langle \mu \rangle _e 
-$\re\ relation for the bulges of elliptical, lenticular, and 
merging/disturbed galaxies, and the two dashed lines represent its $\pm 3 
\sigma$ scatter.}
\label{fig:Kormendy_rel}
\end{figure}

The Kormendy relation (\citealt{Kormendy77}), an inverse correlation between 
the effective radius (\re) of spheroids and their surface brightness ($\mu_e$) 
within \re, provides a useful empirical tool to distinguish bulge types.  At 
a given \re, pseudo bulges have lower $\mu_e$ than classical bulges or 
elliptical galaxies (\citealt{KormendyKennicutt04,Gadotti09,FisherDrory10}).  Figure~\ref{fig:Kormendy_rel} (top) examines the Kormendy relation of 
our sample using best-fit parameters from the $I_{\rm WFC3}$ band.  We fit the 
relation only for the classical bulges (i.e. ellipticals and the bulges of 
merging/disturbed and lenticular galaxies), which is denoted by the solid
line.  It is surprising that the late-type galaxies, which, as argued above, 
contain pseudo bulges, do {\it not}\  depart from the relation of classical
bulges.  This grossly deviates from the behavior of inactive galaxies, for which
pseudo bulges scatter systematically below the locus of classical bulges and 
ellipticals.  Why do pseudo bulges not appear in the Kormendy relation of 
type 2 quasar host galaxies?  The simplest and most plausible explanation is 
that the bulge effective surface brightnesses have been enhanced because of 
excess light from recent or ongoing star formation.  Together with 
the color gradients previously discussed, we conclude that the central regions 
of these obscured quasars have a characteristically young stellar population, 
presumably associated with a recent episode of star formation.
A similar trend is also reported in the hosts of type 1 AGNs (\citealt{KimHo19}).

\subsection{Are Mergers Important for Triggering AGNs?}
\label{sec:discuss_merging}

The role of mergers in governing AGN activity remains a vexingly 
controversial topic.  From a theoretical point of view, major external 
triggers are thought to be necessary to supply the high mass accretion rates 
needed to sustain the luminosities of the most powerful quasars (\citealt{Shlosman90}).  The less stringent fueling requirements of weaker AGNs can 
be met through internal secular processes, such as angular momentum transported 
by bars (e.g., \citealt{Ho97, Somerville08, HopkinsHernquist09}).
A substantial body of observational evidence broadly supports this thesis.  From their 
analysis of a large sample of AGNs spanning a wide range of bolometric 
luminosities and redshifts, \citet{Treister12} report a strong correlation 
between the fraction of host galaxies experiencing major mergers and AGN 
luminosity.   A high incidence of merger features is frequenctly found for 
luminous AGNs, be they unobscured (e.g., \citealt{Letawe10, Liu12, Hong15}), 
dust-reddened (\citealt{Urrutia08, Kocevski15, Fan16}), or highly obscured 
(\citealt{VillarMartin11,Bessiere12,Wylezalek16,Donley18,UrbanoMayorgas18}).  
But not everyone agrees.  A significant number of studies of luminous AGNs at 
moderate ($0.5 < z < 0.8$; \citealt{Villforth14}) and high ($z\lesssim 2$; 
\citealt{Schawinski11, Schawinski12, Mechtley16, Villforth17}) redshifts 
dismiss the importance of major mergers in driving nuclear activity.

One of the most surprising results of this study is the sheer diversity
of the morphologies of the host galaxies of obscured quasars.  As summarized 
in Section~\ref{sec:introduction}, the traditional gas-rich major merger scenario for quasar evolution predicts that type~2 quasars should be hosted by morphologically 
highly disturbed galaxies.  This basic expectation is not supported by our 
observations, at least not for a sizable fraction of our sample.  Among the 29 
objects studied, only 10 (34\%) show clear morphological signatures of 
interactions or mergers, rising at most to 13 (45\%) if we regard the three 
ellipticals as post-merger products.  The majority (55\%; 11 spirals and 5 
lenticulars) possess normal disks.  Even if we accept that lenticulars may be 
remnants of major mergers (\citealt{ElicheMoral18})---by no means a 
universal view (\citealt{Kormendy09})---a substantial fraction (38\%) are 
incontrovertibly ordinary unbarred or barred late-type spiral galaxies. 
Although it is stated that disks can survive from gas-rich major mergers under some circumstances 
(e.g., \citealt{BarnesHernquist96, SpringelHernquist05, Hopkins09}), the 
remnants cannot significantly fuel central black holes 
(\citealt{HopkinsHernquist09}), and the regrowth of the disk 
(\citealt{Hopkins09, Bundy10}) takes much longer that the typical quasar 
lifetime (e.g., \citealt{Porciani04, Hopkins05, Shen07}). Therefore, the 
late-type host galaxies of type 2 quasars, which most likely possess pseudo 
bulges (Section~\ref{sec:results_SF}), probably never experienced major 
mergers during their lifetimes.  Secular processes presumably were responsible 
for their black hole growth. Interestingly, most of the spirals (8 out of 11) 
contain a bar (Figure~\ref{fig:morph_class}).

The subset of our type 2 quasars hosted by spirals has a typical [O~III] 
$\lambda$5007 luminosity of $10^9\,L_\odot$, not dissimilar from those hosted 
by earlier type galaxies or mergers.   For an [O~III] bolometric correction of 
600 (\citealt{KauffmannHeckman09}), this corresponds to a bolometric luminosity 
of $L_{\rm bol} = 2.3\times10^{45}\,{\rm erg}\,{\rm s}^{-1}$, or a mass 
accretion rate of $\dot{M} = (\epsilon c^2)^{-1} L_{\rm bol} \approx 0.4\,
{M_\odot}\,{\rm yr}^{-1}$, where $c$ is the speed of light and $\epsilon=0.1$ 
is the radiative efficiency.  This level of fueling is modest, reflecting the 
fact that our sample of low-redshift obscured AGNs, although technically 
considered as ``quasars'', are, in fact, quite modest in power.  Bar-driven 
gravitational torques may suffice to transport cold gas at this rate to the 
central regions of spiral galaxies (\citealt{Haan09}).

\section{Conclusions}
\label{sec:conculde}
We observed 29 local ($z\approx 0.04 - 0.4$), optically selected  type 2 
quasars in rest-frame $B$ and $I$  using WFC3 on {\it HST}, to study the 
stellar properties of their host galaxies and explore their triggering 
mechanism.  We classify the morphologies, perform detailed two-dimensional decomposition to study structural properties of the hosts, analyze their optical colors and color gradients, and derive bulge stellar masses.  Our principal findings can be summarized as follows:

\begin{itemize}

\item Only a minority (34\%) of the host galaxies exhibit clear merging or disturbed features.  A significant fraction ($38\%$) of the hosts are late-type, mostly disk-dominated spiral galaxies. Major mergers do not seem to play a dominant role in triggering nuclear activity in nearby obscured quasars, but the luminosities of these sources, and hence their mass accretion rates, are not sufficiently high to severely challenge the major merger model for quasar evolution.  Indeed, we argue that secular processes alone may suffice to supply their modest 
fueling rates. 

\item The central regions of most of the host galaxies are bluer than their outer parts, indicating nearly ubiquitous recent or ongoing star formation.

\item While merging/disturbed systems and early-type (lenticular and elliptical)
hosts tend to have \sersic\ indices ($n\geqslant 2$) and bulge-to-total ratios 
($B/T \gtrsim 0.2$) expected of classical bulges, the late-type hosts possess 
pseudo bulges with $n < 2$ and $B/T\lesssim 0.2$.  However, unlike inactive 
galaxies, the pseudo bulges hosting type 2 quasars have systematically higher
central surface brightnesses because of the excess light from young stars.

\end{itemize}

\acknowledgments
We thank an anonymous referee for many helpful comments. We are grateful to MinZhi Kong and Hua Gao for fruitful discussions and suggestions. This work was supported by the National Key R\&D Program of China (2016YFA0400702) and the National Science Foundation of China (11473002, 11721303). Minjin Kim was supported by the National Research Foundation of Korea (NRF) grant funded by the Korea government (MSIT) (No. 2017R1C1B2002879).

\appendix
\setcounter{figure}{0}
\renewcommand{\thefigure}{\arabic{figure}}

\section{Best-fit Decompositions for the Sample}
Here are the best-fit results of structure decomposition for the host galaxies of our 29 type 2 quasar by GALFIT. For each quasar, the 2-D original $I_{\rm WFC3}$ image, model, and residual images are illustrated in the upper panel, and the 1-D distributions of ellipticity, position angle, surface density, and residual of intensity along with radius are shown in the bottom panel. 

\begin{figure*}
\center{\includegraphics[scale=0.94]{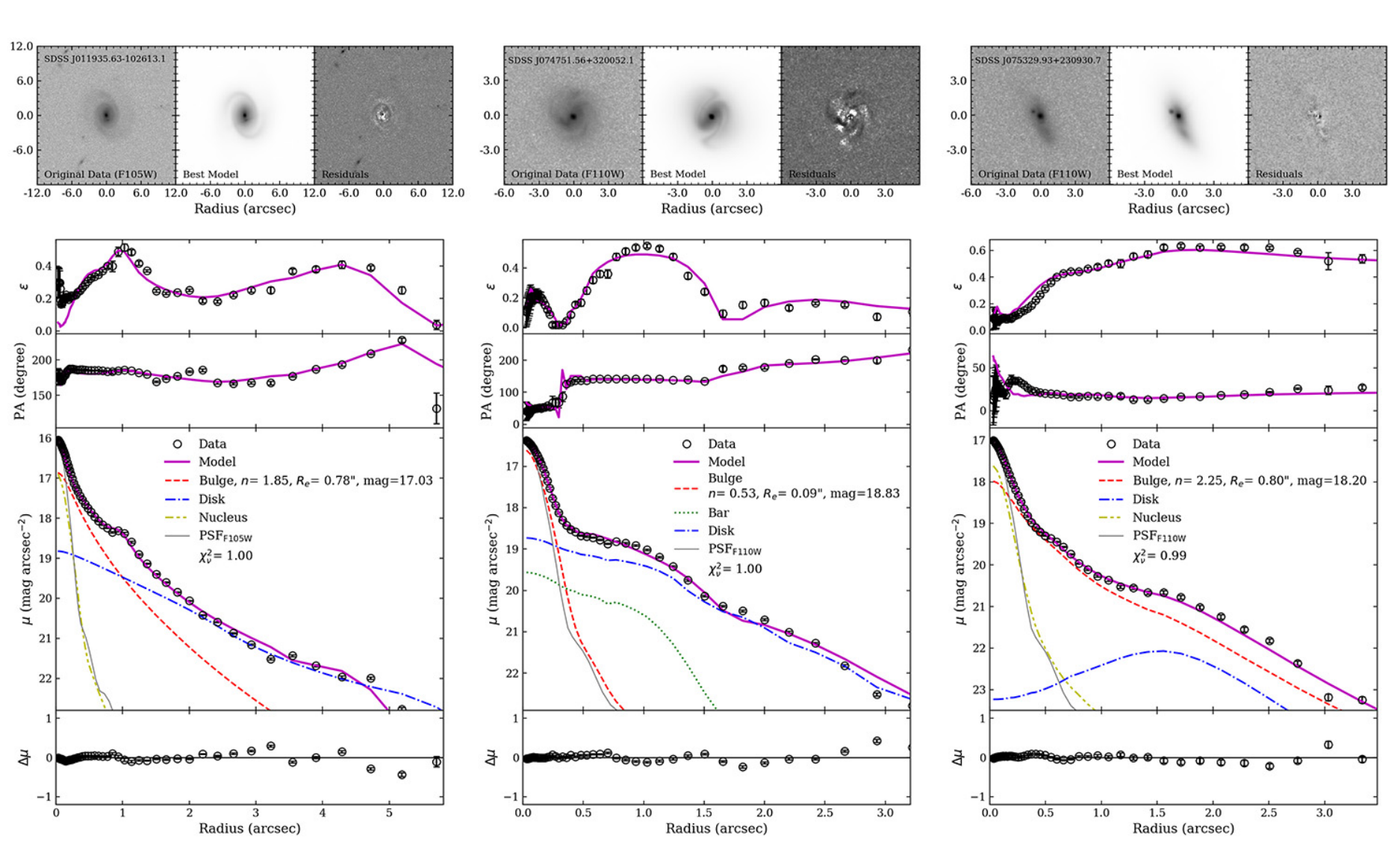}}
\caption{GALFIT decompositions for the 29 type 2 quasars.  For each object, the top row 
shows (left) the observed $I_{\rm WFC3}$ image, (middle) the best-fit 2-D 
model, and (right) the residual image.  The bottom panels give, respectively, 
the radial profile of ellipticity, position angle, surface brightness, are 
the residuals between the best-fit model and the observed surface brightness 
profile.  The best-fit 1-D model (magenta line) was extracted from the 
best-fit 2-D model. The individual subcomponents are listed in the legend, 
which also gives the final reduced chi-squared of the fit.  The profile for 
the PSF is plotted as a gray line.}
\end{figure*}
\renewcommand{\thefigure}{\arabic{figure} (Cont.)}
\addtocounter{figure}{-1}

\begin{figure*}
\center{\includegraphics[scale=0.94]{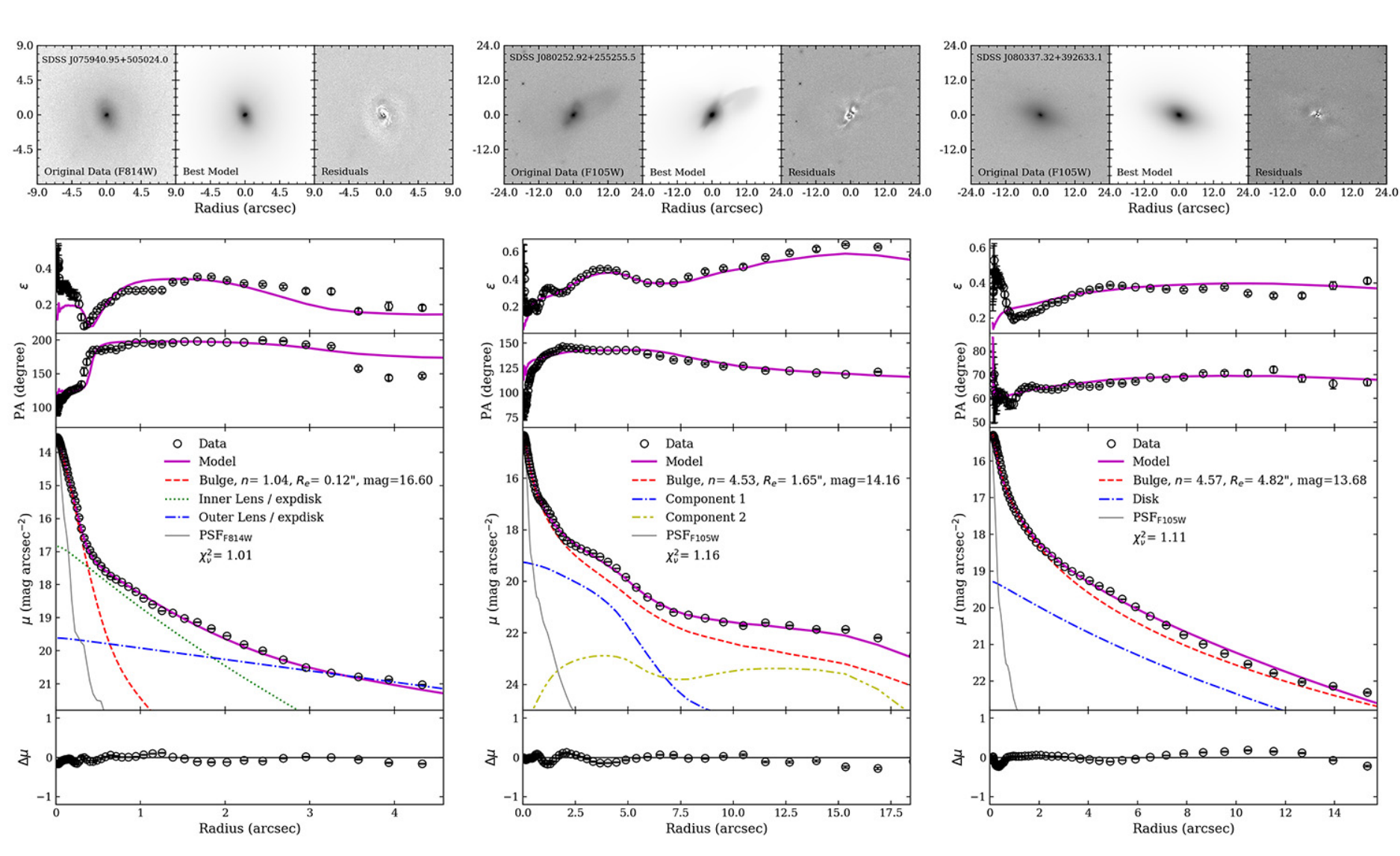}}
\caption{GALFIT decompositions for type 2 quasars.}
\end{figure*}
\renewcommand{\thefigure}{\arabic{figure} (Cont.)}
\addtocounter{figure}{-1}

\begin{figure*}
\center{\includegraphics[scale=0.46]{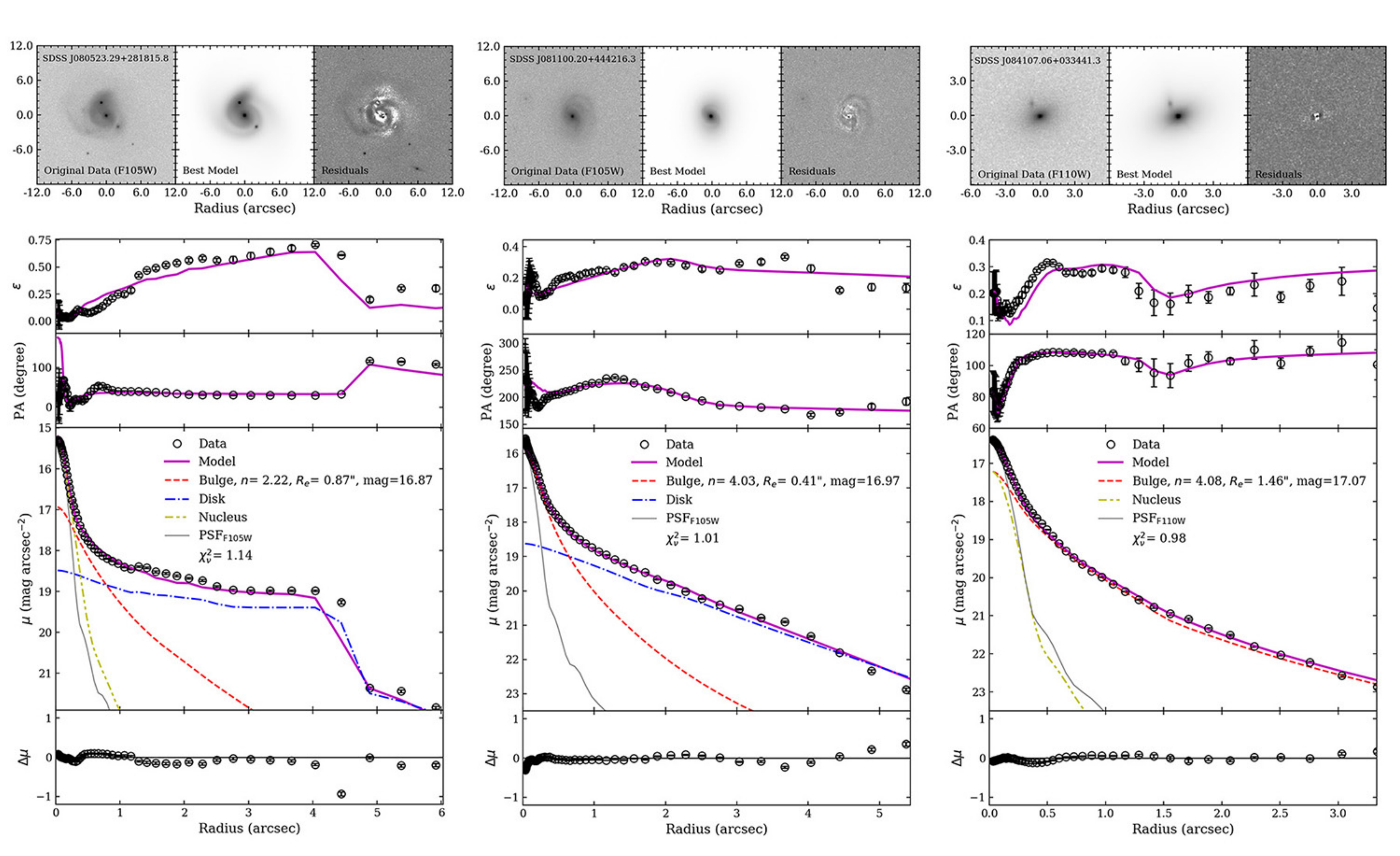}}
\caption{GALFIT decompositions for  type 2 quasars.}
\end{figure*}
\renewcommand{\thefigure}{\arabic{figure} (Cont.)}
\addtocounter{figure}{-1}

\begin{figure*}
\center{\includegraphics[scale=0.46]{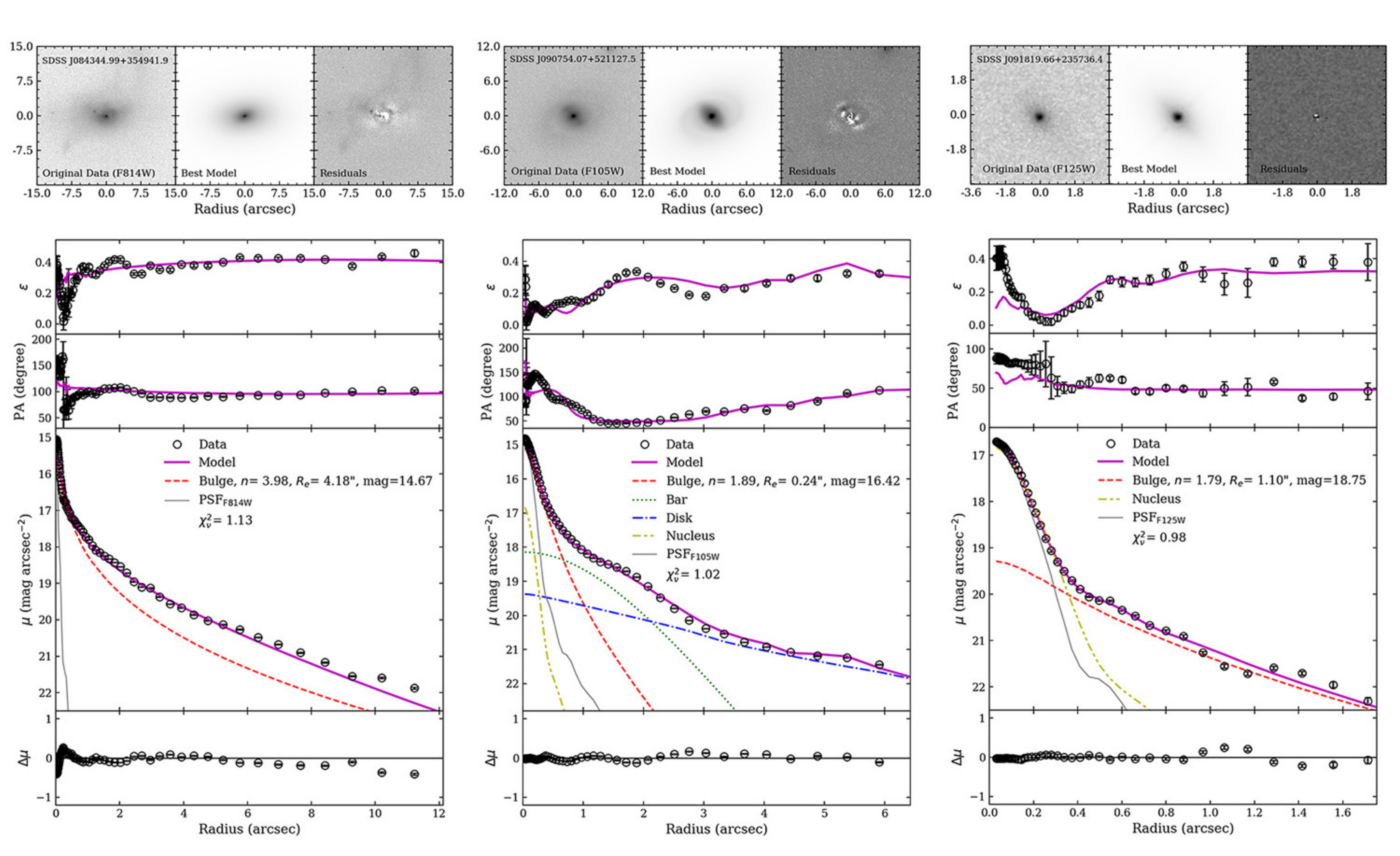}}
\caption{GALFIT decompositions for  type 2 quasars.}
\end{figure*}
\renewcommand{\thefigure}{\arabic{figure} (Cont.)}
\addtocounter{figure}{-1}

\begin{figure*}
\center{\includegraphics[scale=0.46]{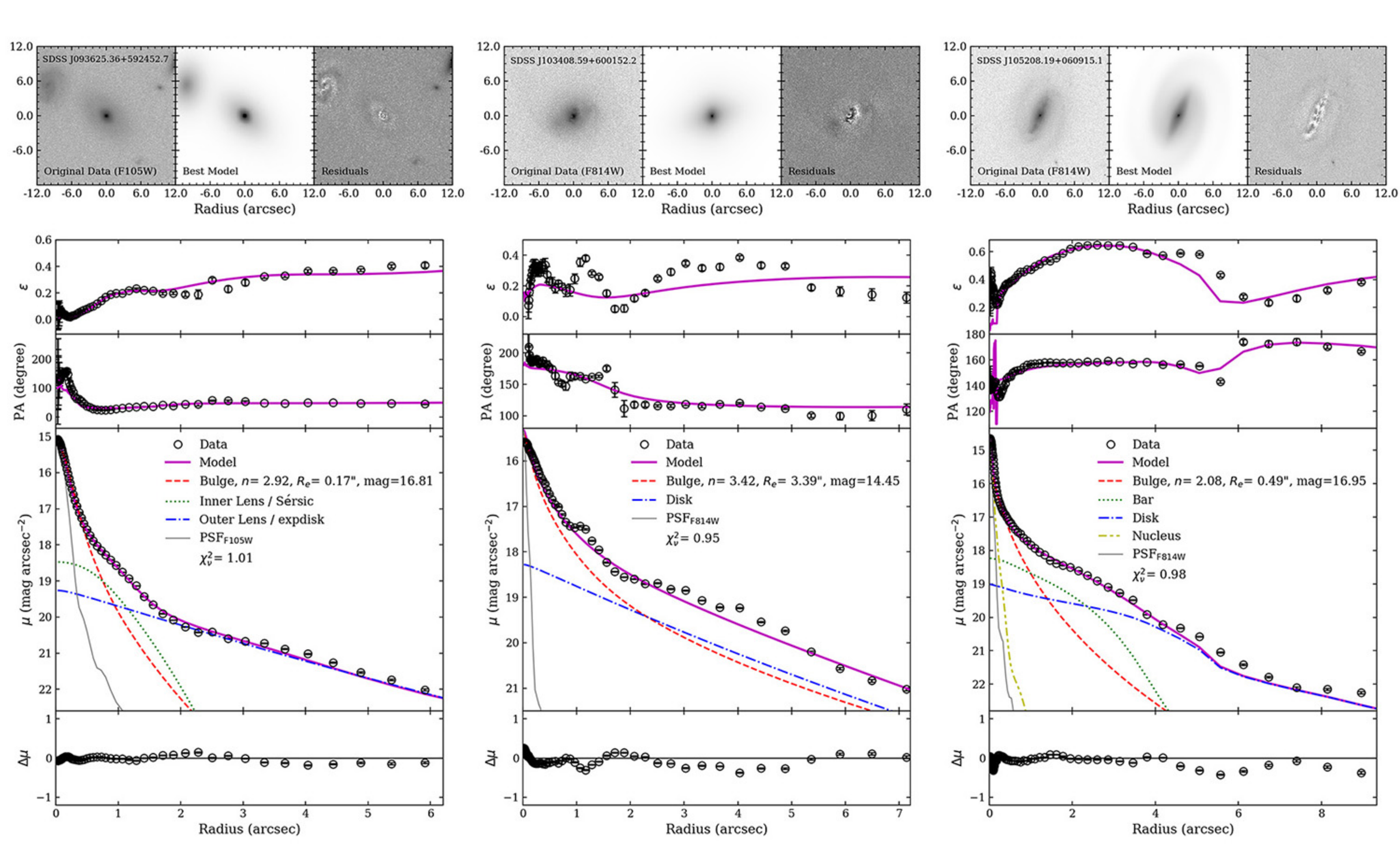}}
\caption{GALFIT decompositions for  type 2 quasars.}
\end{figure*}
\renewcommand{\thefigure}{\arabic{figure} (Cont.)}
\addtocounter{figure}{-1}

\begin{figure*}
\center{\includegraphics[scale=0.46]{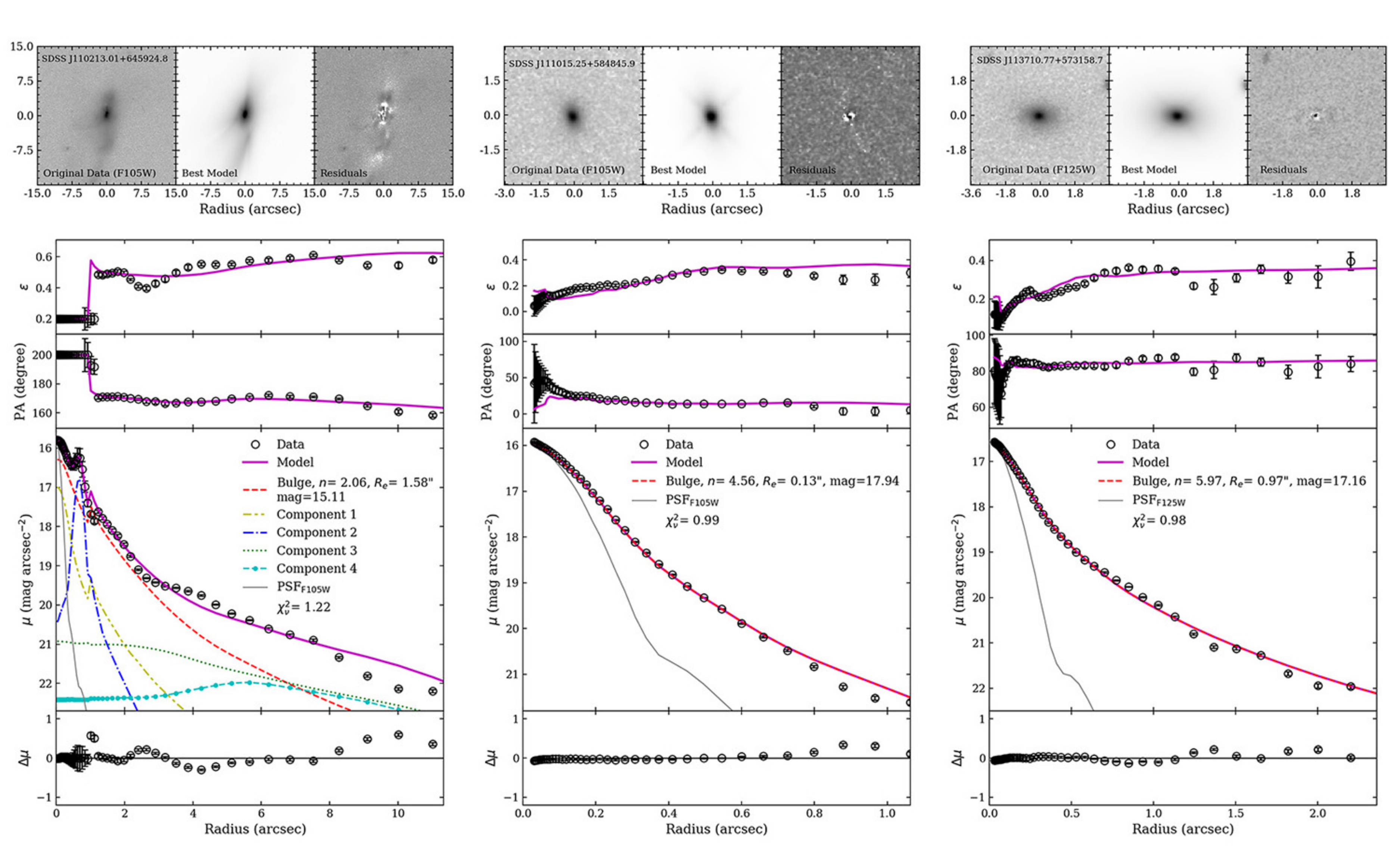}}
\caption{GALFIT decompositions for  type 2 quasars.}
\end{figure*}
\renewcommand{\thefigure}{\arabic{figure} (Cont.)}
\addtocounter{figure}{-1}

\begin{figure*}
\center{\includegraphics[scale=0.46]{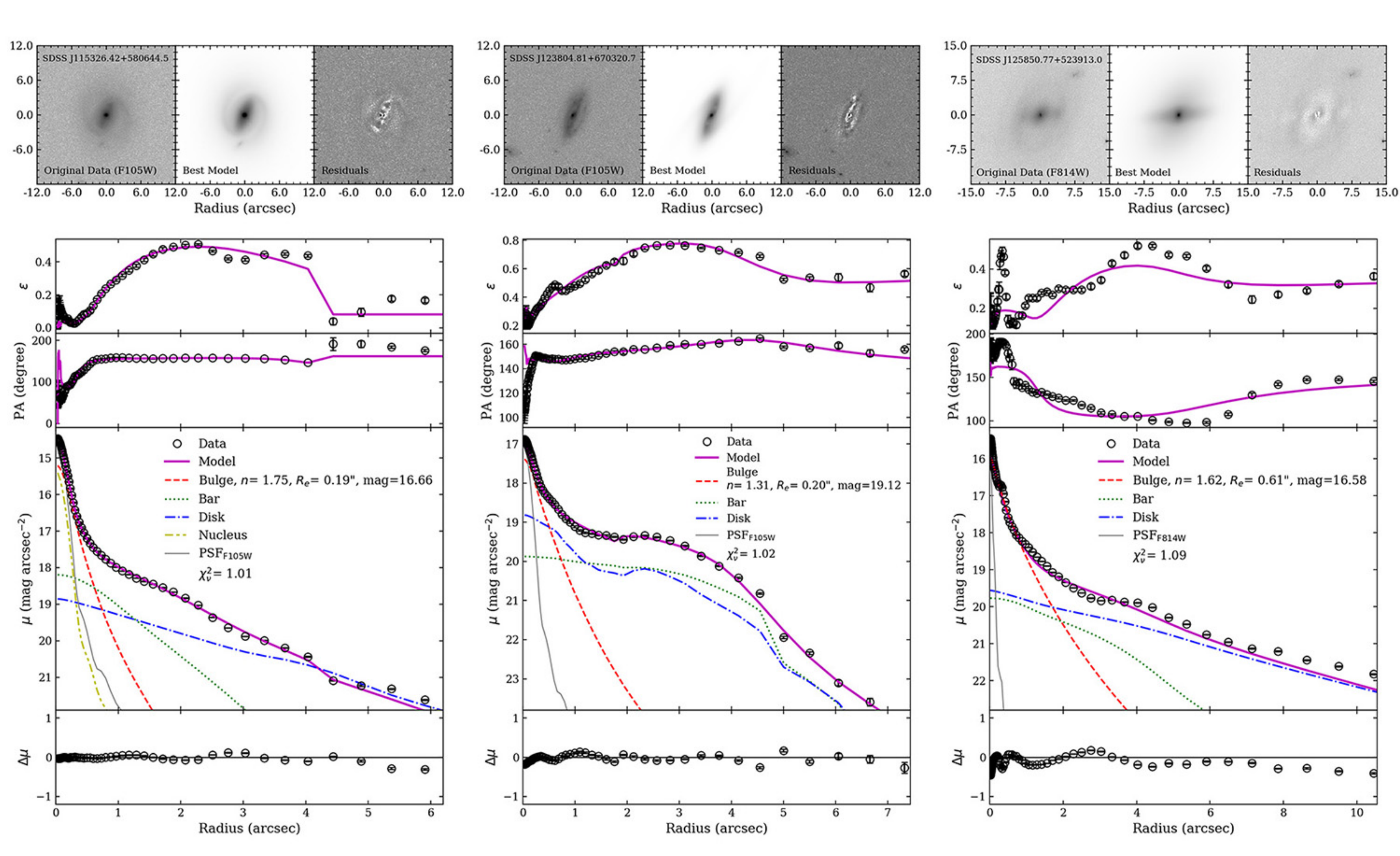}}
\caption{GALFIT decompositions for  type 2 quasars.}
\end{figure*}
\renewcommand{\thefigure}{\arabic{figure} (Cont.)}
\addtocounter{figure}{-1}

\begin{figure*}
\center{\includegraphics[scale=0.46]{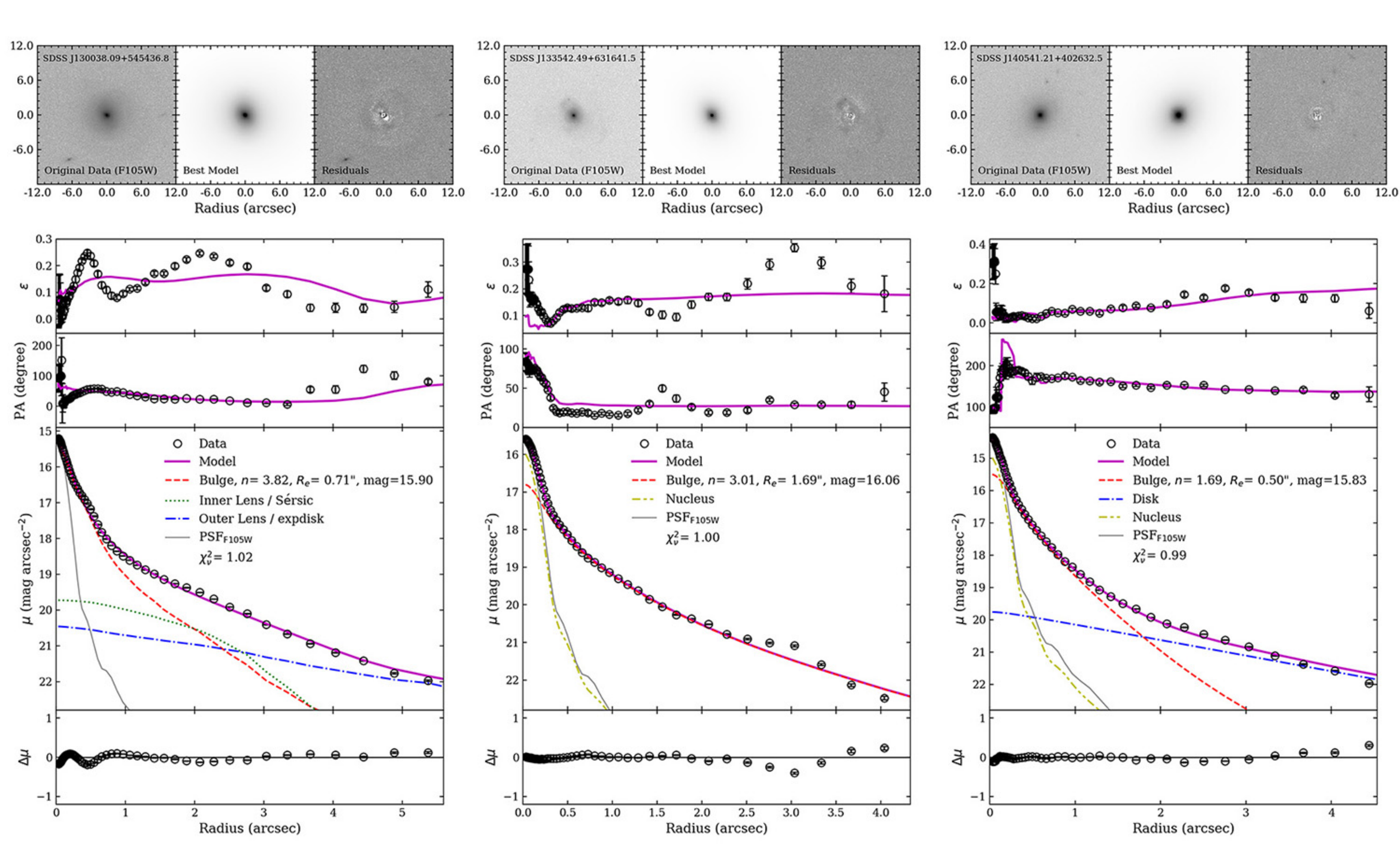}}
\caption{GALFIT decompositions for  type 2 quasars.}
\end{figure*}
\renewcommand{\thefigure}{\arabic{figure} (Cont.)}
\addtocounter{figure}{-1}

\begin{figure*}
\center{\includegraphics[scale=0.46]{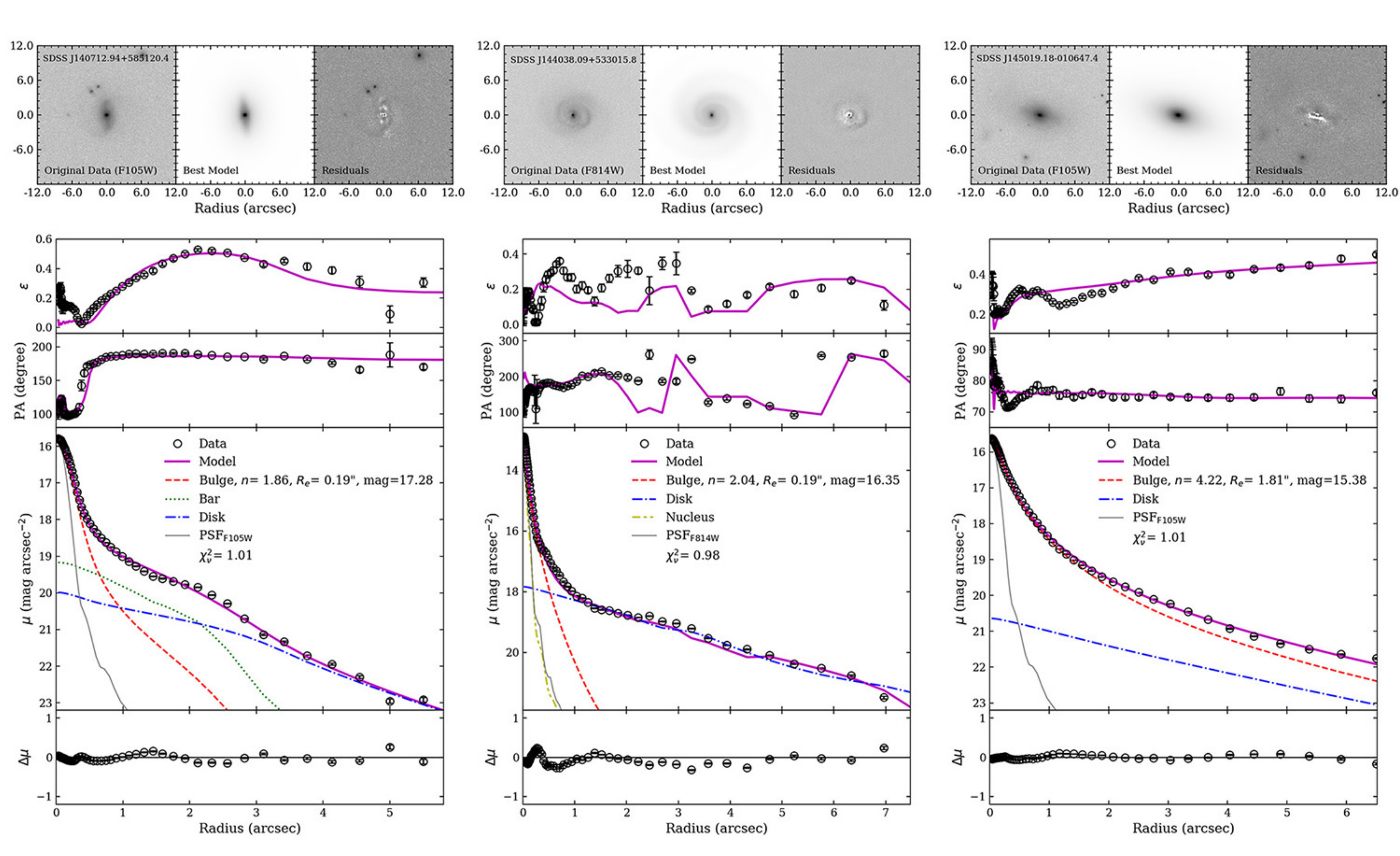}}
\caption{GALFIT decompositions for  type 2 quasars.}
\end{figure*}
\renewcommand{\thefigure}{\arabic{figure} (Cont.)}
\addtocounter{figure}{-1}

\begin{figure*}
\center{\includegraphics[scale=0.46]{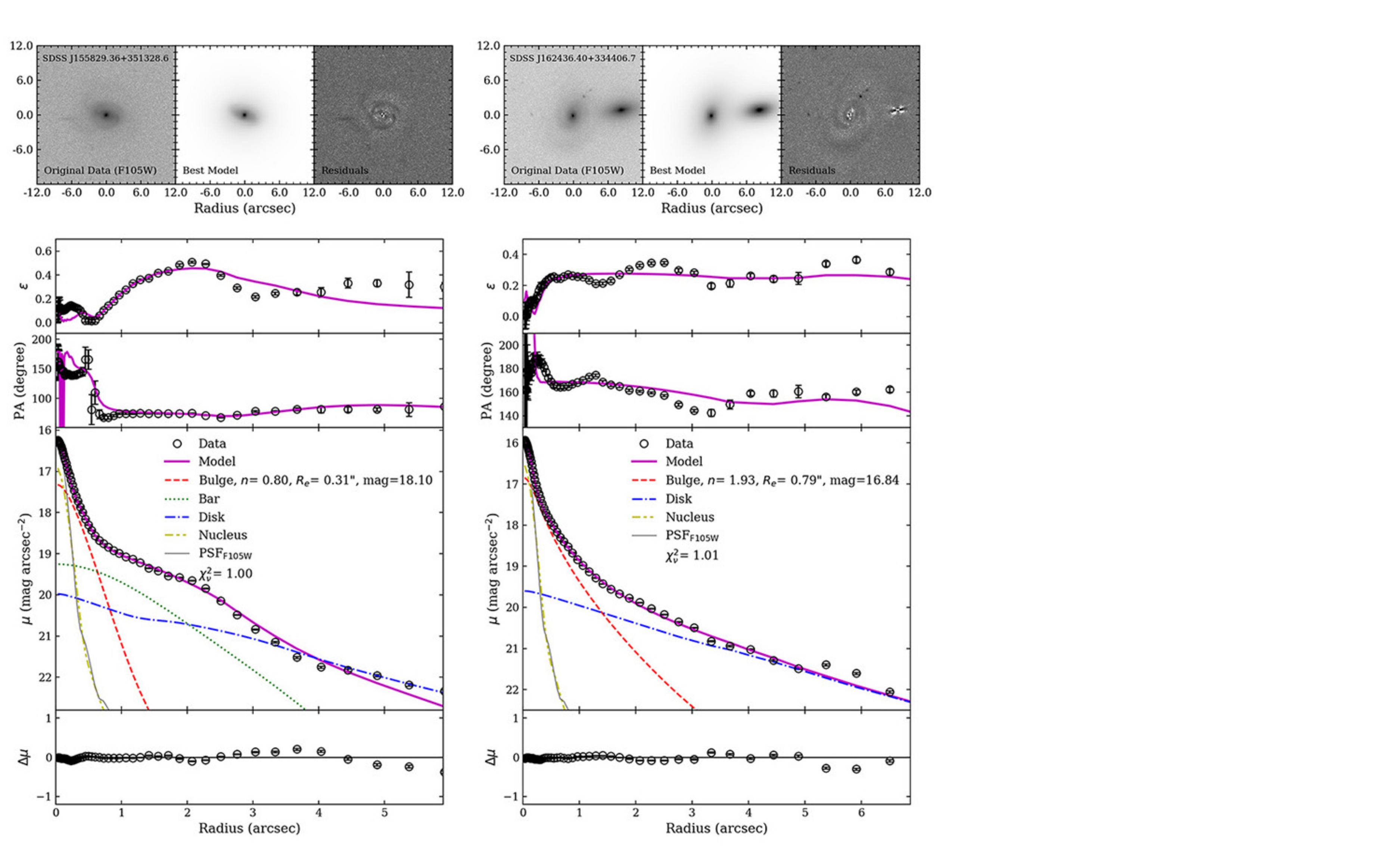}}
\caption{GALFIT decompositions for  type 2 quasars.}
\end{figure*}


\bibliography{QSO2s_paper1}

\begin{thebibliography}{}
\expandafter\ifx\csname natexlab\endcsname\relax\def\natexlab#1{#1}\fi

\bibitem[{{Adelman-McCarthy} {et~al.}(2008){Adelman-McCarthy}, {Ag{\"u}eros},
  {Allam}, {Allende Prieto}, {Anderson}, {Anderson}, {Annis}, {Bahcall},
  {Bailer-Jones}, {Baldry}, {Barentine}, {Bassett}, {Becker}, {Beers}, {Bell},
  {Berlind}, {Bernardi}, {Blanton}, {Bochanski}, {Boroski}, {Brinchmann},
  {Brinkmann}, {Brunner}, {Budav{\'a}ri}, {Carliles}, {Carr}, {Castander},
  {Cinabro}, {Cool}, {Covey}, {Csabai}, {Cunha}, {Davenport}, {Dilday}, {Doi},
  {Eisenstein}, {Evans}, {Fan}, {Finkbeiner}, {Friedman}, {Frieman},
  {Fukugita}, {G{\"a}nsicke}, {Gates}, {Gillespie}, {Glazebrook}, {Gray},
  {Grebel}, {Gunn}, {Gurbani}, {Hall}, {Harding}, {Harvanek}, {Hawley},
  {Hayes}, {Heckman}, {Hendry}, {Hindsley}, {Hirata}, {Hogan}, {Hogg}, {Hyde},
  {Ichikawa}, {Ivezi{\'c}}, {Jester}, {Johnson}, {Jorgensen}, {Juri{\'c}},
  {Kent}, {Kessler}, {Kleinman}, {Knapp}, {Kron}, {Krzesinski}, {Kuropatkin},
  {Lamb}, {Lampeitl}, {Lebedeva}, {Lee}, {French Leger}, {L{\'e}pine}, {Lima},
  {Lin}, {Long}, {Loomis}, {Loveday}, {Lupton}, {Malanushenko}, {Malanushenko},
  {Mandelbaum}, {Margon}, {Marriner}, {Mart{\'{\i}}nez-Delgado}, {Matsubara},
  {McGehee}, {McKay}, {Meiksin}, {Morrison}, {Munn}, {Nakajima}, {Neilsen},
  {Newberg}, {Nichol}, {Nicinski}, {Nieto-Santisteban}, {Nitta}, {Okamura},
  {Owen}, {Oyaizu}, {Padmanabhan}, {Pan}, {Park}, {Peoples}, {Pier}, {Pope},
  {Purger}, {Raddick}, {Re Fiorentin}, {Richards}, {Richmond}, {Riess}, {Rix},
  {Rockosi}, {Sako}, {Schlegel}, {Schneider}, {Schreiber}, {Schwope}, {Seljak},
  {Sesar}, {Sheldon}, {Shimasaku}, {Sivarani}, {Allyn Smith}, {Snedden},
  {Steinmetz}, {Strauss}, {SubbaRao}, {Suto}, {Szalay}, {Szapudi}, {Szkody},
  {Tegmark}, {Thakar}, {Tremonti}, {Tucker}, {Uomoto}, {Vanden Berk},
  {Vandenberg}, {Vidrih}, {Vogeley}, {Voges}, {Vogt}, {Wadadekar}, {Weinberg},
  {West}, {White}, {Wilhite}, {Yanny}, {Yocum}, {York}, {Zehavi}, \&
  {Zucker}}]{Adelman08}
{Adelman-McCarthy}, J.~K., {Ag{\"u}eros}, M.~A., {Allam}, S.~S., {et~al.} 2008,
  \apjs, 175, 297

\bibitem[{{Alexander} \& {Hickox}(2012)}]{AlexanderHickox12}
{Alexander}, D.~M., \& {Hickox}, R.~C. 2012, \nar, 56, 93

\bibitem[{{Alexandroff} {et~al.}(2016){Alexandroff}, {Zakamska}, {van Velzen},
  {Greene}, \& {Strauss}}]{Alexandroff16}
{Alexandroff}, R.~M., {Zakamska}, N.~L., {van Velzen}, S., {Greene}, J.~E., \&
  {Strauss}, M.~A. 2016, \mnras, 463, 3056

\bibitem[{{Antonucci} \& {Miller}(1985)}]{AntonucciMiller85}
{Antonucci}, R.~R.~J., \& {Miller}, J.~S. 1985, \apj, 297, 621

\bibitem[{{Balcells} {et~al.}(2003){Balcells}, {Graham},
  {Dom{\'{\i}}nguez-Palmero}, \& {Peletier}}]{Balcells03}
{Balcells}, M., {Graham}, A.~W., {Dom{\'{\i}}nguez-Palmero}, L., \& {Peletier},
  R.~F. 2003, \apjl, 582, L79

\bibitem[{{Barden} {et~al.}(2012){Barden}, {H{\"a}u{\ss}ler}, {Peng},
  {McIntosh}, \& {Guo}}]{Barden12}
{Barden}, M., {H{\"a}u{\ss}ler}, B., {Peng}, C.~Y., {McIntosh}, D.~H., \&
  {Guo}, Y. 2012, \mnras, 422, 449

\bibitem[{{Barden} {et~al.}(2008){Barden}, {Jahnke}, \&
  {H{\"a}u{\ss}ler}}]{Barden08}
{Barden}, M., {Jahnke}, K., \& {H{\"a}u{\ss}ler}, B. 2008, \apjs, 175, 105

\bibitem[{{Barnes} \& {Hernquist}(1996)}]{BarnesHernquist96}
{Barnes}, J.~E., \& {Hernquist}, L. 1996, \apj, 471, 115

\bibitem[{{Bell} \& {de Jong}(2001)}]{Bell01}
{Bell}, E.~F., \& {de Jong}, R.~S. 2001, \apj, 550, 212

\bibitem[{{Bell} {et~al.}(2003){Bell}, {McIntosh}, {Katz}, \&
  {Weinberg}}]{Bell03}
{Bell}, E.~F., {McIntosh}, D.~H., {Katz}, N., \& {Weinberg}, M.~D. 2003, \apjs,
  149, 289

\bibitem[{{Bennett} {et~al.}(2014){Bennett}, {Larson}, {Weiland}, \&
  {Hinshaw}}]{Bennett14}
{Bennett}, C.~L., {Larson}, D., {Weiland}, J.~L., \& {Hinshaw}, G. 2014, \apj,
  794, 135

\bibitem[{{Bertin} \& {Arnouts}(1996)}]{BA96}
{Bertin}, E., \& {Arnouts}, S. 1996, \aaps, 117, 393

\bibitem[{{Bessiere} {et~al.}(2012){Bessiere}, {Tadhunter}, {Ramos Almeida}, \&
  {Villar Mart{\'{\i}}n}}]{Bessiere12}
{Bessiere}, P.~S., {Tadhunter}, C.~N., {Ramos Almeida}, C., \& {Villar
  Mart{\'{\i}}n}, M. 2012, \mnras, 426, 276

\bibitem[{{Bessiere} {et~al.}(2014){Bessiere}, {Tadhunter}, {Ramos Almeida}, \&
  {Villar Mart{\'{\i}}n}}]{Bessiere14}
---. 2014, \mnras, 438, 1839

\bibitem[{{Bessiere} {et~al.}(2017){Bessiere}, {Tadhunter}, {Ramos Almeida},
  {Villar Mart{\'{\i}}n}, \& {Cabrera-Lavers}}]{Bessiere17}
{Bessiere}, P.~S., {Tadhunter}, C.~N., {Ramos Almeida}, C., {Villar
  Mart{\'{\i}}n}, M., \& {Cabrera-Lavers}, A. 2017, \mnras, 466, 3887

\bibitem[{{Binney} \& {Merrifield}(1998)}]{BinneyMerrifield98}
{Binney}, J., \& {Merrifield}, M. 1998, {Galactic Astronomy}

\bibitem[{{B{\"o}ker} {et~al.}(2002){B{\"o}ker}, {Laine}, {van der Marel},
  {Sarzi}, {Rix}, {Ho}, \& {Shields}}]{Boker02}
{B{\"o}ker}, T., {Laine}, S., {van der Marel}, R.~P., {et~al.} 2002, \aj, 123,
  1389

\bibitem[{{Boroson} \& {Green}(1992)}]{Boroson92}
{Boroson}, T.~A., \& {Green}, R.~F. 1992, \apjs, 80, 109

\bibitem[{{Bruzual} \& {Charlot}(2003)}]{BC03}
{Bruzual}, G., \& {Charlot}, S. 2003, \mnras, 344, 1000

\bibitem[{{Bundy} {et~al.}(2010){Bundy}, {Scarlata}, {Carollo}, {Ellis},
  {Drory}, {Hopkins}, {Salvato}, {Leauthaud}, {Koekemoer}, {Murray}, {Ilbert},
  {Oesch}, {Ma}, {Capak}, {Pozzetti}, \& {Scoville}}]{Bundy10}
{Bundy}, K., {Scarlata}, C., {Carollo}, C.~M., {et~al.} 2010, \apj, 719, 1969

\bibitem[{{Canalizo} \& {Stockton}(2013)}]{CanalizoStockton13}
{Canalizo}, G., \& {Stockton}, A. 2013, \apj, 772, 132

\bibitem[{{Cardelli} {et~al.}(1989){Cardelli}, {Clayton}, \&
  {Mathis}}]{Cardelli89}
{Cardelli}, J.~A., {Clayton}, G.~C., \& {Mathis}, J.~S. 1989, \apj, 345, 245

\bibitem[{{Cisternas} {et~al.}(2011){Cisternas}, {Jahnke}, {Inskip},
  {Kartaltepe}, {Koekemoer}, {Lisker}, {Robaina}, {Scodeggio}, {Sheth},
  {Trump}, {Andrae}, {Miyaji}, {Lusso}, {Brusa}, {Capak}, {Cappelluti},
  {Civano}, {Ilbert}, {Impey}, {Leauthaud}, {Lilly}, {Salvato}, {Scoville}, \&
  {Taniguchi}}]{Cisternas11}
{Cisternas}, M., {Jahnke}, K., {Inskip}, K.~J., {et~al.} 2011, \apj, 726, 57

\bibitem[{{Conroy} {et~al.}(2009){Conroy}, {Gunn}, \& {White}}]{Conroy09}
{Conroy}, C., {Gunn}, J.~E., \& {White}, M. 2009, \apj, 699, 486

\bibitem[{{de Jong}(1996)}]{deJong96}
{de Jong}, R.~S. 1996, \aaps, 118, 557

\bibitem[{{Di Matteo} {et~al.}(2005){Di Matteo}, {Springel}, \&
  {Hernquist}}]{DiMatteo05}
{Di Matteo}, T., {Springel}, V., \& {Hernquist}, L. 2005, \nat, 433, 604

\bibitem[{{Dicken} {et~al.}(2014){Dicken}, {Tadhunter}, {Morganti}, {Axon},
  {Robinson}, {Magagnoli}, {Kharb}, {Ramos Almeida}, {Mingo}, {Hardcastle},
  {Nesvadba}, {Singh}, {Kouwenhoven}, {Rose}, {Spoon}, {Inskip}, \&
  {Holt}}]{Dicken14}
{Dicken}, D., {Tadhunter}, C., {Morganti}, R., {et~al.} 2014, \apj, 788, 98

\bibitem[{{Donley} {et~al.}(2018){Donley}, {Kartaltepe}, {Kocevski}, {Salvato},
  {Santini}, {Suh}, {Civano}, {Koekemoer}, {Trump}, {Brusa}, {Cardamone},
  {Castro}, {Cisternas}, {Conselice}, {Croton}, {Hathi}, {Liu}, {Lucas},
  {Nair}, {Rosario}, {Sanders}, {Simmons}, {Villforth}, {Alexander}, {Bell},
  {Faber}, {Grogin}, {Lotz}, {McIntosh}, \& {Nagao}}]{Donley18}
{Donley}, J.~L., {Kartaltepe}, J., {Kocevski}, D., {et~al.} 2018, \apj, 853, 63

\bibitem[{{Dunlop} {et~al.}(2003){Dunlop}, {McLure}, {Kukula}, {Baum}, {O'Dea},
  \& {Hughes}}]{Dunlop03}
{Dunlop}, J.~S., {McLure}, R.~J., {Kukula}, M.~J., {et~al.} 2003, \mnras, 340,
  1095

\bibitem[{{Eliche-Moral} {et~al.}(2018){Eliche-Moral},
  {Rodr{\'{\i}}guez-P{\'e}rez}, {Borlaff}, {Querejeta}, \&
  {Tapia}}]{ElicheMoral18}
{Eliche-Moral}, M.~C., {Rodr{\'{\i}}guez-P{\'e}rez}, C., {Borlaff}, A.,
  {Querejeta}, M., \& {Tapia}, T. 2018, \aap, 617, A113

\bibitem[{{Fan} {et~al.}(2016){Fan}, {Han}, {Fang}, {Gao}, {Zhang}, {Jiang},
  {Wu}, {Yang}, \& {Li}}]{Fan16}
{Fan}, L., {Han}, Y., {Fang}, G., {et~al.} 2016, \apjl, 822, L32

\bibitem[{{Ferrarese} \& {Merritt}(2000)}]{FerrareseMerritt00}
{Ferrarese}, L., \& {Merritt}, D. 2000, \apjl, 539, L9

\bibitem[{{Fisher} \& {Drory}(2008)}]{FisherDrory08}
{Fisher}, D.~B., \& {Drory}, N. 2008, \aj, 136, 773

\bibitem[{{Fisher} \& {Drory}(2010)}]{FisherDrory10}
---. 2010, \apj, 716, 942

\bibitem[{{Floyd} {et~al.}(2004){Floyd}, {Kukula}, {Dunlop}, {McLure},
  {Miller}, {Percival}, {Baum}, \& {O'Dea}}]{Floyd04}
{Floyd}, D.~J.~E., {Kukula}, M.~J., {Dunlop}, J.~S., {et~al.} 2004, \mnras,
  355, 196

\bibitem[{{Freeman}(1966)}]{Freeman66}
{Freeman}, K.~C. 1966, \mnras, 133, 47

\bibitem[{{Gadotti}(2008)}]{Gadotti08}
{Gadotti}, D.~A. 2008, \mnras, 384, 420

\bibitem[{{Gadotti}(2009)}]{Gadotti09}
---. 2009, \mnras, 393, 1531

\bibitem[{{Gao} \& {Ho}(2017)}]{GaoHo17}
{Gao}, H., \& {Ho}, L.~C. 2017, \apj, 845, 114

\bibitem[{{Gao} {et~al.}(2018){Gao}, {Ho}, {Barth}, \& {Li}}]{Gao18}
{Gao}, H., {Ho}, L.~C., {Barth}, A.~J., \& {Li}, Z.-Y. 2018, \apj, 862, 100

\bibitem[{{Gao} {et~al.}(2019){Gao}, {Ho}, {Barth}, \& {Li}}]{Gao19}
---. 2019, \apjs, submitted

\bibitem[{{Gebhardt} {et~al.}(2000){Gebhardt}, {Bender}, {Bower}, {Dressler},
  {Faber}, {Filippenko}, {Green}, {Grillmair}, {Ho}, {Kormendy}, {Lauer},
  {Magorrian}, {Pinkney}, {Richstone}, \& {Tremaine}}]{Gebhardt00}
{Gebhardt}, K., {Bender}, R., {Bower}, G., {et~al.} 2000, \apjl, 539, L13

\bibitem[{{Greene} {et~al.}(2008){Greene}, {Ho}, \& {Barth}}]{Greene08}
{Greene}, J.~E., {Ho}, L.~C., \& {Barth}, A.~J. 2008, \apj, 688, 159

\bibitem[{{Haan} {et~al.}(2009){Haan}, {Schinnerer}, {Emsellem},
  {Garc{\'{\i}}a-Burillo}, {Combes}, {Mundell}, \& {Rix}}]{Haan09}
{Haan}, S., {Schinnerer}, E., {Emsellem}, E., {et~al.} 2009, \apj, 692, 1623

\bibitem[{{Heckman} {et~al.}(2005){Heckman}, {Ptak}, {Hornschemeier}, \&
  {Kauffmann}}]{Heckman05}
{Heckman}, T.~M., {Ptak}, A., {Hornschemeier}, A., \& {Kauffmann}, G. 2005,
  \apj, 634, 161

\bibitem[{{Ho} {et~al.}(1997){Ho}, {Filippenko}, \& {Sargent}}]{Ho97}
{Ho}, L.~C., {Filippenko}, A.~V., \& {Sargent}, W.~L.~W. 1997, \apj, 487, 591

\bibitem[{{Ho} {et~al.}(2011){Ho}, {Li}, {Barth}, {Seigar}, \& {Peng}}]{Ho11}
{Ho}, L.~C., {Li}, Z.-Y., {Barth}, A.~J., {Seigar}, M.~S., \& {Peng}, C.~Y.
  2011, \apjs, 197, 21

\bibitem[{{Hong} {et~al.}(2015){Hong}, {Im}, {Kim}, \& {Ho}}]{Hong15}
{Hong}, J., {Im}, M., {Kim}, M., \& {Ho}, L.~C. 2015, \apj, 804, 34

\bibitem[{{Hopkins} {et~al.}(2009){Hopkins}, {Cox}, {Younger}, \&
  {Hernquist}}]{Hopkins09}
{Hopkins}, P.~F., {Cox}, T.~J., {Younger}, J.~D., \& {Hernquist}, L. 2009,
  \apj, 691, 1168

\bibitem[{{Hopkins} \& {Hernquist}(2009)}]{HopkinsHernquist09}
{Hopkins}, P.~F., \& {Hernquist}, L. 2009, \apj, 694, 599

\bibitem[{{Hopkins} {et~al.}(2005){Hopkins}, {Hernquist}, {Cox}, {Di Matteo},
  {Martini}, {Robertson}, \& {Springel}}]{Hopkins05}
{Hopkins}, P.~F., {Hernquist}, L., {Cox}, T.~J., {et~al.} 2005, \apj, 630, 705

\bibitem[{{Hopkins} {et~al.}(2008){Hopkins}, {Hernquist}, {Cox}, \& {Kere{\v
  s}}}]{Hopkins08}
{Hopkins}, P.~F., {Hernquist}, L., {Cox}, T.~J., \& {Kere{\v s}}, D. 2008,
  \apjs, 175, 356

\bibitem[{{Hopkins} {et~al.}(2014){Hopkins}, {Kocevski}, \&
  {Bundy}}]{Hopkins14}
{Hopkins}, P.~F., {Kocevski}, D.~D., \& {Bundy}, K. 2014, \mnras, 445, 823

\bibitem[{{Hopkins} {et~al.}(2006){Hopkins}, {Somerville}, {Hernquist}, {Cox},
  {Robertson}, \& {Li}}]{Hopkins06}
{Hopkins}, P.~F., {Somerville}, R.~S., {Hernquist}, L., {et~al.} 2006, \apj,
  652, 864

\bibitem[{{Huang} {et~al.}(2013){Huang}, {Ho}, {Peng}, {Li}, \&
  {Barth}}]{Huang13}
{Huang}, S., {Ho}, L.~C., {Peng}, C.~Y., {Li}, Z.-Y., \& {Barth}, A.~J. 2013,
  \apj, 766, 47

\bibitem[{{Ilbert} {et~al.}(2005){Ilbert}, {Tresse}, {Zucca}, {Bardelli},
  {Arnouts}, {Zamorani}, {Pozzetti}, {Bottini}, {Garilli}, {Le Brun}, {Le
  F{\`e}vre}, {Maccagni}, {Picat}, {Scaramella}, {Scodeggio}, {Vettolani},
  {Zanichelli}, {Adami}, {Arnaboldi}, {Bolzonella}, {Cappi}, {Charlot},
  {Contini}, {Foucaud}, {Franzetti}, {Gavignaud}, {Guzzo}, {Iovino},
  {McCracken}, {Marano}, {Marinoni}, {Mathez}, {Mazure}, {Meneux}, {Merighi},
  {Paltani}, {Pello}, {Pollo}, {Radovich}, {Bondi}, {Bongiorno}, {Busarello},
  {Ciliegi}, {Lamareille}, {Mellier}, {Merluzzi}, {Ripepi}, \&
  {Rizzo}}]{Ilbert05}
{Ilbert}, O., {Tresse}, L., {Zucca}, E., {et~al.} 2005, \aap, 439, 863

\bibitem[{{Jiang} {et~al.}(2011){Jiang}, {Greene}, {Ho}, {Xiao}, \&
  {Barth}}]{Jiang11}
{Jiang}, Y.-F., {Greene}, J.~E., {Ho}, L.~C., {Xiao}, T., \& {Barth}, A.~J.
  2011, \apj, 742, 68

\bibitem[{{Jogee}(2006)}]{Jogee06}
{Jogee}, S. 2006, in Lecture Notes in Physics, Berlin Springer Verlag, Vol.
  693, Physics of Active Galactic Nuclei at all Scales, ed. D.~{Alloin}, 143

\bibitem[{{Kauffmann} \& {Heckman}(2009)}]{KauffmannHeckman09}
{Kauffmann}, G., \& {Heckman}, T.~M. 2009, \mnras, 397, 135

\bibitem[{{Kim} \& {Ho}(2019)}]{KimHo19}
{Kim}, M., \& {Ho}, L.~C. 2019, arXiv e-prints, arXiv:1903.08796

\bibitem[{{Kim} {et~al.}(2008{\natexlab{a}}){Kim}, {Ho}, {Peng}, {Barth}, \&
  {Im}}]{Kim08a}
{Kim}, M., {Ho}, L.~C., {Peng}, C.~Y., {Barth}, A.~J., \& {Im}, M.
  2008{\natexlab{a}}, \apjs, 179, 283

\bibitem[{{Kim} {et~al.}(2017){Kim}, {Ho}, {Peng}, {Barth}, \& {Im}}]{Kim17}
---. 2017, \apjs, 232, 21

\bibitem[{{Kim} {et~al.}(2008{\natexlab{b}}){Kim}, {Ho}, {Peng}, {Barth}, {Im},
  {Martini}, \& {Nelson}}]{Kim08b}
{Kim}, M., {Ho}, L.~C., {Peng}, C.~Y., {et~al.} 2008{\natexlab{b}}, \apj, 687,
  767

\bibitem[{{Kocevski} {et~al.}(2015){Kocevski}, {Brightman}, {Nandra},
  {Koekemoer}, {Salvato}, {Aird}, {Bell}, {Hsu}, {Kartaltepe}, {Koo}, {Lotz},
  {McIntosh}, {Mozena}, {Rosario}, \& {Trump}}]{Kocevski15}
{Kocevski}, D.~D., {Brightman}, M., {Nandra}, K., {et~al.} 2015, \apj, 814, 104

\bibitem[{{Kong} \& {Ho}(2018)}]{KongHo18}
{Kong}, M., \& {Ho}, L.~C. 2018, \apj, 859, 116

\bibitem[{{Kormendy}(1977)}]{Kormendy77}
{Kormendy}, J. 1977, \apj, 218, 333

\bibitem[{{Kormendy} {et~al.}(2009){Kormendy}, {Fisher}, {Cornell}, \&
  {Bender}}]{Kormendy09}
{Kormendy}, J., {Fisher}, D.~B., {Cornell}, M.~E., \& {Bender}, R. 2009, \apjs,
  182, 216

\bibitem[{{Kormendy} \& {Ho}(2013)}]{KormendyHo13}
{Kormendy}, J., \& {Ho}, L.~C. 2013, \araa, 51, 511

\bibitem[{{Kormendy} \& {Kennicutt}(2004)}]{KormendyKennicutt04}
{Kormendy}, J., \& {Kennicutt}, Jr., R.~C. 2004, \araa, 42, 603

\bibitem[{{Krist} {et~al.}(2011){Krist}, {Hook}, \& {Stoehr}}]{Krist11}
{Krist}, J.~E., {Hook}, R.~N., \& {Stoehr}, F. 2011, in \procspie, Vol. 8127,
  Optical Modeling and Performance Predictions V, 81270J

\bibitem[{{Kron}(1980)}]{Kron80}
{Kron}, R.~G. 1980, \apjs, 43, 305

\bibitem[{{Kroupa}(2001)}]{Kroupa01}
{Kroupa}, P. 2001, \mnras, 322, 231

\bibitem[{{LaMassa} {et~al.}(2009){LaMassa}, {Heckman}, {Ptak},
  {Hornschemeier}, {Martins}, {Sonnentrucker}, \& {Tremonti}}]{LaMassa09}
{LaMassa}, S.~M., {Heckman}, T.~M., {Ptak}, A., {et~al.} 2009, \apj, 705, 568

\bibitem[{{Laurikainen} {et~al.}(2005){Laurikainen}, {Salo}, \&
  {Buta}}]{Laurikainen05}
{Laurikainen}, E., {Salo}, H., \& {Buta}, R. 2005, \mnras, 362, 1319

\bibitem[{{Laurikainen} {et~al.}(2004){Laurikainen}, {Salo}, {Buta}, \&
  {Vasylyev}}]{Laurikainen04}
{Laurikainen}, E., {Salo}, H., {Buta}, R., \& {Vasylyev}, S. 2004, \mnras, 355,
  1251

\bibitem[{{Letawe} {et~al.}(2010){Letawe}, {Letawe}, \& {Magain}}]{Letawe10}
{Letawe}, Y., {Letawe}, G., \& {Magain}, P. 2010, \mnras, 403, 2088

\bibitem[{{Li} {et~al.}(2011){Li}, {Ho}, {Barth}, \& {Peng}}]{Li2011}
{Li}, Z.-Y., {Ho}, L.~C., {Barth}, A.~J., \& {Peng}, C.~Y. 2011, \apjs, 197, 22

\bibitem[{{Liu} {et~al.}(2012){Liu}, {Shen}, \& {Strauss}}]{Liu12}
{Liu}, X., {Shen}, Y., \& {Strauss}, M.~A. 2012, \apj, 745, 94

\bibitem[{{Magorrian} {et~al.}(1998){Magorrian}, {Tremaine}, {Richstone},
  {Bender}, {Bower}, {Dressler}, {Faber}, {Gebhardt}, {Green}, {Grillmair},
  {Kormendy}, \& {Lauer}}]{Magorrian98}
{Magorrian}, J., {Tremaine}, S., {Richstone}, D., {et~al.} 1998, \aj, 115, 2285

\bibitem[{{Mart{\'{\i}}nez-Sansigre} {et~al.}(2005){Mart{\'{\i}}nez-Sansigre},
  {Rawlings}, {Lacy}, {Fadda}, {Marleau}, {Simpson}, {Willott}, \&
  {Jarvis}}]{MartinezSansigre05}
{Mart{\'{\i}}nez-Sansigre}, A., {Rawlings}, S., {Lacy}, M., {et~al.} 2005,
  \nat, 436, 666

\bibitem[{{Martini}(2004)}]{Martini04}
{Martini}, P. 2004, Coevolution of Black Holes and Galaxies, 169

\bibitem[{{Martini} \& {Weinberg}(2001)}]{MartiniWeinberg01}
{Martini}, P., \& {Weinberg}, D.~H. 2001, \apj, 547, 12

\bibitem[{{Mechtley} {et~al.}(2016){Mechtley}, {Jahnke}, {Windhorst}, {Andrae},
  {Cisternas}, {Cohen}, {Hewlett}, {Koekemoer}, {Schramm}, {Schulze},
  {Silverman}, {Villforth}, {van der Wel}, \& {Wisotzki}}]{Mechtley16}
{Mechtley}, M., {Jahnke}, K., {Windhorst}, R.~A., {et~al.} 2016, \apj, 830, 156

\bibitem[{{Peng} {et~al.}(2002){Peng}, {Ho}, {Impey}, \& {Rix}}]{Peng02}
{Peng}, C.~Y., {Ho}, L.~C., {Impey}, C.~D., \& {Rix}, H.-W. 2002, \aj, 124, 266

\bibitem[{{Peng} {et~al.}(2010){Peng}, {Ho}, {Impey}, \& {Rix}}]{Peng10}
---. 2010, \aj, 139, 2097

\bibitem[{{Porciani} {et~al.}(2004){Porciani}, {Magliocchetti}, \&
  {Norberg}}]{Porciani04}
{Porciani}, C., {Magliocchetti}, M., \& {Norberg}, P. 2004, \mnras, 355, 1010

\bibitem[{{Ramos Almeida} {et~al.}(2011){Ramos Almeida}, {Tadhunter}, {Inskip},
  {Morganti}, {Holt}, \& {Dicken}}]{RamosAlmeida11}
{Ramos Almeida}, C., {Tadhunter}, C.~N., {Inskip}, K.~J., {et~al.} 2011,
  \mnras, 410, 1550

\bibitem[{{Ramos Almeida} {et~al.}(2012){Ramos Almeida}, {Bessiere},
  {Tadhunter}, {P{\'e}rez-Gonz{\'a}lez}, {Barro}, {Inskip}, {Morganti}, {Holt},
  \& {Dicken}}]{RamosAlmeida12}
{Ramos Almeida}, C., {Bessiere}, P.~S., {Tadhunter}, C.~N., {et~al.} 2012,
  \mnras, 419, 687

\bibitem[{{Reyes} {et~al.}(2008){Reyes}, {Zakamska}, {Strauss}, {Green},
  {Krolik}, {Shen}, {Richards}, {Anderson}, \& {Schneider}}]{Reyes08}
{Reyes}, R., {Zakamska}, N.~L., {Strauss}, M.~A., {et~al.} 2008, \aj, 136, 2373

\bibitem[{{Schawinski} {et~al.}(2012){Schawinski}, {Simmons}, {Urry},
  {Treister}, \& {Glikman}}]{Schawinski12}
{Schawinski}, K., {Simmons}, B.~D., {Urry}, C.~M., {Treister}, E., \&
  {Glikman}, E. 2012, \mnras, 425, L61

\bibitem[{{Schawinski} {et~al.}(2011){Schawinski}, {Treister}, {Urry},
  {Cardamone}, {Simmons}, \& {Yi}}]{Schawinski11}
{Schawinski}, K., {Treister}, E., {Urry}, C.~M., {et~al.} 2011, \apjl, 727, L31

\bibitem[{{Schmidt} \& {Green}(1983)}]{SchmidtGreen83}
{Schmidt}, M., \& {Green}, R.~F. 1983, \apj, 269, 352

\bibitem[{{S{\'e}rsic}(1968)}]{Sersic68}
{S{\'e}rsic}, J.~L. 1968, {Atlas de Galaxias Australes}

\bibitem[{{Shen} {et~al.}(2007){Shen}, {Mulchaey}, {Raychaudhury}, {Rasmussen},
  \& {Ponman}}]{Shen07}
{Shen}, Y., {Mulchaey}, J.~S., {Raychaudhury}, S., {Rasmussen}, J., \&
  {Ponman}, T.~J. 2007, \apjl, 654, L115

\bibitem[{{Shlosman} {et~al.}(1990){Shlosman}, {Begelman}, \&
  {Frank}}]{Shlosman90}
{Shlosman}, I., {Begelman}, M.~C., \& {Frank}, J. 1990, \nat, 345, 679

\bibitem[{{Somerville} {et~al.}(2008){Somerville}, {Hopkins}, {Cox},
  {Robertson}, \& {Hernquist}}]{Somerville08}
{Somerville}, R.~S., {Hopkins}, P.~F., {Cox}, T.~J., {Robertson}, B.~E., \&
  {Hernquist}, L. 2008, \mnras, 391, 481

\bibitem[{{Springel} {et~al.}(2005){Springel}, {Di Matteo}, \&
  {Hernquist}}]{Springel05}
{Springel}, V., {Di Matteo}, T., \& {Hernquist}, L. 2005, \apjl, 620, L79

\bibitem[{{Springel} \& {Hernquist}(2005)}]{SpringelHernquist05}
{Springel}, V., \& {Hernquist}, L. 2005, \apjl, 622, L9

\bibitem[{{Tadhunter} {et~al.}(2011){Tadhunter}, {Holt}, {Gonz{\'a}lez
  Delgado}, {Rodr{\'{\i}}guez Zaur{\'{\i}}n}, {Villar-Mart{\'{\i}}n},
  {Morganti}, {Emonts}, {Ramos Almeida}, \& {Inskip}}]{Tadhunter11}
{Tadhunter}, C., {Holt}, J., {Gonz{\'a}lez Delgado}, R., {et~al.} 2011, \mnras,
  412, 960

\bibitem[{{Treister} {et~al.}(2012){Treister}, {Schawinski}, {Urry}, \&
  {Simmons}}]{Treister12}
{Treister}, E., {Schawinski}, K., {Urry}, C.~M., \& {Simmons}, B.~D. 2012,
  \apjl, 758, L39

\bibitem[{{Urbano-Mayorgas} {et~al.}(2018){Urbano-Mayorgas}, {Villar
  Mart{\'{\i}}n}, {Buitrago}, {L{\'o}pez}, {Rodr{\'{\i}}guez del Pino},
  {Koekemoer}, {Huertas-Company}, {Dom{\'{\i}}nguez-Tenreiro}, {Carrera}, \&
  {Tadhunter}}]{UrbanoMayorgas18}
{Urbano-Mayorgas}, J.~J., {Villar Mart{\'{\i}}n}, M., {Buitrago}, F., {et~al.}
  2018, \mnras, arXiv:1810.11240

\bibitem[{{Urrutia} {et~al.}(2008){Urrutia}, {Lacy}, \& {Becker}}]{Urrutia08}
{Urrutia}, T., {Lacy}, M., \& {Becker}, R.~H. 2008, \apj, 674, 80

\bibitem[{{Veilleux} {et~al.}(2009){Veilleux}, {Kim}, {Rupke}, {Peng},
  {Tacconi}, {Genzel}, {Lutz}, {Sturm}, {Contursi}, {Schweitzer}, {Dasyra},
  {Ho}, {Sanders}, \& {Burkert}}]{Veilleux09}
{Veilleux}, S., {Kim}, D.-C., {Rupke}, D.~S.~N., {et~al.} 2009, \apj, 701, 587

\bibitem[{{Vika} {et~al.}(2015){Vika}, {Vulcani}, {Bamford}, {H{\"a}u{\ss}ler},
  \& {Rojas}}]{Vika15}
{Vika}, M., {Vulcani}, B., {Bamford}, S.~P., {H{\"a}u{\ss}ler}, B., \& {Rojas},
  A.~L. 2015, \aap, 577, A97

\bibitem[{{Villar-Mart{\'{\i}}n} {et~al.}(2012){Villar-Mart{\'{\i}}n}, {Cabrera
  Lavers}, {Bessiere}, {Tadhunter}, {Rose}, \& {de Breuck}}]{VillarMartin12}
{Villar-Mart{\'{\i}}n}, M., {Cabrera Lavers}, A., {Bessiere}, P., {et~al.}
  2012, \mnras, 423, 80

\bibitem[{{Villar-Mart{\'{\i}}n} {et~al.}(2011){Villar-Mart{\'{\i}}n},
  {Tadhunter}, {Humphrey}, {Encina}, {Delgado}, {Torres}, \&
  {Mart{\'{\i}}nez-Sansigre}}]{VillarMartin11}
{Villar-Mart{\'{\i}}n}, M., {Tadhunter}, C., {Humphrey}, A., {et~al.} 2011,
  \mnras, 416, 262

\bibitem[{{Villforth} {et~al.}(2014){Villforth}, {Hamann}, {Rosario},
  {Santini}, {McGrath}, {van der Wel}, {Chang}, {Guo}, {Dahlen}, {Bell},
  {Conselice}, {Croton}, {Dekel}, {Faber}, {Grogin}, {Hamilton}, {Hopkins},
  {Juneau}, {Kartaltepe}, {Kocevski}, {Koekemoer}, {Koo}, {Lotz}, {McIntosh},
  {Mozena}, {Somerville}, \& {Wild}}]{Villforth14}
{Villforth}, C., {Hamann}, F., {Rosario}, D.~J., {et~al.} 2014, \mnras, 439,
  3342

\bibitem[{{Villforth} {et~al.}(2017){Villforth}, {Hamilton}, {Pawlik},
  {Hewlett}, {Rowlands}, {Herbst}, {Shankar}, {Fontana}, {Hamann}, {Koekemoer},
  {Pforr}, {Trump}, \& {Wuyts}}]{Villforth17}
{Villforth}, C., {Hamilton}, T., {Pawlik}, M.~M., {et~al.} 2017, \mnras, 466,
  812

\bibitem[{{Wild} {et~al.}(2010){Wild}, {Heckman}, \& {Charlot}}]{Wild10}
{Wild}, V., {Heckman}, T., \& {Charlot}, S. 2010, \mnras, 405, 933

\bibitem[{{Wylezalek} {et~al.}(2016){Wylezalek}, {Zakamska}, {Liu}, \&
  {Obied}}]{Wylezalek16}
{Wylezalek}, D., {Zakamska}, N.~L., {Liu}, G., \& {Obied}, G. 2016, \mnras,
  457, 745

\bibitem[{{York} {et~al.}(2000){York}, {Adelman}, {Anderson}, {Anderson},
  {Annis}, {Bahcall}, {Bakken}, {Barkhouser}, {Bastian}, {Berman}, {Boroski},
  {Bracker}, {Briegel}, {Briggs}, {Brinkmann}, {Brunner}, {Burles}, {Carey},
  {Carr}, {Castander}, {Chen}, {Colestock}, {Connolly}, {Crocker}, {Csabai},
  {Czarapata}, {Davis}, {Doi}, {Dombeck}, {Eisenstein}, {Ellman}, {Elms},
  {Evans}, {Fan}, {Federwitz}, {Fiscelli}, {Friedman}, {Frieman}, {Fukugita},
  {Gillespie}, {Gunn}, {Gurbani}, {de Haas}, {Haldeman}, {Harris}, {Hayes},
  {Heckman}, {Hennessy}, {Hindsley}, {Holm}, {Holmgren}, {Huang}, {Hull},
  {Husby}, {Ichikawa}, {Ichikawa}, {Ivezi{\'c}}, {Kent}, {Kim}, {Kinney},
  {Klaene}, {Kleinman}, {Kleinman}, {Knapp}, {Korienek}, {Kron}, {Kunszt},
  {Lamb}, {Lee}, {Leger}, {Limmongkol}, {Lindenmeyer}, {Long}, {Loomis},
  {Loveday}, {Lucinio}, {Lupton}, {MacKinnon}, {Mannery}, {Mantsch}, {Margon},
  {McGehee}, {McKay}, {Meiksin}, {Merelli}, {Monet}, {Munn}, {Narayanan},
  {Nash}, {Neilsen}, {Neswold}, {Newberg}, {Nichol}, {Nicinski}, {Nonino},
  {Okada}, {Okamura}, {Ostriker}, {Owen}, {Pauls}, {Peoples}, {Peterson},
  {Petravick}, {Pier}, {Pope}, {Pordes}, {Prosapio}, {Rechenmacher}, {Quinn},
  {Richards}, {Richmond}, {Rivetta}, {Rockosi}, {Ruthmansdorfer}, {Sandford},
  {Schlegel}, {Schneider}, {Sekiguchi}, {Sergey}, {Shimasaku}, {Siegmund},
  {Smee}, {Smith}, {Snedden}, {Stone}, {Stoughton}, {Strauss}, {Stubbs},
  {SubbaRao}, {Szalay}, {Szapudi}, {Szokoly}, {Thakar}, {Tremonti}, {Tucker},
  {Uomoto}, {Vanden Berk}, {Vogeley}, {Waddell}, {Wang}, {Watanabe},
  {Weinberg}, {Yanny}, {Yasuda}, \& {SDSS Collaboration}}]{York00}
{York}, D.~G., {Adelman}, J., {Anderson}, Jr., J.~E., {et~al.} 2000, \aj, 120,
  1579

\bibitem[{{Yu} \& {Tremaine}(2002)}]{YuTremaine02}
{Yu}, Q., \& {Tremaine}, S. 2002, \mnras, 335, 965

\bibitem[{{Zakamska} {et~al.}(2003){Zakamska}, {Strauss}, {Krolik}, {Collinge},
  {Hall}, {Hao}, {Heckman}, {Ivezi{\'c}}, {Richards}, {Schlegel}, {Schneider},
  {Strateva}, {Vanden Berk}, {Anderson}, \& {Brinkmann}}]{Zakamska03}
{Zakamska}, N.~L., {Strauss}, M.~A., {Krolik}, J.~H., {et~al.} 2003, \aj, 126,
  2125

\bibitem[{{Zakamska} {et~al.}(2006){Zakamska}, {Strauss}, {Krolik}, {Ridgway},
  {Schmidt}, {Smith}, {Heckman}, {Schneider}, {Hao}, \&
  {Brinkmann}}]{Zakamska06}
---. 2006, \aj, 132, 1496

\end{thebibliography}

\label{lastpage}
\end{document}